\documentclass[fleqn,usenatbib]{mnras}
\usepackage[utf8]{inputenc} % Kodierung
\usepackage[T1]{fontenc}
\usepackage{ae,aecompl}
\usepackage{graphicx}
\usepackage{amsmath,amsfonts,amssymb,bm}
\usepackage{mathtools}
\usepackage{orcidlink}

\usepackage{txfonts}
\usepackage{color, soul}
\sethlcolor{yellow}
\soulregister\cite7
\soulregister\citep7
\soulregister\ref7
\soulregister\eqref7
\newcommand{\highlighttext}[1] {#1}
  
\usepackage{pgfplots}
\usepgfplotslibrary{patchplots}
\pgfplotsset{compat=1.15}

\usepackage{float}
\usepackage{adjustbox}

\def\spirou#1{{\bf}}

\newcommand{\ltsima}{$\; \buildrel < \over \sim \;$}
\newcommand{\lsim}{\lower.47ex\hbox{\ltsima}}
\newcommand{\gtsima}{$\; \buildrel > \over \sim \;$}
\newcommand{\gsim}{\lower.47ex\hbox{\gtsima}}

\newcommand{\ci}{\mathrm{i}}

\title[lensing and fuzzy dark matter]
{A weak lensing perspective on nonlinear structure formation with fuzzy dark matter}
\author[A. Kunkel, T. Chiueh, B.M. Sch{\"a}fer]
{Alexander Kunkel$^{1,3}$\thanks{e-mail: alexanderkunkel@ntu.edu.tw}, Tzihong Chiueh$^{2,3,4}$\orcidlink{0000-0003-2654-8763}, Bj{\"o}rn Malte Sch{\"a}fer$^1$\thanks{e-mail: bjoern.malte.schaefer@uni-heidelberg.de}\\
$^1$Zentrum f{\"u}r Astronomie der Universit{\"a}t Heidelberg, Astronomisches Rechen-Institut, Philosophenweg 12, 69120 Heidelberg, Germany\\
$^2$Department of Physics at National Taiwan University, Taipei 10617, Taiwan\\
$^3$Institute of Astrophysics at National Taiwan University, Taipei 10617, Taiwan\\
$^4$Center for Theoretical Physics at National Taiwan University, Taipei 10617, Taiwan}

\begin{document}
\pagerange{\pageref{firstpage}--\pageref{lastpage}}
\pubyear{2022}
\maketitle
\label{firstpage}

% --- abstract --- %
\begin{abstract}
We investigate nonlinear structure formation in the fuzzy dark matter (FDM) model in comparison to cold dark matter (CDM) models from a weak lensing perspective using perturbative methods. We use Eulerian perturbation theory (PT) up to fourth order to compute the tree-level matter trispectrum and the one-loop matter spectrum and bispectrum from consistently chosen initial conditions. \highlighttext{In addition, we predict the non-linear matter power spectra using $N$-body simulations with CDM and FDM initial conditions.} We go on to derive the respective lensing spectra, bispectra and trispectra in CDM and FDM in the context of a Euclid-like weak lensing survey. Finally, we compute the attainable cumulative signal-to-noise ratios and an estimate of the attainable $\chi^2$-functionals for distinguishing FDM from CDM at particle masses $m=10^{-21}$~eV, $m = 10^{-22}$~eV and $m = 10^{-23}$~eV. \highlighttext{We find that PT predictions cannot be used to reliably distinguish the three models in a weak lensing survey. Assuming that $N$-body simulations overestimate the late-time small-scale power in the FDM model, future weak lensing survey might be used to distinguish between the FDM and CDM cases up to a mass of $m = 10^{-23}$ eV. However, observations probing the local high-$z$ universe are probably more suited to constrain the FDM mass.}
\end{abstract}

% --- keywords --- %

% --- section:  --- %
\section{introduction}
The Fuzzy Dark Matter (FDM) model first proposed by \cite{Hu2000} describes dark matter as a bosonic, scalar field composed of very light particles with typical masses $m \sim 10^{-22}$ eV  with negligible self interactions and a macroscopic de Broglie wavelength. The dynamics of FDM in the non-relativistic limit are governed by the Schrödinger-Poisson system (SPS) of equations. Because of its small mass, the FDM boson has an astrophysically relevant de Broglie wavelength at the order of a few kiloparsecs. Its wave-like behaviour suppresses structure formation on small scales while one recovers CDM behaviour on large scales; as such FDM constitutes a small-scale modification of CDM, governed by the particle mass. Therefore, the FDM model has the potential to solve the small-scale crisis of CDM. 

\highlighttext{The FDM bosons are also called axions or axion-like particles. Axion-like particles with exponentially suppressed masses are naturally generated in supersymmetric theories and theories with extra-dimensions including string theory \citep{Marsh2016}.}
A currently popular FDM model is one where all of dark matter is composed of axions with $m \approx 10^{-22}$ eV \citep{Hu2000, Marsh2016}. This mass has also been suggested by \cite{Schive2014} who fitted the ground state density of a halo obtained in a numerical simulation of FDM to the mass distribution of the Fornax dwarf spheroidal galaxy. \highlighttext{However, the allowed range of FDM masses that could solve the small-scale crisis of CDM is increasingly narrowed down by a number of different observations:

%Observations of the \textit{Lyman-$\alpha$ forest} have been used to derive strong bounds on the FDM mass. The Lyman-$\alpha$ forest iss sensitive to the power on FDM scales and by suitable modelling, the spatial fluctuations can be turned into statements about the dark matter power spectrum \citep{Hui2021}. 
\cite{Kobayashi2017, Irsic2017, Armengaud2017} constrained the FDM mass using measurements of the \textit{Lyman-$\alpha$ forest} flux spectrum from the XQ-100, HIRES/MIKE, SDSS/BOSS and XSHOOTER data sets and derive bounds between $m > 7.1 \times 10^{-22}$ eV and $m > 2.3 \times 10^{-21}$ eV. The biggest uncertainties in their approach stem from the modelling of the observed fluctuations in the neutral hydrogen and the use of hydrodynamical CDM simulations using FDM initial conditions (IC), thereby neglecting FDM dynamics at late times. \cite{Nori2018a} improved on the simulation side by including the effect of the quantum pressure term in their $N$-body simulations and derive the bound $m > 2.1 \times 10^{-21}$ eV. \cite{Rogers2021} found a stronger bound of $2\times 10^{-20}$ eV at 95\% confidence using a new, robust $N$-body-based modelling pipeline marginalising over a IGM model which allows for a wide range of heating and ionization histories. The fact that the different groups' results are consistent indicates that the effect of late-time FDM dynamics on the Lyman-$\alpha$ flux power spectrum may be negligible.  At the same time, \cite{Zhang2019} carefully examined the Lyman-$\alpha$ forest constraints on FDM and concluded that full FDM simulation are required for reliable constraints based on the Lyman-$\alpha$ forest. They also pointed out that the simulation uncertainties of hydrodynamic simulations may have been underestimated in previous studies. They nonetheless find that FDM with masses smaller than $m = 2.5 \times 10^{-23} eV$ can be rejected based on the Lyman-$\alpha$ forest.}

\highlighttext{Order unity density fluctuations as a result of wave interference in FDM halos are a robust and well-understood prediction of the FDM model. They are expected to dynamically heat stellar orbits in \textit{ultra-faint dwarf galaxies (UFDs)}. \cite{Marsh2018} compute the gravitational heating of a star cluster in the UFD Eridanus II and find that the range $0.8 \times 10^{-21}$ eV $ < m < 10^{-19}$ eV is disfavoured. Further, they find that the formation of Eridanus II as a subhalo demands $m \gtrsim 0.8 \times 10^{-21}$ eV. \cite{Schive2020} use FDM simulations to demonstrate that the random walk which the central soliton in the FDM halo undergoes represents an even bigger challenge for the survival of the central star cluster in Eridanus II. As a possible solution, they argue that the star cluster could possibly survive the soliton random walk if the FDM subhalo had been tidally stripped by the influence of the Milky way's gravitational potential. 

%Using $N$-body simulations of test particles in an approximated FDM halo potential, 
\cite{Dalal2022} derive the bound $m > 3 \times 10^{-19}$ eV by showing that the dynamic stellar heating predicted by FDM is not consistent with the observed stellar velocity dispersion in the UFDs Segue 1 and Segue 2. \cite{Amorisco2018} study the dynamical heating of thin stellar streams through the density granulation in FDM halos and find the limit $m > 1.5 \times 10^{-22}$ eV. In a similar spirit, \cite{Church2019} study the dynamical heating of galactic disks in CDM and FDM and find the lower bound $m > 0.6 \times 10^{-22}$ eV.} 

\highlighttext{The \textit{inner density profile of dwarf galaxies} provides another test bed for FDM predictions.  \cite{Chen2016, Calabrese2016} compare the flat, inner density profile of Milky way dwarf satellites to the inner solitonic profile of FDM halos and find $m \approx 1 - 6 \times 10^{-22}$ eV. \cite{Safarzadeh_2020} point out that while the FDM prediction for $m \gtrsim 10^{-21}$ eV matches the profiles of the known UFD Milky Way satellites, the density profiles of the dwarf spheroidals Fornax and Sculptor originally considered by \cite{Schive2014} disagrees with the FDM prediction. In turn, a mass around $10^{-22}$ eV seems to suggest that the mass of the UFDs is too large. In other words, no single FDM mass seems to be able to accomodate for the density profiles of all known dwarf galaxies.}

\highlighttext{Another constraint on the FDM mass can be derived from the \textit{circular velocity in disk galaxies}. Assuming the soliton-host halo mass relation empirically found by \cite{Schive2014}, \cite{Bar2018} argue that the peak circular velocity characterising the host halo on large scales should repeat itself in the central region and rule out FDM masses in the range $m = 10^{-21} - 10^{-22}$ eV. The biggest uncertainty in their approach stems from the empirical soliton host-halo relation that has only been studied numerically for a limited range of FDM and halo masses.}

\highlighttext{Using the \textit{subhalo mass function} inferred from the observed luminosity function of Milky Way Satellites, \cite{Nadler2021} constrain the warm dark matter, self-interacting dark matter and fuzzy dark matter models alike and obtain the bound $m > 2.9 \times 10^{-21}$ eV at $95$ \% on the FDM mass. Uncertainty comes from the prediction of halo mass function from the linear power spectrum. A further mass constraint based on early halo formation is given by \cite{Lidz2018} who use the \textit{hyperfine $21$ cm transition} of hydrogen at redshift around $z = 15$-$20$ to constrain the FDM mass to $m > 5\times 10^{-21}$ eV.} 

\highlighttext{Future surveys of the \textit{$21$-cm power spectrum} such as HERA are expected to be sensitive to FDM masses up to $10^{-18}$ eV if all dark matter is fuzzy as well as to very small FDM fractions around 1 \% in the mass range $10^{-25} $ eV $ \lesssim m \lesssim 10^{-23}$ eV \citep{Hotinli2022}. }

\highlighttext{\textit{Superradiance} rules out FDM masses in the mass range from $m = 10^{-21} - 10^{-17}$ eV \citep{Uenal2021}. \cite{Davies2020} study solitons with supermassive black holes at their center and exclude $10^{-22.12}$ eV $\lesssim m \lesssim 10^{-22.06}$ eV. \cite{GonzalezMorales2017} find the upper bound of $m<0.4\times 10^{-22}$~eV ($97.5 \%$ C.L.) by fitting the luminosity-averaged \textit{velocity dispersion of dwarf spheroidal galaxies}; \cite{Sarkar2021} use the Lyman-$\alpha$ effective opacity to find $m>0.2 \times 10^{-22}$ eV ($1 \sigma$) and \cite{Maleki2019} find $m>0.7\times 10^{-22}$ eV by studying the \textit{$X$-ray emissions during solitons mergers}.}

\highlighttext{\textit{Gravitational lensing} provides yet another complementary probe in addition to the above constraints. \cite{Powell2023} study the granular structure in the main dark matter halo of a single gravitational lens system and derive the bound $m > 4.4 \times 10^{-21}$ eV. For lower masses, \cite{Hlozek2018} identify the possible mass range for FDM via CMB lensing based on the full Planck data set and find no evidence for an FDM component in the mass range $10^{-33} \mathrm{eV} \leq m \leq 10^{-24}$ eV. 
More recently, \cite{Rogers2023} combine the Planck CMB data and galaxy power spectrum and bispectrum data from the BOSS survey to conclude that fuzzy dark matter with $m \leq 10^{-26}$ eV makes up less than $10$ \% and fuzzy dark matter with $10^{-30}$ eV $\leq m \leq 10^{-28}$ eV makes up less than $1$ \% of dark matter today.}

\highlighttext{To sum up, essentially the entire mass range from $m = 10^{-26}$ eV to $m = 10^{-16}$ eV is under tension with a number of different observations. The  $m \approx 10^{-22}$ model is particularly constrained by the Lyman-$\alpha$ forest, galactic rotation curves, dynamical heating arguments, the subhalo mass function and strong lensing.}

\indent In this work, we estimate whether future \textit{weak lensing surveys} have the constraining power to further complement some of the existing constraints on the FDM mass. Weak lensing surveys such as Euclid measure the lensing shear spectrum and allow to infer the shape and amplitude of the matter spectrum $P_{\delta}$, thereby giving constraints on cosmological parameters such as the axion mass. Gravitational lensing probes the matter distribution without any biasing assumption. This is important because the preferred value of the galaxy bias $b$ and therefore the normalisation of the galaxy spectrum are different for axion cosmologies than for $\Lambda$CDM. As a consequence, the constraining power of galaxy surveys on the FDM mass is lower than that of weak lensing.
\newline
\indent 
In previous weak lensing studies on FDM, \cite{Marsh2011} analysed whether adding a small fraction of axions of mass in the range $m = 10^{-29}$ eV would be detectable via the lensing convergence spectrum. For modelling nonlinearities, they neglected the quantum pressure and employed the CDM halofit model implemented in the \textsc{CAMB} code \citep{CAMB}. More recently, \cite{Dentler2021} combined CMB Planck data with shear correlation data from the Dark Energy Survey year 1 to find a $95$\% C.L. lower limit $m > 10^{-23}$ eV. They modelled the nonlinear FDM spectra using the adapted halo model \textsc{HMCODE}. Lensing can also serve as a tool to investigate axion-related isocurvature fluctuations \citep{Feix:2019lpo, Feix:2020txt}, with a sensitivity to axion masses at the scale $10^{-19}$~eV. In contrast, we estimate the spectrum, bispectrum and trispectrum using fluid perturbation theory. To this end, we employ a recasting of SPS into a hydrodynamical form using the so-called \textit{Madelung transform}. The Madelung equations are Euler-Poisson equations with an additional scale-dependent modification, the so-called quantum pressure term. On large scales, this term vanishes and one recovers the ideal fluid equations for CDM. One of the appeals of the Madelung transform lies in being able to apply standard cosmological perturbation theory to the SPS. Its range of validity is higher than that of wave perturbation theory and one can easily contrast it with perturbation theory in CDM \citep{Woo2009, Li2018}. Whether the inherently higher degree of small-scale anisotropy in FDM \citep{Dome2022} can differentiate between dark matter models remains to be investigated.
\newline
\indent 
Building on and extending the work of \cite{Li2018} who computed the one-loop matter spectrum for FDM, we use time-dependent nonlinear Eulerian perturbation theory up to fourth order to compute the tree- and loop-level bispectrum as well as the tree-level trispectrum. \highlighttext{To estimate the range of validity of our results, we compare the perturbation theory predictions with a set of $N$-body simulations with CDM and FDM initial conditions. } We go on to derive the corresponding lensing spectra for a Euclid-like lensing survey. We estimate the attainable signal-to-noise ratios as well as the $\chi^2$-functional for distinguishing axions of the masses $m = 10^{-21}$~eV, $m = 10^{-22}$~eV and $m = 10^{-23}$~eV from standard CDM. \highlighttext{These masses are interesting because they cover the range of masses that could possibly solve the small-scale crisis of CDM}.
\newline
\indent 
The outline of this paper is as follows: Section~\ref{sec:NonlinearStructureFormation} introduces the SPS, presents linear and nonlinear Eulerian perturbation theory for FDM and computes the matter spectra, bispectra and trispectra. Section~\ref{sec:WeakLensingPredictions} introduces the basics of weak lensing surveys, computes the respective lensing spectra and estimates the attainable weak lensing signal-to-noise ratios and $\chi^2$-functionals. Section~\ref{sect_summary} summarises and discusses the results. Appendices~\ref{appendix:NLEPT} through~\ref{appendix:InfraredSafeIntegrands} list the technical and numerical details of the FDM perturbation theory. Appendix~\ref{appendix:NBody} lists the details of the $N$-body simulations.

% --- subsection:  --- %
\section{nonlinear structure formation with fuzzy dark matter}
\label{sec:NonlinearStructureFormation}
Consider a (pseudo-)scalar field $\varphi$ minimally coupled to gravity:
\begin{equation}
\label{eq:AxionAction}
    S = \frac{1}{\hbar c^2} \int \mathrm{d}^4x \sqrt{-g} \left[\frac{1}{2} g^{\mu\nu} \nabla_\mu\varphi\nabla_\nu\varphi -\frac{1}{2} \frac{m^2 c^2}{\hbar^2} \varphi^2\right]
\end{equation} 
where we follow the convention in \cite{Hui2017}. $m$ is the axion mass, which naturally appears in the Compton-scale $\lambda = \hbar / (mc)$. This action is invariant under parity- and time-inversion because it is quadratic in $\varphi$. For QCD axions, this action is applicable after symmetry breaking and after non-perturbative effects have been switched on. Further, we neglect possible non-gravitational self-interactions of the axion in the form of an axion potential $V(\varphi)$. In our discussion, it will not be important that the FDM particle is actually an axion or an axion-like particle. We only assume that the particle is bosonic, non-relativistic, has negligible self-interaction and makes up the entirety of dark matter. We can study solutions of the relativistic axion equation of motion for a perturbed Friedmann-Lema{\^i}tre-Robertson-Walker (FLRW) metric including a cosmological constant $\Lambda$, all in Newtonian gauge. This is appropriate for studying structure formation because the virial velocity in a typical galaxy $\upsilon_\mathrm{vir} \sim 100 \frac{km}{s} \ll c$ and galaxies are much smaller than the Hubble horizon. On the lower scale end, we are concerned with scales above the axion Compton wave length which would correspond to relativistic scales in the Klein-Gordon equation. Except in the vicinity of black holes, the Newtonian potential $\Phi$ obeys $\left|\Phi\right|/c^2 \ll 1$. To study the clustering of axions on nonlinear scales, we further take the WKB-approximation of the form
\begin{equation}
    \varphi = \sqrt{\frac{\hbar^3c}{2m}}(\psi e^{-\frac{i m c^2 t}{\hbar}} + \psi^*e^{\frac{i m c^2t}{\hbar}})
\end{equation}
where $\psi$ is a complex scalar field because axions that cluster on galactic scales began oscillating in the very early universe. We apply the previous considerations by taking $\Phi\sim \epsilon^2$, $k/m \sim \epsilon$ and $H/m \sim \epsilon$ and work to order $\mathcal{O}(\epsilon^2)$. Further, we assume the non-relativistic approximation $|\dot{\psi}| \ll \frac{mc^2}{\hbar} |\psi|$. The assumption $\partial_t \ll m$ is non-relativistic because we have $\partial_t \sim \Delta/m \sim k^2/m$ and therefore $k^2/m \ll m$. With these simplifications, we obtain the comoving Schrödinger-Poisson system of equations
\begin{align}
    i\hbar\left(\partial_t \psi(\bm{x}, t) + \frac{3}{2} H \psi(\bm{x}, t)\right) &= \left(-\frac{\hbar^2}{2m a^2}\Delta + m \Phi(\bm{x}, t)\right) \psi(\bm{x}, t), \label{eq:ComovingSchroedinger}\\
    \Delta \Phi(\bm{x}, t) &= 4\pi G a^2 (|\psi(\bm{x}, t)|^2 - \rho_b(t)) \label{eq:ComovingPoisson}
\end{align}
where $\rho_b(t)$ is the background density and $|\psi|^2$ measures the density in a proper volume.
We supplement the Schrödinger-Poisson equations with the normalisation condition
\begin{equation}
    \int\mathrm{d}^3x\: \rho(\bm{x}, t)  = N m
\end{equation}
which fixes the density $\rho = |\psi|^2$ for $N$ axions of mass $m$. In Eq. \eqref{eq:ComovingSchroedinger}, positions of particles are described using comoving coordinates $\bm{x}$.

% --- subsection: quantum pressure and Jeans scale --- %
\subsection{quantum pressure and Jeans scale}
\label{sec:QPandJeans}
\begin{figure}
\centering
\includegraphics[width=0.47\textwidth]{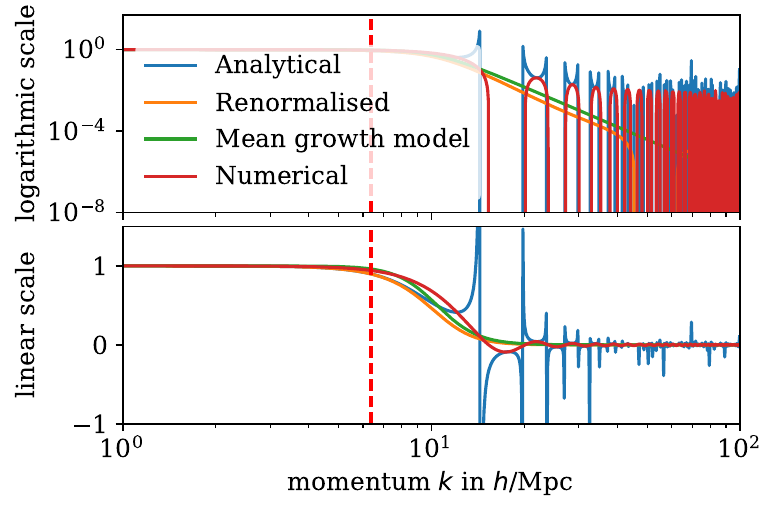}
\caption{Asymptotic behaviour of growing modes represented via the quotient of FDM and CDM growth factors $D_{+,FDM}(k, a)/D_{+, CDM}(a)$ in a FDM-dominated EdS universe for $m = 10^{-23}$ eV. The graphs show the analytical, divergent expression $\eqref{eq:FDMLinearGrowthASolution}$, the renormalised expression using the fifth-order Taylor expansion, the numerical solution obtained by integrating the linear growth equation with initial conditions given by Eq. \eqref{eq:CorrectIC} as well as a mean growth model $D(k, a) \approx \Bigl(1 + \alpha \left(\frac{k}{k_J}\right)^{\beta}\Bigl) D_{\mathrm{CDM}}(a)$ with the fit parameters $\alpha = 0.17$, $\beta = 6.50$ obtained via fitting to the analytical solution.
}
\label{fig:FDMSuppression}
\end{figure}

The Madelung transform \citep{Madelung1927} follows by substituting
\begin{equation}
\label{eq:MadelungPsi}
    \psi(\bm{x}, t) =: \sqrt{\frac{\rho(\bm{x}, t)}{m}} e^{\ci S(\bm{x}, t)}
\end{equation}
for real fields $\rho(x, t)$ and $S(x, t)$ into Eq. \eqref{eq:ComovingSchroedinger} and defining the velocity field
\begin{equation}
\label{eq:MadelungV}
    \bm{\upsilon} = \frac{\hbar}{m a}\nabla S = \frac{\ci \hbar}{2 a m|\psi|^2}(\psi \nabla \psi^* - \psi^*\nabla\psi),
\end{equation}
we obtain the Madelung equations
\begin{equation}
\label{eq:MadelungEquations}
\begin{split}
    \partial_t  \rho + 3 H \rho + \frac{1}{a} \nabla \cdot (\rho \bm{\upsilon})  &= 0,\\
    \partial_t \bm{\upsilon} + H \bm{\upsilon} + \frac{1}{a}(\bm{\upsilon} \cdot \nabla)\bm{\upsilon} + \frac{1}{a} \nabla \Phi+ \frac{1}{a^3} \nabla Q_P &= 0, \\
    \Delta \Phi(\bm{x}, t) - 4\pi G a^{2}\rho_b(t) \delta(\bm{x}, t)  &= 0.
\end{split}
\end{equation}
The Madelung equations describe the Schrödinger equation via a system of fluid equations for frictionless, compressible flow in an external gravitational potential $\Phi$. The flow gets modified by the quantum pressure $Q_P$ with
\begin{equation}
    \label{eq:QuantumPressure}
   \frac{2m^2}{\hbar^2}  Q_P = - \frac{\Delta \sqrt{\rho}}{\sqrt{\rho}} = -\frac{1}{2} \Delta \log{\rho} - \frac{1}{4}(\nabla \log{\rho})^2.
\end{equation}
The quantum pressure accounts for the underlying wave dynamics in FDM. For a narrowly located source, the quantum pressure is large and reflects the Heisenberg uncertainty principle in quantum mechanics  \footnote{Note, however, that in the context of axion cosmology the quantum pressure follows from the non-relativistic equation of motion of a classical field and is therefore not a quantum mechanical result.}. On large scales, the Madelung equations reduce to the Euler equations of a pressureless fluid and we recover the dynamics of standard cold dark matter. The quantum pressure $Q_P$ is equivalent to an anisotropic pressure stress $P_{ij}$ where
\begin{equation}
    \frac{1}{m} \nabla Q_P = \frac{a^2}{\rho} \partial_j P_{ij} = \left(\frac{a^2}{\rho} \partial_j -\frac{\hbar^4}{2m^2a^2} \rho \partial_i \partial_j\right) \log(\rho).
\end{equation}
We can describe the linearised evolution of the density contrast by assuming $\delta \ll 1$ and $\left|\bm{\upsilon}\right| \ll 1$ for the fluctuation fields and neglecting higher-order perturbations of the form $\mathcal{O}(\delta^2, \upsilon \delta, \upsilon^2)$ to obtain
\begin{equation}
    \label{eq:FDMLinearGrowthFourierSpace}
    \ddot{\delta_k} + 2H\dot{\delta_k} - \left(4\pi G\rho_b - \frac{\hbar^2k^4}{4 m^2 a^4}\right) \delta_k = 0.
\end{equation}
For each mode $k$, Eq. \eqref{eq:FDMLinearGrowthFourierSpace} describes a harmonic oscillator with time-dependent dampening $H(t)$ and frequency $\omega(k, t)$ as
\begin{equation}
    \omega(k, t) = \sqrt{\frac{\hbar^2k^4}{4m^2a^4} - \frac{4\pi G \rho_b}{a^3}}.
\end{equation}
Unlike in the CDM case, linear growth in FDM is scale-dependent because of the quantum pressure term. The condition $\omega = 0$ defines the comoving quantum Jeans scale $k_J$ \citep[see][]{10.1093/mnras/215.4.575}:
\begin{equation}
\begin{split}
\label{eq:ComovingQuantumJeansScale}
    k_J &= \left(\frac{16\pi G m^2 \rho_b a}{\hbar^2}\right)^{\frac{1}{4}}\\
    &= 44.7\,\mathrm{Mpc}^{-1} \left(6 a \frac{\Omega_{m, 0}}{0.3}\right)^{\frac{1}{4}} \left(\frac{H_0}{70  \frac{\mathrm{km}}{\mathrm{s}}\frac{1}{\mathrm{Mpc}}} \frac{m}{10^{-22}\mathrm{eV}}\right)^{\frac{1}{2}}.
\end{split}
\end{equation}
The Jeans scale describes a force balance between gravity and quantum pressure. For $k < k_J$, i.e. scales larger than $\lambda_J$, $\omega(k, t)$ becomes imaginary and we recover a growing and a decaying mode just like for CDM. Perturbations on these scales are unstable and will gravitationally collapse. For $k > k_J$, i.e. scales smaller than $\lambda_J$, the frequency $\omega(k, t)$ becomes real. Perturbations on small scales therefore undergo oscillations and are stabilised against gravitational collapse. The comoving Jeans wavelength decreases with time as $\lambda_J \sim a^{-\frac{1}{4}}$ which is why more small-scale features can develop as the universe expands.
We can find an analytical solution of the linear FDM growth equation \eqref{eq:FDMLinearGrowthFourierSpace} in the Einstein-de Sitter case for $\Omega_m = 1$. To this end, we substitute the ansatz $\delta(\bm{x}, a) = \delta(\bm{x}, a=a_0) D(a, a_0)$ into Eq. \eqref{eq:FDMLinearGrowthFourierSpace} and rewrite the resulting ODE for the growth factor $D$ in terms of the scale factor $a$
\begin{equation}
\begin{split}
\label{eq:GeneralFDMGrowthEquation}
    D''(a) &+ \frac{1}{a} \left(3 + \frac{\mathrm{d} \ln H}{\mathrm{d} \ln a}\right) D'(a) \\
    &- \frac{1}{a^2 H(a)^2} \left(\frac{3}{2} \frac{\Omega_{m,0} H_0^2}{a^3} - \frac{\hbar^2 k^4}{2m^2 a^4}\right) D(a) = 0.
\end{split}
\end{equation}
Simplifying with $\Omega_{m, 0} = 1$ yields
\begin{equation}
    \label{eq:FDMLinearGrowthScaleFactor}
    D''(a) + \frac{3}{2a} D'(a) - \left(\frac{3}{2a^2} -  \frac{\hbar^2k^4}{4m^2 a^3 H_0^2}\right) D(a) = 0.
\end{equation}
\cite{Chavanis2011b} and \cite{Chavanis2015} study solutions of Eq. \eqref{eq:FDMLinearGrowthScaleFactor} in terms of Bessel functions in the context of Bose-Einstein condensate dark matter with self-interaction and the weak-field limit of the Klein-Gordon-Einstein equation. They find 
\begin{equation}
    \label{eq:FDMLinearGrowthScaleFactorProportional}
     D_{\pm}(k, a) \propto \frac{1}{a^{\frac{1}{4}}} J_{\mp5/2}\left(\frac{\hbar k^2}{m H_0 \sqrt{a}}\right)
\end{equation}
where $J_{\mp 5/2}$ are cylindrical Bessel functions of fractional order
\begin{align}
    J_{-5/2} = \sqrt{\frac{2}{\pi z}} \left(\frac{3 \cos z}{z^2} + \frac{3 \sin z}{z} - \cos z\right),\\
    \qquad J_{+5/2} = \sqrt{\frac{2}{\pi z}} \left(\frac{3 \sin z}{z^2} - \frac{3 \cos z}{z} - \sin z\right).
\end{align}
\cite{Li2018} give a solution of Eq. \eqref{eq:FDMLinearGrowthScaleFactor} with $D_{\pm}(k, a_0, a_0) = 1$ in terms of the solutions Eq. \eqref{eq:FDMLinearGrowthScaleFactorProportional} as
\begin{align}
\label{eq:FDMLinearGrowthASolution}
    D_{\pm}(k, a, a_0) &= D_{\pm}(k, a) / D_{\pm}(k, a_0).
\end{align}
We immediately notice that Eq. \eqref{eq:FDMLinearGrowthASolution} can exhibit unphysical divergences at the roots of the Bessel function. \cite{Lague2020} argue that this is simply a matter of choice of normalisation, but this is not quite true. Eq. \eqref{eq:FDMLinearGrowthASolution} is not a solution of the linear growth equation \eqref{eq:FDMLinearGrowthScaleFactor} if it diverges in the domain of interest. We are not aware of a general analytical solution to Eq. \eqref{eq:FDMLinearGrowthScaleFactor}, even in the EdS case. \cite{Lague2020} suggest two ways to remedy this issue:
One way is to approximate the denominator of Eq. \eqref{eq:FDMLinearGrowthASolution} by a fifth-order Taylor expansion of the Bessel function which removes divergences from Eq. \eqref{eq:FDMLinearGrowthASolution} and gives fewer oscillations. In contrast, a different number of terms in the Taylor expansion leads to fast oscillations and/or divergences.
An alternative that \cite{Lague2020} adopt is to use a model for the mean growth of the FDM growth function. They describe the growing mode $D_+$ in terms of a smoothed Heaviside step function where free parameters are determined via a fit to the \textsc{axionCAMB} transfer function \citep{axionCAMB}. For large scales, this approach exactly recovers the CDM growth function, but has the disadvantage that oscillations are entirely neglected. Moreover, the smoothed Heaviside step function falls off exponentially for large $k$ which is not a correct description of the asymptotic behaviour of $D_+$. 
We therefore opt for integrating Eq. \eqref{eq:FDMLinearGrowthScaleFactor} numerically. As initial conditions, we choose
\begin{equation}
\label{eq:CorrectIC}
    D(a_0) = 1, \qquad D'(a_0) = D_{\mathrm{CDM}}'(a_0)
\end{equation}
where $D_{\mathrm{CDM}}'(a_0)$ is obtained via numerical integration of the linear growth equation \eqref{eq:GeneralFDMGrowthEquation} for $\hbar = 0$ corresponding to CDM. This approach has several advantages: We ensure $D(k, a_0, a_0) = 1$ via the initial conditions and obtain the correct CDM evolution in a general cosmology in the limit $k \rightarrow 0$. Further and most importantly, the growth factor obtained in this way is in fact a solution to the linear growth equation \eqref{eq:FDMLinearGrowthScaleFactor}. It does not exhibit unphysical divergences as shown in Fig. \ref{fig:FDMSuppression}. We see that the numerical solution of the growth equation also exhibits the correct asymptotic behaviour $\lim_{k\rightarrow \infty} \frac{D_{\mathrm{FDM}}(k, a)}{D_{\mathrm{CDM}}(a)} = 0.$

\begin{figure}
\centering
\includegraphics[width=.47\textwidth]{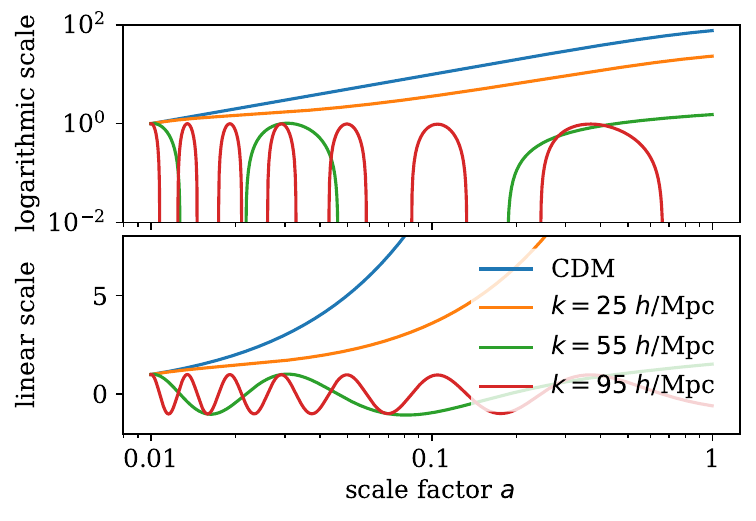}
\caption{Growing modes $D_+(a)$ in CDM and $D_+(k, a)$ in FDM in fiducial cosmology obtained by numerical integration of Eq. \eqref{eq:FDMLinearGrowthScaleFactor} for $\hbar = 0$ (CDM) and a mass of $m = 10^{-22}$ eV at three different scales $k$ (FDM). Growth factors are normalised to $D_{+}(a_0) = 1$ at $a_0=0.01$.}
\label{fig:GrowingModes}
\end{figure}

In addition, the numerical solution captures the oscillatory behaviour of the analytical solution for high $k$. At the same time, there are also several disadvantages to numerically integrating the linear growth equation in FDM. Firstly, we do not capture the growing mode in the oscillating regime. This is because we do not know the correct initial conditions for the growing mode. Therefore, the numerical solution in the oscillating regime will in general be a linear combination of the two modes $D_+$ and $D_-$. However, since the correct initial conditions for the linear fluctuation fields are unknown as well, this does not add any uncertainty to the linear growth model in terms of initial conditions. In any case, we recover the correct modes for $a > a_{osc}$. This is because any component proportional to $D_-$ in the initial conditions quickly decays away for $a > a_{osc}$. Fig. \ref{fig:GrowingModes} shows the growing solutions obtained for a fiducial cosmology with $\Omega_{m, 0} = 0.3159$ and a dark energy equation of state $w = -0.9$ used in the rest of this work.
 
To sum up, the key difference between CDM and FDM in linear perturbation theory is the existence of a unique length scale in FDM. Whereas all scales are gravitationally unstable in CDM, the perturbations below the quantum Jeans scale are stabilised in FDM. Nonetheless, nonlinear perturbation theory may alter these conclusions since the quantum Jeans scale is a concept only valid within linear perturbation theory. This can be seen by taking the first few terms of the Taylor expansion of the quantum pressure Eq. \eqref{eq:QuantumPressure}: $\frac{2m^2}{\hbar^2}  \Delta Q_P = - \Delta \frac{\Delta \sqrt{\rho}}{\sqrt{\rho}} = \Delta\left(- \Delta \delta + \frac{1}{4} \Delta \delta^2 + \frac{1}{2} \delta \Delta \delta + \mathcal{O}(\delta^2) \right)$. The linear contribution $- \Delta^2 \delta$ counteracts gravity. However, the quadratic terms acts in the same direction as gravity and could therefore potentially enhance gravitational collapse \citep{Li2018}. This is related to the fact that the interference of waves can lead to structures that are smaller than their wavelength. In order to estimate the effect of nonlinearities on cosmological structure formation, we develop nonlinear perturbation theory for CDM and FDM in the next section.

% --- subsection:  --- %
\subsection{tree- and loop level perturbation theory}
The goal of cosmological perturbation theory is to describe the departure of matter evolution from the homogeneous Hubble expansion perturbatively. In Eulerian perturbation theory (PT), one describes the nonlinear gravitational dynamics in terms of solutions $\delta^{(1)}$ and $\bm{\upsilon}^{(1)}$ of the linearised fluid equations in a fixed laboratory frame:
\begin{equation}
    \delta(\bm{x}, a) = \sum_{n = 1}^\infty \delta^{(n)}(\bm{x}, a), \qquad \theta(\bm{x}, a) = \sum_{n = 1}^\infty \theta^{(n)}(\bm{x}, a)
\end{equation}
where $\theta = \nabla \cdot \bm{\upsilon}$ and the $n^{th}$-order fluctuation fields $\delta^{(n)}$ and $\theta^{(n)}$ are proportional to the $n^{th}$ power of the linear fluctuation fields:
\begin{equation}
    \delta^{(n)} \sim \left(\delta^{(1)}\right)^n, \qquad \theta^{(n)} \sim \left(\theta^{(1)}\right)^n.
\end{equation}
In addition, we have $\partial_a \delta^{(1)} = \theta^{(1)}$ from the linearised continuity equation. Therefore, both the velocity and the density field are fully determined by the linear density fluctuations. The coupling of linear modes in the nonlinear theory is described by the nonlinear coupling kernels $F_n$ and $G_n$:
\begin{align}
    \delta^{(n)}(\bm{k}) &= \int \frac{\mathrm{d}^3\bm{q}_1}{(2\pi)^3} \hdots \int \frac{\mathrm{d}^3\bm{q}_n}{(2\pi)^3} \, \delta_D(\bm{k} - \bm{q}_{1,\hdots, n})\nonumber\\
    &\times F_n(\bm{q}_1, \hdots, \bm{q}_n)\, \delta^{(1)}(\bm{q}_1) \hdots  \delta^{(1)}(\bm{q}_n), \label{eq:DefinitionFn} \\
    \theta^{(n)}(\bm{k}) &= \int \frac{\mathrm{d}^3\bm{q}_1}{(2\pi)^3} \hdots \int \frac{\mathrm{d}^3\bm{q}_n}{(2\pi)^3} \,  \delta_D(\bm{k} - \bm{q}_{1,\hdots, n})\nonumber \\
    &\times G_n(\bm{q}_1, \hdots, \bm{q}_n)\, \delta^{(1)}(\bm{q}_1) \hdots \delta^{(1)}(\bm{q}_n) \label{eq:DefinitionGn}
\end{align}
where $F_n$ and $G_n$ are homogeneous functions of the wave vectors $\bm{q}_1, \hdots, \bm{q}_n$ with degree zero and $\bm{q}_{1, \hdots, n} \equiv \bm{q}_1 + \hdots + \bm{q}_n$. A series of papers \citep{Fry1984, Goroff1986, Jain1993} developed a method for deriving the time-independent CDM mode coupling kernels in terms of algebraic recursion relations. They rely on the Euler-Poisson system being homogeneous in the scale factor $a$ in the EdS case. Including the quantum pressure breaks this homogeneity and just like for a general cosmology in CDM, the solutions at each order become non-separable functions of time and scale.

A method to nonetheless obtain the PT kernels in this case is described in the series of papers \citep{Scoccimarro1998, Scoccimarro2006} that developed time-dependent Eulerian PT with the aid of Feynman diagrams. Effectively, \cite{Li2018} use this method to compute the kernels $F_2$ and $F_3$ in FDM. The nonlinear terms in the Madelung equations are treated as inhomogeneity $g(\eta)$ while solving the linear growth equation. This leads to an integral equation that can be represented as a Dyson series. The Dyson series allows for a diagrammatic representation and can be recursively solved up to a given order in PT. In appendix \ref{appendix:NLEPT}, we retrace the steps taken by \cite{Li2018} and extend them to higher order in perturbation theory for consistent tree- and loop-level computation of bi- and trispectra in FDM, which necessitates the kernels $F_2$, $F_3$ and $F_4$; recent developments in perturbation theory recast this straightforward perturbative approach into more powerful formalisms \citep{Pietroni:2008jx, Bartelmann2016b, Kozlikin:2020exj} including FDM \citep{Littek:2018ljy}.

% --- subsection: spectra, bispectra and trispectra --- %
\subsection{spectra, bispectra and trispectra at tree- and loop-level}
We derive explicit expressions for the kernels $F_2$, $F_3$ and $F_4$ in the FDM and CDM case in appendix \ref{appendix:NLEPT}.
These kernels $F_n$ can then be used to perturbatively expand $n$-point-matter correlation functions $P_n(\bm{k}_1, \bm{k}_2, \ldots, \bm{k}_n)$. The latter are defined as Fourier transform of the connected correlation functions:
\begin{equation}
\begin{split}
\langle \delta(\bm{k}_1), \bm{k}_2 \ldots, \delta(\bm{k}_n)\rangle_c &\equiv  (2\pi)^{3(n-1)}\delta_D(\bm{k}_1 + \bm{k}_2 + \ldots + \bm{k}_n)\\
&\times P_n(\bm{k}_1, \bm{k}_2, \ldots, \bm{k}_n).
\end{split}
\end{equation}
In the following, we are interested in the equal-time matter spectrum $P \equiv P_2$, bispectrum $B  \equiv P_3$ and trispectrum $T  \equiv P_4$. We express the $n$-point correlations of non-Gaussian perturbations as integrals over higher-order correlations of Gaussian perturbations. We then apply Wick's theorem to express higher-order correlations of Gaussian perturbations as two-point correlations of Gaussian perturbations. The latter are completely specified by the initial conditions. 
Expanding the two-point correlation as a perturbation series $P(k, a) = P^{(0)}(k, a) + P^{(1)}(k, a) + \hdots$, we find that the time evolution of the linear spectrum $P^{(0)}$ can be expressed as linear scaling of the initial spectrum:
\begin{equation}
\label{eq:LinearMatterPowerSpectrumEvolution}
    P^{(0)}(k, a) = \frac{D_+(a)^2}{D_+(a_0)^2}P^{(0)}(k, a_0).
\end{equation}
In the following, we omit the explicit time dependencies. 
The first higher-order contribution $P^{(1)}$ to the spectrum comes at loop-level \footnote{
Note that a consistent truncation of series in PT is obtained by including terms up to a certain power $m$ in the linear spectrum. This corresponds to grouping the PT contributions in terms of the number of loops in a diagrammatic representation.}. The two loop-level contributions to the one-loop spectrum $P^{(1)}$ are given by 
\begin{equation}
\label{eq:PowerSpectrumCorrections} P^{(1)}(k) = P_{22}(k) + P_{13}(k)
\end{equation}
with
\begin{align}
    P_{22}(k) &= 2 \int \mathrm{d}^3q\: [F_2^{(s)}(\bm{k - q}, \bm{q})]^2 P^{(0)}(|\bm{k - q}|) P^{(0)}(q),\\ 
    P_{13}(k) &= 6 \int \mathrm{d}^3q\: F_3^{(s)}(\bm{k}, \bm{q}, \bm{-q}) P^{(0)}(k) P^{(0)}(q).
\end{align}
The bispectrum at tree-level corresponding to second-order PT is given by
\begin{equation}
\begin{split}
    B^{(0)}(\bm{k}_1, \bm{k}_2, \bm{k}_3) &= 2 F_2^{(s)}(\bm{k}_1, \bm{k}_1 - \bm{k}_2) P^{(0)}(\bm{k}_1) P^{(0)}(\bm{k}_2)\\
    &+ \text{2 permutations}.
\end{split}
\end{equation}
The one-loop contribution to the bispectrum consists of four distinct diagram involving up to fourth-order PT kernels:
\begin{equation}
\label{eq:BispectrumCorrections}
    B^{(1)} = B_{222} + B^I_{321} + B^{II}_{321} + B_{411}
\end{equation}
whose explicit expressions are given by
\begin{equation}
    \begin{split}
        B_{222} &= 8\int \mathrm{d}^3\bm{q}\: P^{(0)}(q) P^{(0)}(|\bm{q} + \bm{k}_1|) P^{(0)}(|\bm{q} - \bm{k}_2|)\\
        &\times F_2^{(s)}(-\bm{q}, \bm{q} + \bm{k}_1)\\
        &\times F_2^{(s)}(\bm{q} + \bm{k}_1, -\bm{q} + \bm{k}_2)\\
        &\times F_2^{(s)}(\bm{k}_2 - \bm{q}, \bm{q}),\\
        B^I_{321} &= 6 \int \mathrm{d}^3\bm{q}\: P^{(0)}(k_3) P^{(0)}(q) P^{(0)}(|\bm{q} - \bm{k}_2|)\\
        &\times F_2^{(s)}(\bm{q}, \bm{k}_2 - \bm{q}) F_3^{(s)}(-\bm{q}, \bm{q} - \bm{k}_2, -\bm{k}_3)\\
        & + 5 \text{ permutations},\\
        B^{II}_{321} &= 6 \int \mathrm{d}^3\bm{q}\: P^{(0)}(k_2) P^{(0)}(k_3) P^{(0)}(q)\\
        &\times F_2^{(s)}(\bm{k}_2, \bm{k}_3) F_3^{(s)}(\bm{k}_3, \bm{q}, -\bm{q})\\
        & + 5 \text{ permutations},\\
        B_{411} &= 12 \int \mathrm{d}^3\bm{q}\: P^{(0)}(k_2) P^{(0)}(k_3) P^{(0)}(q)\\
        &\times F_4^{(s)}(\bm{q}, -\bm{q}, -\bm{k}_2, -\bm{k}_3)\\
        & + 2 \text{ cyclic permutations}.
    \end{split}
\end{equation}
In the following, we also examine the reduced bispectrum
\begin{equation}
    Q(\bm{k}_1, \bm{k}_2, \bm{k}_3) = \frac{B(\bm{k}_1, \bm{k}_2, \bm{k}_3)}{P(k_1)P(k_2) + P(k_2)P(k_3) + P(k_3)P(k_1)}.
\end{equation}
It is independent of time and normalisation.  For scale-free initial conditions $P^{(0)}$, i.e. $P^{(0)}\propto k^n$ with the spectral index $n$, $Q^{(0)}$ is also independent of overall scale and for equilateral configurations it is also independent of the spectral index. Finally, the diagrams for the trispectrum involve vertices connecting two and three lines and therefore correspond to second- and third-order PT. 
One can decompose the tree-level trispectrum into the contributions 
\begin{equation}
T^{(0)} =  T_{1221} + T_{3111}    
\end{equation} 
where
\begin{equation}
\begin{split}
    T_{1221}(\bm{k}_1, \hdots, \bm{k}_4) &= 4 P^{(0)}(k_3) P^{(0)}(k_4)\\ 
    &\times \bigl(F_2^{(s)}(\bm{k}_{13}, -\bm{k}_3) F_2^{(s)}(\bm{k}_{24}, -\bm{k}_4) P^{(0)}(k_{13})\\
    &+F_2^{(s)}(\bm{k}_{14}, -\bm{k}_4) F_2^{(s)}(\bm{k}_{23}, -\bm{k}_3) P^{(0)}(k_{14})\bigl),\\
    T_{3111}(\bm{k}_1, \hdots, \bm{k}_4) &= 6 P^{(0)}(k_1) P^{(0)}(k_2) P^{(0)}(k_3) F_3^{(s)}(\bm{k}_{1}, \bm{k}_{2}, \bm{k}_3).
\end{split}
\end{equation}
We now present the different spectra computed with time-dependent PT in CDM and FDM for a dark energy cosmology with $w=-0.9$ and $\Omega_{m, 0} = 0.3159$ at $z = 0$. 
For computing the nonlinear PT kernels $F_n$, we use the EdS growth factors in both CDM and FDM \footnote{We were unable to compute the PT kernels in a general cosmology in FDM. This would require integrating two linearly independent solutions to the linear growth equation Eq. \eqref{eq:FDMLinearGrowthScaleFactor}.}. The initial spectra are computed using the \textsc{CAMB} and \textsc{axionCAMB} codes for CDM and FDM respectively \citep{CAMB, axionCAMB}.
\highlighttext{
Importantly, the nonlinear CDM matter power spectra obtained via N-body simulations and via PT are only expected to agree to percent-level up to $k \sim 0.1$ $h$/Mpc at $z=0$ \citep{Carrasco2014}. Therefore, we also run a total of $8$ $N$-body simulations in two boxes of the sizes $L = 256$ Mpc/$h$ and $L = 30$ Mpc/$h$ with $N = 512^3$ particles using the \textsc{GADGET-2} code \citep{Springel2005}. The $L = 256$ Mpc/$h$ box provides a cross-check of the large-scale loop-level PT matter power spectra. The $L = 30$ Mpc/$h$ simulations are used to obtain a rough estimate of the full nonlinear power spectra on small scales. Since we probe the matter power spectrum at late times and on scales smaller than the Jeans length, we expect the effect of the quantum pressure to be important. \cite{Armengaud2017} note that especially for masses smaller than $m \sim 10^{-22}$ eV, the quantum pressure affects a significant portion of the dark matter particles on scales $k \gtrsim 1$ $h$/Mpc. Therefore, the $N$-body power spectra for the $m = 10^{-22}$ eV and $m = 10^{-23}$ eV cases are expected to exhibit large errors on small scales that could only be eliminated with large-scale FDM simulations including wave dynamics which are arguably beyond the reach of current supercomputers. To the authors' knowledge, \cite{May2021, May2022} perform the largest high-resolution wave-based simulation of the FDM model to date. They employ a mass of $m=7\times10^{-23}$ eV in a $L = 10$ Mpc/$h$ box and compare CDM and FDM simulations both with CDM and FDM IC at different times up to $z=3$. They observe that the effect of quantum pressure generally leads to a a further suppression of small-scale power, except for a feature at $k \sim 1000$ $h$/Mpc where the FDM spectrum slightly exceeds the CDM spectrum - likely as a result of wave interference. \cite{Nori2018a} also compare $N$-body simulations with and without quantum pressure at $z=0$ and find that the quantum pressure reduces small-scale power. 
Therefore, we believe that the $N$-body simulations serve as an important cross-validation. Technical details can be found in appendix \ref{appendix:NBody}.}

\begin{figure}
\centering
\includegraphics[width=.47\textwidth]{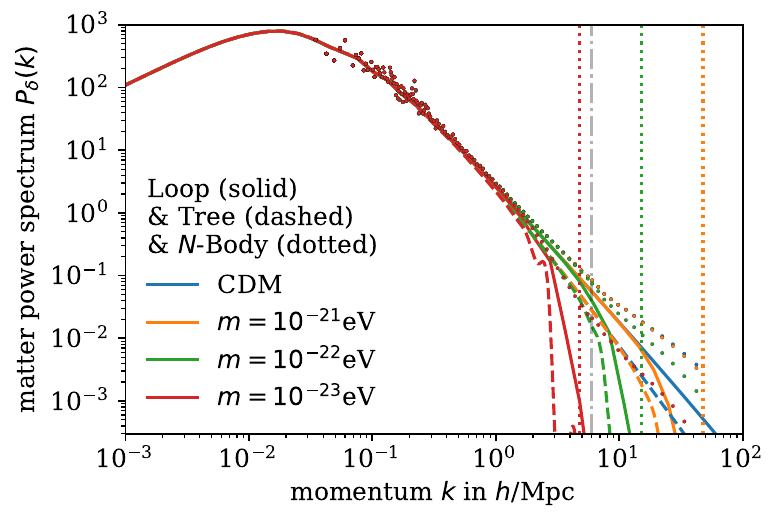}
\caption{\highlighttext{Dark matter spectra $P_\delta(k)$ at tree- and loop-level as well as from $N$-body simulation at $z=6.1$. The dashed, grey line indicates the transition between the different simulation boxes. The dotted, vertical lines denote the respective Jeans scales from Eq. \eqref{eq:ComovingQuantumJeansScale} at $z=99$.}} 
\label{fig:MatterPowerSpectrumZ6}
\end{figure}

Fig. \ref{fig:MatterPowerSpectrum} shows the CDM and FDM spectra at tree-level and with loop-level corrections \highlighttext{as well as the spectra from several $N$-body simulations with FDM and CDM IC at $z=0$. At tree-level, power is strongly suppressed below the Jeans scale in the FDM model. Nonlinear corrections at loop-level transfer power to small scales, but suppression is still dominant. As previously noted by \cite{Nori2018, Nori2018a}, the initial suppression of power on small scales in the $N$-body simulations is almost completely overcome by late-time nonlinear CDM dynamics. Note that the low-$k$ modes in the $N$-body simulations suffer from large variations that stem from a low number of samples.  Moreover, there is a mismatch between the high-$k$ modes of the  $L = 256$ Mpc/$h$ and the low-$k$ modes of the $L = 30$ Mpc/$h$ simulations at $k = 6$ $h$/Mpc indicated by the vertical gray line due to the fact that the  $L = 30$ Mpc/$h$ box is too small for the large-scale modes to evolve correctly. Fig. \ref{fig:MatterPowerSpectrumZ6} shows the respective spectra at $z=6.1$ for reference. The loop-level PT and the $N$-body results agree up to $k = 1$ $h$/Mpc, but still start to differ on smaller scales. The different $N$-body power spectra demonstrate that late-time CDM dynamics have not yet overcome the initial suppression of power on small-scales at $z \sim 6$. }

\begin{figure}
\centering
\includegraphics[width=.47\textwidth]{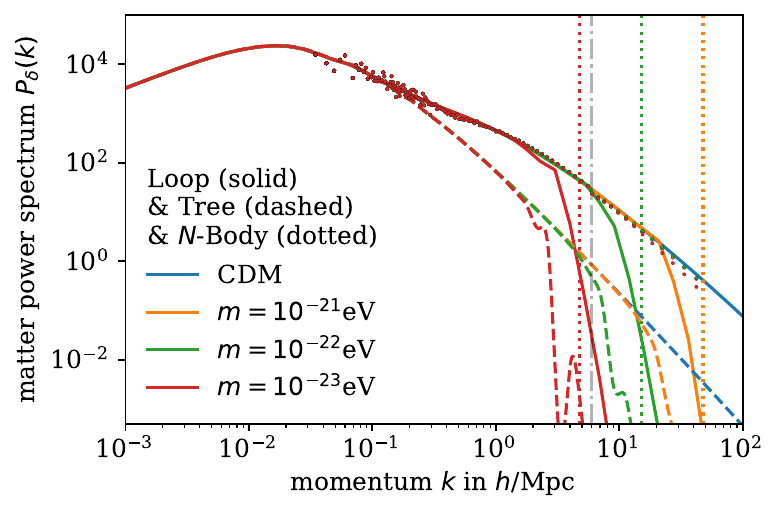}
\caption{Dark matter spectra $P_\delta(k)$ \highlighttext{at tree- and loop-level as well as from $N$-body simulation at $z=0$. The $N$-body results overlap. The dashed, grey line indicates the transition between the different simulation boxes.} The dotted, vertical lines denote the respective Jeans scales from Eq. \eqref{eq:ComovingQuantumJeansScale} at $z=99$.}
\label{fig:MatterPowerSpectrum}
\end{figure}

\begin{figure}
\centering
\includegraphics[width=.47\textwidth]{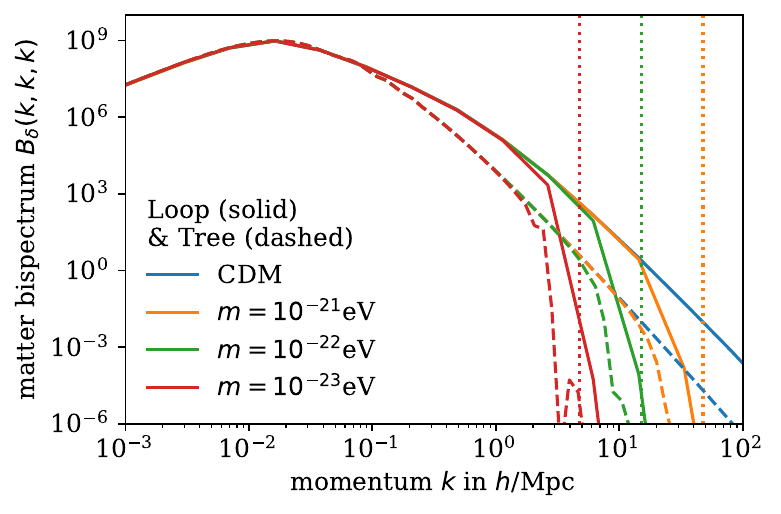}
\caption{Equilateral dark matter bispectra $B_\delta(k,k,k)$ at tree- and loop-level at $z=0$. The dotted, vertical lines denote the respective Jeans scales from Eq. \eqref{eq:ComovingQuantumJeansScale} at $z=99$.}
\label{fig:EquilateralMatterBispectrum}
\end{figure}

\begin{figure}
\centering
\includegraphics[width=.47\textwidth]{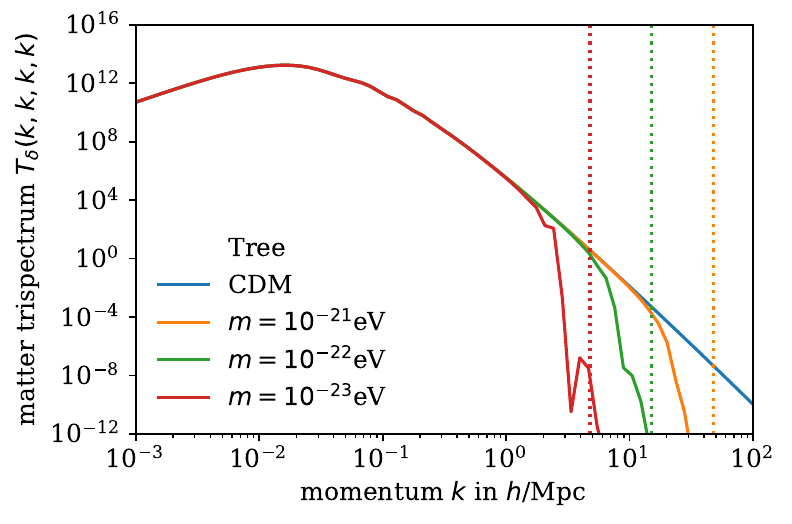}
\caption{Equilateral square matter trispectra $T_\delta(k,k,k,k)$ at tree-level at $z=0$. Dashed lines indicate respective Jeans scales from Eq. \eqref{eq:ComovingQuantumJeansScale} at $z = 99$.}
\label{fig:EquilateralMatterTrispectrum}
\end{figure}

\begin{figure}
\centering
\includegraphics[width=.47\textwidth]{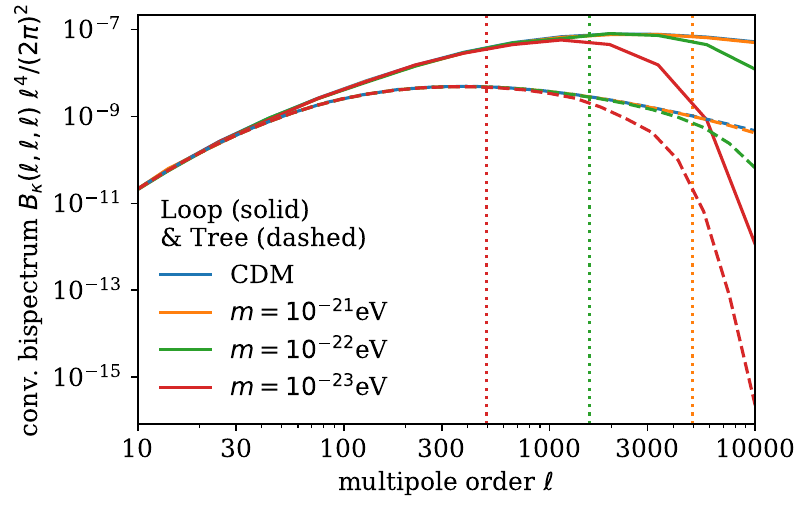}
\caption{Dimensionless equilateral convergence bispectrum configurations. The vertical, dotted lines correspond to $0.1\times\ell_J$, where the quantum Jeans multipole order $\ell_J$ is defined in Eq. \eqref{eq:JeansMultipoleOrder}.}
\label{fig:EquilateralConvergenceBispectrum}
\end{figure}

\begin{figure}
\centering
\includegraphics[width=.47\textwidth]{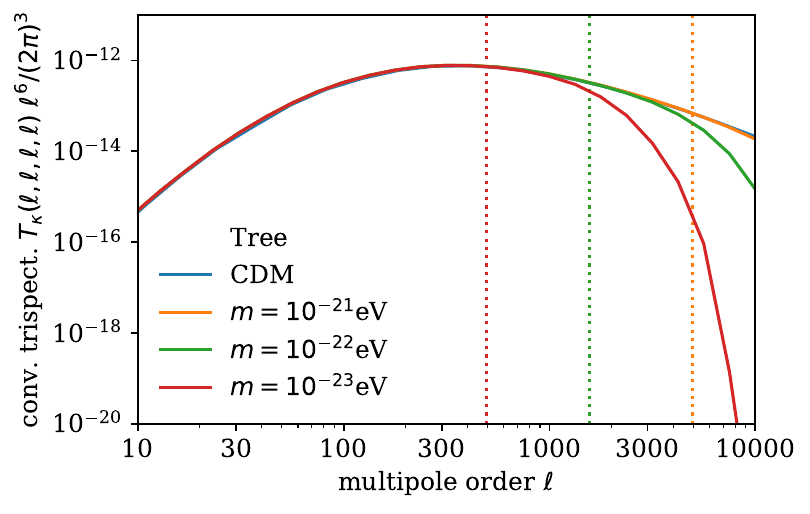}
\caption{Dimensionless equilateral square convergence trispectrum configurations. The vertical, dotted lines correspond to $0.1\times\ell_J$, where the quantum Jeans multipole order $\ell_J$ is defined in Eq. \eqref{eq:JeansMultipoleOrder}.}
\label{fig:EquilateralConvergenceTrispectrum}
\end{figure}

Figs. \ref{fig:EquilateralMatterBispectrum} and \ref{fig:EquilateralMatterTrispectrum} show the equilateral matter bispectra and equilateral square matter trispectra and Figs. \ref{fig:EquilateralConvergenceBispectrum} and \ref{fig:EquilateralConvergenceTrispectrum} the respective convergence spectra. As in the case of the spectrum, loop-level corrections for the bispectrum have the effect of adding power on small scales in both CDM and FDM. At both tree- and loop-level, however, suppression below the Jeans scale is still the dominant effect in FDM.

The loop-level corrections given by Eq.~\eqref{eq:PowerSpectrumCorrections} were computed in a form free of infrared divergences using the \textsc{CUBA}-library \cite{Hahn2004}. Details of the numerical integration can be found in appendix \ref{appendix:InfraredSafeIntegrands}. Fig. \ref{fig:BispectrumAngle} shows the angular dependence of the reduced matter bispectrum $Q^{(0)}$ at tree-level. The fact that $Q^{(0)}$ is enhanced for $\theta = 0,\pi$ reflects the fact that large scale flows generated by gravitational instability are mostly parallel to density gradients. As discussed in section \ref{sec:QPandJeans}, the kernel $F_2^{(s)}$ includes the first higher-order correction from the quantum pressure term. Therefore, it does not only counteract gravitational collapse but can also enhance it as exemplified by the graphs for $m = 10^{-21}$ eV and $m = 10^{-22}$ eV at scales $k$ around or below the Jeans scale for the respective masses. 
\begin{figure}
\centering
\includegraphics[width=.47\textwidth]{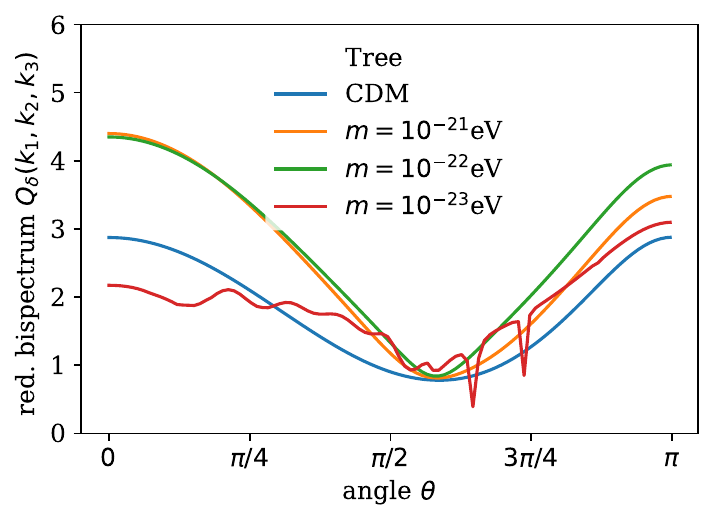}
\caption{Angular dependence of reduced bispectrum at tree-level with $\theta = \sphericalangle(\bm{k}_1, \bm{-k_2})$ with $k_1 = 10$ $h$/Mpc and $k_2 = 0.1$ $h$/Mpc at $z=0$. The graph for $m = 10^{-23}$ eV is in the oscillating regime and its shape depends on how the growth factors and the initial spectra approximate oscillations.}
\label{fig:BispectrumAngle}
\end{figure}

% --- section:  --- %
\section{weak lensing predictions}
\label{sec:WeakLensingPredictions}
\begin{figure} 
\centering
\includegraphics[page=1, height=4cm]{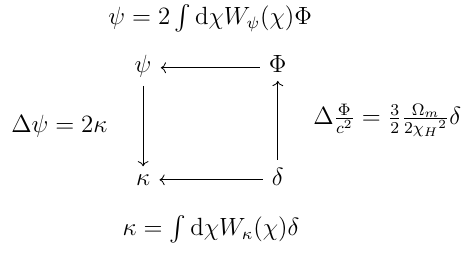}
%\begin{tikzcd}[sep=huge]
%\psi  \arrow[d, "\Delta \psi = 2\kappa", swap, outer %sep=10pt] & \Phi \arrow[l,"\psi = 2 \int \mathrm{d}  %\chi W_{\psi}(\chi)\Phi", swap, outer sep=10pt] \\
%\kappa & \delta \arrow[l,"\kappa = \int \mathrm{d} %\chi W_{\kappa}(\chi)\delta", outer sep=10pt] %\arrow[u, "\Delta  \frac{\Phi}{c^2} = %\frac{3}{2}\frac{\Omega_m}{2{\chi_H}^2}\delta", swap, %outer sep=10pt]\\
%\end{tikzcd}
    \caption{Relationship between density contrast, gravitational potential, lensing potential and effective convergence.}
    \label{fig:RelationshipConvergenceDensity}
\end{figure}

The observational quantity of interest in a weak lensing survey is the shear $\gamma$ or equivalently the convergence $\kappa$.  The weak lensing convergence is given by a line-of-sight integration over the density contrast:
\begin{equation}
\label{eq:EffectiveConvergence}
    \kappa(\theta, \chi) = \int_0^{\chi_H} \mathrm{d}\chi\, W_\kappa(\chi) \delta(\chi)
\end{equation}
where the weight function $W_{\kappa}(\chi)$ in the integral  is also called weak lensing efficiency and can be modelled as
\begin{equation}
\label{eq:WeakLensingEfficiency}
    W_{\kappa}(\chi) = \frac{2\Omega_m}{2a}\frac{1}{\chi_H^2}G(\chi)\chi.
\end{equation}
$G(\chi)$ is the weighted distance distribution of the lensed galaxies:
\begin{equation}
    G(\chi) = \int_{\chi}^{\chi_H}\mathrm{d}\chi'\;q(z)\frac{\mathrm{d}z}{\mathrm{d}\chi'}\frac{\chi' - \chi}{\chi'}
\end{equation}
where $q(z)$ is the galaxy-redshift distribution measured in a weak lensing survey. 
We assume a simple model for the galaxy redshift distribution
\begin{equation}
\label{eq:RedshiftDistribution}
    q(z) = q_0\left(\frac{z}{z_0}\right)^2\exp\left(-\left(\frac{z}{z_0}\right)^{\beta}\right)\mathrm{d}z\quad\text{with}\quad q_0^{-1} = \frac{z_0}{\beta}\Gamma\left(\frac{3}{\beta}\right)
\end{equation}
used in the Euclid survey \citep{Laureijs2011} with a median redshift $z_0 = 0.9$ and $\beta = 1.47$. If we know the correlation functions of the convergence, this will give us a way to infer correlation functions of the density contrast.  
In practice, weak lensing surveys do not measure the convergence $\kappa$, but the shear $\gamma$ by fitting models to or measuring the quadrupole moments of the surface brightness of distant galaxies. The convergence $\kappa$ is harder to measure since we do not know the intrinsic luminosity of the background galaxies. However, we find $|\gamma|^2 = |\kappa|^2$ in Fourier space and conclude that the respective convergence and shear spectra are equivalent. Schematically, the relationship between the density contrast and the effective convergence is represented in Fig. \ref{fig:RelationshipConvergenceDensity}.
 
Our lensing survey parameters are those of a Euclid-like survey. We consider a half-sky survey with  $f_{sky} = 0.5$, a standard deviation of intrinsic ellipticities of galaxies of $\sigma_{\epsilon} = 0.4$ and an average number of galaxies per steradian $\bar{n} = 4.727 \times 10^8$. Perhaps most importantly, we list results up to a maximum multipole moment of $\ell_{max} = 10^4$. Such a high multipole moment is not accessible in a weak lensing survey. In order to see the effect of axion DM in the considered mass range, we need to resolve scales at the order of $k = 1$ $h$/Mpc which roughly corresponds to multipole orders $\ell \gtrsim 10^3$ in the case of Euclid. In practice, the highest multipole moment measurable in a weak lensing survey is limited by the shape noise. The signal-to-noise ratio of a single multipole moment drops below $1$ at  $\ell \sim 3\times10^3$. However, we may still gain information by summing over these noise-dominated multipole moments as we make a relatively conservative estimate for the shape noise. At even higher multipole moments of around $\ell \sim 5\times10^3$, baryonic feedback becomes a important, a process which we neglect entirely in PT. At this point, our results become completely unreliable. We still show the plots for multipole moments  $\ell \gtrsim 5\times10^3$ in order visually compare the different masses. 

\highlighttext{More importantly, PT disagrees with $N$-body simulations on the scales considered and significantly underestimates the full nonlinearity for large $k$. We therefore compare with an estimate of the lensing spectrum signal obtained by integrating the power spectra obtained from the $L = 30$ Mpc/$h$  $N$-body simulations from $k=6$ $h$/Mpc to $k=30$ $h$/Mpc well below the Nyquist frequency $k_{Nyq} = \pi \frac{N}{L} = 53.6$ $h$/Mpc. In addition, we demonstrate how the lensing PT results for CDM and the $m=10^{-23}$ eV case change when implementing a hard cutoff on the spectra, that is, we set the input matter spectra to zero above a cutoff scale and study how the PT lensing results depend on the cutoff scale.}

% --- subsection: lensing spectra, bispectra and trispectra --- %
\subsection{lensing spectra, bispectra and trispectra}
\begin{figure}
\centering
\includegraphics[width=.47\textwidth]{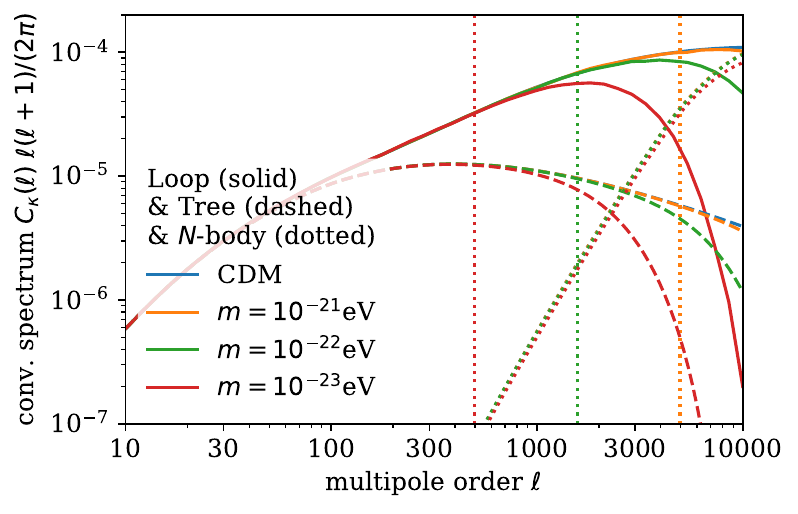}
\caption{Dimensionless convergence spectra at tree-level, with loop-level corrections \highlighttext{and from $N$-body simulations}. The $N$-body results overlap. The vertical, dotted lines correspond to $0.1\times\ell_J$, where the quantum Jeans multipole order $\ell_J$ is defined in Eq. \eqref{eq:JeansMultipoleOrder}.}
\label{fig:ConvergenceSpectrum}
\end{figure}

The angular correlation function for a quantity such as the convergence $\kappa(\bm{\theta})$ measured on the sky is given by
\begin{equation}
    \xi_\kappa(\bm{\varphi}) \equiv \left\langle \kappa(\bm{\theta}) \kappa(\bm{\theta} + \bm{\varphi})\right\rangle
\end{equation}
where the expectation value is computed as ensemble average over statistically equivalent realisations of the field $\kappa(\bm{\theta})$. We take the Fourier transform to obtain the angular spectrum in the flat-sky approximation:
\begin{equation}
    C_\kappa(\bm{\ell}) = \int \mathrm{d}^2 \varphi\: \xi(\varphi) \exp(-\ci\bm{\ell} \cdot \bm{\varphi}).
\end{equation}
In order to simplify the computation of angular correlation functions we employ Limber's approximation \citep{Limber1953}. It asserts that if the quantity $x(\bm{\theta})$ defined in two dimensions is a projection
\begin{equation}
\label{eq:Limber}
    x(\bm{\theta}) = \int_0^{\chi_S} \mathrm{d}\chi w_x(\chi) y(\chi \bm{\theta}, \chi)
\end{equation}
of a quantity $y(\bm{r})$ defined in three dimensions with a weight function $w_x(\chi)$, then the angular spectrum of $x$ is given by
\begin{equation}
    C_x(\ell) = 
    \int_0^{\chi_S} \frac{\mathrm{d}\chi}{\chi^2}\: w_x^2(\chi)P_y\left(\frac{\ell}{\chi}\right)
\end{equation}
where $P_y(k)$ is the spectrum of $y$, evaluated at the three-dimensional wave number $k = \ell/\chi$. This approximation is applicable if $y$ varies on length scales much smaller than the typical length scale of the weight function $w_x$. Intuitively, we divide $\chi$ by $\ell$ such that we can compare different scales for a given angle. From the Limber approximation, it immediately follows that the convergence spectrum $C_{\kappa}(\ell)$ is determined by a weighted line-of-sight integral over the spectrum of the density contrast $P_{\delta}(k)$. Likewise, we can express the convergence bi- and trispectrum as appropriately weighted line-of-sight integrals over the bi- and trispectra. All in all, we find
\begin{align}
    C_{\kappa}(\ell_1) &= \int_0^{\chi_{\infty}}  \frac{\mathrm{d}\chi}{\chi^2}\: W_{\kappa}^2 (\chi) P_{\delta}\left(\frac{\ell_1}{\chi}\right),\\
    B_{\kappa}(\ell_1, \ell_2, \ell_3) &= \int_0^{\chi_{\infty}}  \frac{\mathrm{d}\chi}{\chi^4}\: W_{\kappa}^3 (\chi) B_{\delta}\left(\frac{\ell_1}{\chi},\frac{\ell_2}{\chi}, \frac{\ell_3}{\chi}\right),\\
    T_{\kappa}(\ell_1, \ell_2, \ell_3, \ell_4) &= \int_0^{\chi_{\infty}}  \frac{\mathrm{d}\chi}{\chi^6}\:W_{\kappa}^4 (\chi) T_{\delta}\left(\frac{\ell_1}{\chi}, \frac{\ell_2}{\chi}, \frac{\ell_3}{\chi}, \frac{\ell_4}{\chi}\right)
\end{align}
where we introduced the subscripts $\delta$ to denote the matter spectra as opposed to the convergence spectra denoted by the subscript $\kappa$ and the weight function $W_{\kappa}$ is the weak lensing efficiency defined in Eq. \eqref{eq:WeakLensingEfficiency}.
Fig. \ref{fig:ConvergenceSpectrum} shows the respective convergence spectra. Nonlinear corrections significantly increase the magnitude of the dimensionless spectra for multipole moments  $\ell \gtrsim 100$.  \highlighttext{In order to highlight where one would naively expect the models to differ}, we translate the comoving quantum Jeans scale into a corresponding multipole order by an order-of-magnitude estimate. We define the quantum Jeans multipole order $\ell_J(m)$ via
\begin{equation}
\label{eq:JeansMultipoleOrder}
    \ell_J(m) \equiv \frac{\pi}{\arctan\left(\frac{k_J(m)}{\chi(z=z_0)}\right)}
\end{equation}
where $\chi(z=z_0)$ is the comoving distance at the redshift $z_0$ amd $k_J$ is as defined in Eq. \eqref{eq:ComovingQuantumJeansScale} at $z = 99$. For the mean redshift $z_0 = 0.9$ of the redshift distribution defined in Eq. \eqref{eq:RedshiftDistribution}, we obtain $\ell_J\left(m = 10^{-21} \mathrm{eV}\right) \approx 5.0\times10^4$, $\ell_J\left(m = 10^{-22} \mathrm{eV}\right)  \approx 1.6\times10^4$ and $\ell_J\left(m = 10^{-23} \mathrm{eV}\right) \approx 5.0\times10^3$ . Since these quantum Jeans multipole orders are too high to be measurable in a weak lensing survey, the vertical lines in Fig. \ref{fig:ConvergenceSpectrum} display $0.1\times\ell_J$ which roughly describes the multipole order where the CDM PT and FDM PT lensing spectra start to differ.

\highlighttext{We also compute the lensing spectra using CDM PT with FDM IC via the CDM coupling kernels obtained from the EdS recursion relations as well as the CDM growth factors in the fiducial cosmology. CDM dynamics at tree- and loop-level yield lensing spectra, bispectra and trispectra that are, within the numerical errors, indistinguishable from the ones computed using FDM PT. This is why we refrain from showing the corresponding figures. We conclude that the reason for the suppression of the lensing spectra below the quantum Jeans multipole order in PT lies mainly in the initial conditions and is not the result of the approximation of late-time FDM dynamics through FDM PT.}

\begin{figure}
\centering
\includegraphics[width=.47\textwidth]{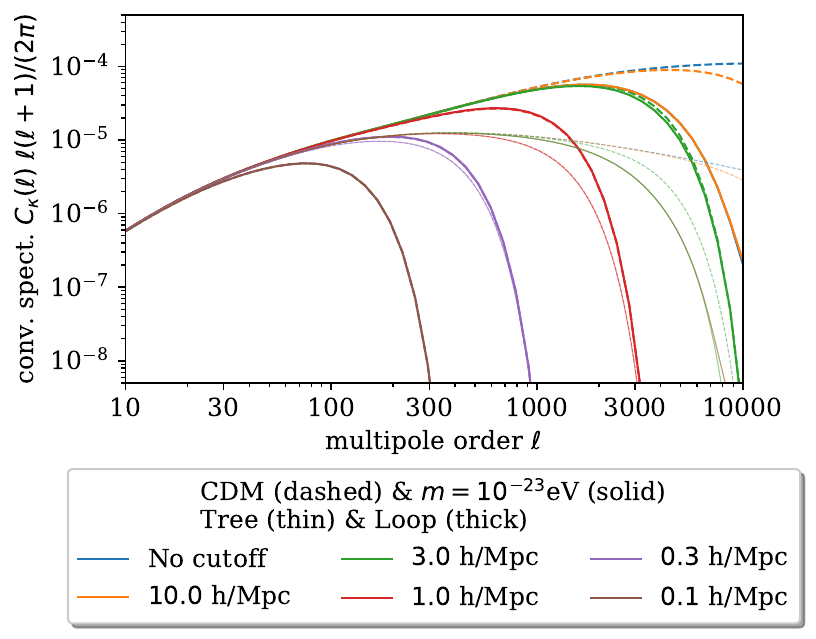}
\caption{Dimensionless convergence spectra at tree-level and loop-level corrections with small-scale cutoff in the input power spectra.}
\label{fig:ConvergenceSpectrumWithCutoff}
\end{figure}

\highlighttext{In contrast, the lensing spectra computed from the $N$-body spectra in the range $k = 6$ $h$/Mpc to $k=30$ $h$/Mpc paint a different picture. The late-time difference in power on small angular scales between the FDM and CDM models is significantly smaller than estimated via PT. Moreover, the FDM models exhibit more power on small angular scales than predicted via PT. Note that the $N$-body lensing spectra underestimate power for small $\ell$ because of the cutoff of $k$.
In order to assess, the impact of the high $k$-modes of the power spectra estimated via PT, we implement a high-$k$ cutoff in the PT lensing results for CDM and $m=10^{-23}$ eV. The tree-level and full loop-level input matter power used as input for the lensing integral are set to zero above the cutoff scale.  Fig. \ref{fig:ConvergenceSpectrumWithCutoff} demonstrates that a significant part of the discerning power of the model survey comes from the modes with $k > 3$ $h$/Mpc. For a smaller cutoff scale, the difference between the CDM and FDM model is not visible and for a cutoff at $k=0.1$ $h$/Mpc where the loop-level PT corrections for CDM are highly accurate, even the difference between the tree- and loop-level PT is not discernible. }


% --- subsection: attainable signal strength --- %
\subsection{attainable signal strength}
The convergence spectrum is measured by analysing the ellipticities of an ensemble of galaxy images.
%Together with the crucial assumption that the intrinsic ellipticities $\epsilon_S$ approach zero when averaged over sufficiently large samples $\langle \epsilon_S \rangle = 0$, one can distinguish intrinsic ellipticities and the ellipticity caused by lensing. Parameterising the latter by the reduced shear $g = \frac{\gamma}{1 - \kappa}$,the total measured ellipticity $\epsilon$ approximately is $\epsilon \sim g + \epsilon_S.$
Averaging over $N$ faint galaxy images, the scatter of the intrinsic ellipticity is reduced to 
\begin{equation}
    \Delta \langle \epsilon_S \rangle \approx \frac{\sigma_\epsilon}{\sqrt{N}}
\end{equation}
where $\sigma_{\epsilon}$ is the standard deviation of the intrinsic ellipticity. The angular resolution of this measurement is limited by
\begin{equation}
    \Delta \theta = \frac{N}{\bar{n}\pi}^{\frac{1}{2}}
\end{equation}
where $\bar{n}$ is the average number of source galaxies per squared arc minute. As a result, the observed convergence spectrum $C_{\kappa}^{(\mathrm{obs})}(\ell)$ can be modelled as the true spectrum with an additional shot noise contribution:
\begin{equation}
C_{\kappa}^{(\mathrm{obs})}(\ell) = C_{\kappa}(\ell) + \frac{\sigma_{\epsilon}^2}{\bar{n}}.
\end{equation}
Assuming that the estimates for the spectra can be approximated by a Gaussian distribution, we can estimate the covariance matrices for the lensing spectra, bispectra and trispectra. The covariance of the lensing spectrum is given by
\begin{equation}
\mathrm{cov}(\ell_1, \ell_2) = \delta_D(\ell_1 - \ell_2) \frac{2}{(2l + 1) f_{sky}} C_{\kappa}^{(\mathrm{obs})}(\ell_1) C_{\kappa}^{(\mathrm{obs})}(\ell_2)
\end{equation}
where $f_{sky}$ denotes the fraction of the observed sky and we neglect a contribution proportional to the lensing trispectrum due to the non-Gaussianity of the weak lensing field \citep{Kaiser1996, Scoccimarro1999}. \cite{Takada2003} provide an expression for the covariance of the weak lensing bispectrum:
\begin{equation}
\mathrm{cov}(\ell_1, \ell_2, \ell_3) = \frac{\Delta(\ell_1, \ell_2, \ell_3)}{f_{sky}}
\prod_{i=1}^3C_{\kappa}^{(\mathrm{obs})}(\ell_i)
\end{equation}
where $\ell_i \leq \ell_{i+1}$ to count every triangle/quadrilateral configuration only once. $\Delta(\ell_1, l_2, l_3)$ counts the multiplicity of triangle configurations and is defined as
\begin{equation}
    \Delta(\ell_1, l_2, l_3) =
    \begin{cases}
       6, &\quad \text{if } \ell_1 = \ell_2 = \ell_3,\\
       2, &\quad \text{if } \ell_i = \ell_j \text{ for } i \neq j,\\
       1 ,&\quad \text{otherwise}.\\ 
     \end{cases}
\end{equation}
Similarly, we have
\begin{equation}
\mathrm{cov}(\ell_1, \ell_2, \ell_3, \ell_4) = 
\frac{\Delta(\ell_1, \ell_2, \ell_3, \ell_4)}{f_{sky}}\prod_{i=1}^4C_{\kappa}^{(\mathrm{obs})}(\ell_i)
\end{equation}
for the covariance of the weak lensing trispectrum where $\Delta(\ell_1, \ell_2, \ell_3, \ell_4)$ counts the multiplicity of quadrilateral configurations. These covariance matrices allow us to understand the statistical uncertainties on the spectrum measurement. With their help, we can calculate the expected cumulative signal-to-noise ratio $\Sigma(\ell)$ for weak lensing measurements of the different spectra up to multipole order $\ell$ \footnote{Note that unlike for lensing bispectrum configurations that are uniquely specified by the three multipole moments $\ell_1, \ell_2, \ell_3$ up to spatial orientation, a lensing trispectrum configuration is not uniquely specified by $\ell_1, \ldots, \ell_4$. In our code, we sum over all configurations by varying the length of three sides as well as the two enclosed angles.}:
\begin{alignat}{3}
\Sigma^2_C(\ell) &=\quad &&\smashoperator[l]{\sum_{\ell_1=\ell_{min}}^{\ell}} &&\frac{C_{\kappa}^2(\ell_1)}{\mathrm{cov}(\ell_1, \ell_1)}, \label{eq:S2NP}\\
\Sigma^2_B(\ell) &=\quad &&\smashoperator[l]{\sum_{\ell_1, \ell_2, \ell_3=\ell_{min}}^{\ell}} &&\frac{B_{\kappa}^2(\ell_1, \ell_2, \ell_3)}{\mathrm{cov}(\ell_1, \ell_2, \ell_3)},\label{eq:S2NB}\\
\Sigma^2_T(\ell) &=\quad &&\smashoperator[l]{\sum_{\ell_1, \ell_2, \ell_3, \ell_4=\ell_{min}}^{\ell}} &&\frac{T_{\kappa}^2(\ell_1, \ell_2, \ell_3, \ell_4)}{\mathrm{cov}(\ell_1, \ell_2, \ell_3, \ell_4)}.\label{eq:S2NT}
\end{alignat}

\begin{figure}
\centering
\includegraphics[width=.47\textwidth]{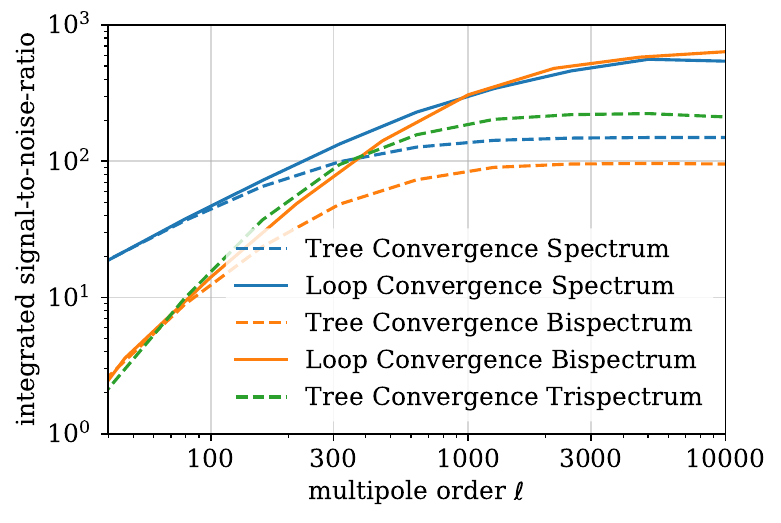}
\caption{Attainable signal to noise-ratio for convergence spectrum at tree- and loop-level, convergence bispectrum at tree- and loop-level and convergence trispectrum at tree-level (CDM and FDM identical).}
\label{fig:S2NRatios}
\end{figure}

Fig. \ref{fig:S2NRatios} shows the signal-to-noise ratios obtained in CDM according to Eqs. \eqref{eq:S2NP}, \eqref{eq:S2NB} and \eqref{eq:S2NT}. We compute the respective covariance matrices using the convergence spectrum with loop-level corrections. This is because the nonvanishing bi- and trispectrum themselves are generated by nonlinear dynamics. Using the tree-level convergence spectrum would therefore underestimate the covariance and overestimate the attainable signal-to-noise ratio. Since the bulk of the cumulative signal comes from the modes with low $\ell$, there are no significant differences for the attainable signal-to-noise ratios in CDM and FDM weak lensing surveys for the considered masses. \highlighttext{This is also why we do not show the respective $N$-body results: They would severely underestimate the attainable S2N ratios because of the lack of large-scale power.}
The sums in Eqs. \eqref{eq:S2NP}, \eqref{eq:S2NB} and \eqref{eq:S2NT} were expressed as integrals and integrated using the \textsc{CUBA}-library \citep{Hahn2004}. We could not compute the respective signal-to-noise ratios for the weak lensing bispectra at loop-level for FDM since the integrals involved proved computationally intractable.

\begin{figure}
\centering
\includegraphics[width=.47\textwidth]{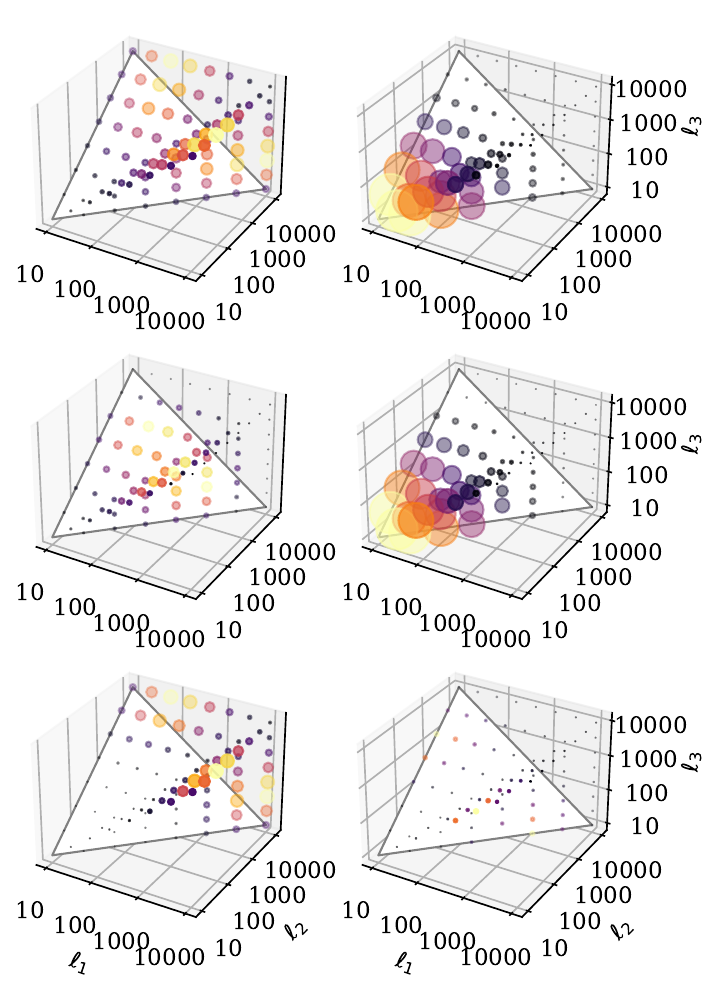}
\caption{Configuration dependence (first column) and signal-to-noise ratio (second column) of weak lensing bispectrum  at tree-level. Color and size both represent magnitudes; Same, arbitrary normalisation across rows. Top to bottom: CDM, FDM for $m = 10^{-23}$ and difference between the two. Left column: Dimensionless lensing bispectrum  $(\ell_1\ell_2\ell_3)^{\frac{3}{4}}B_{\kappa}(\ell_1, \ell_2, \ell_3)$ at $z=0$. Right column: Signal-to-noise ratio $B_{\kappa}(\ell_1, \ell_2, \ell_3)/\sqrt{\text{cov}(\ell_1, \ell_2, \ell_3)}$ at $z=0$.}
\label{fig:FDMWeakLensingBispectrum}
\end{figure}

\begin{figure}
\centering
\includegraphics[width=.47\textwidth]{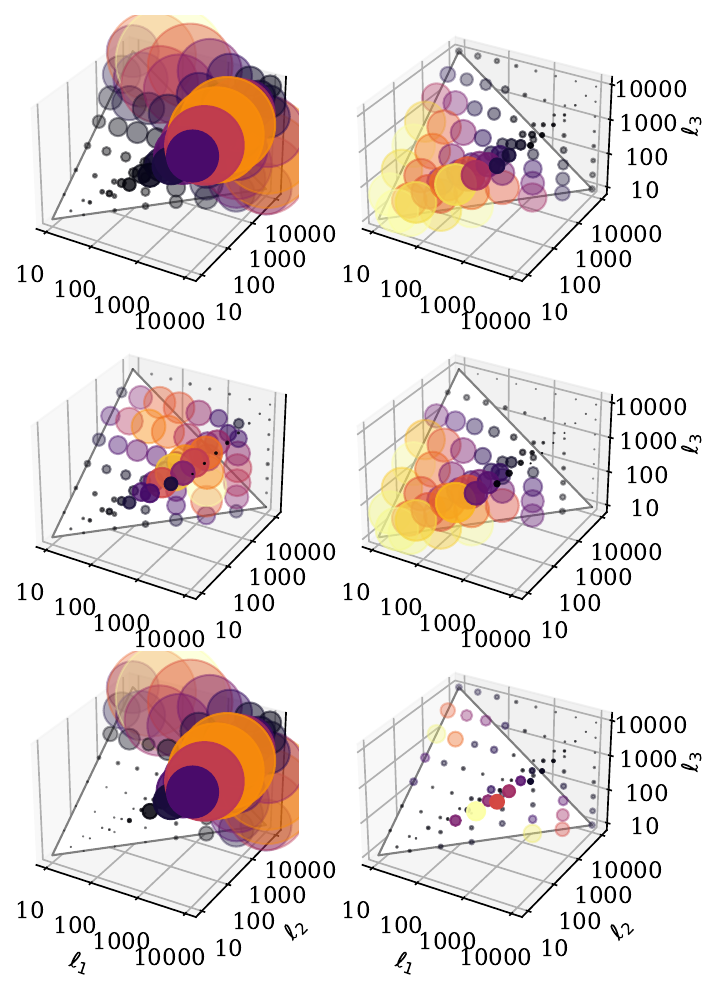}
\caption{Configuration dependence (first column) and signal-to-noise ratio (second column) of weak lensing bispectrum at loop-level where FDM dynamics are approximated by CDM PT with FDM IC. Color and size both represent magnitudes; Same normalisation as in Fig. \ref{fig:FDMWeakLensingBispectrum}. Top to bottom: CDM, CDM with FDM IC for $m = 10^{-23}$ and difference between the two. Left column: Dimensionless lensing bispectrum  $(\ell_1\ell_2\ell_3)^{\frac{3}{4}}B_{\kappa}(\ell_1, \ell_2, \ell_3)$ at $z=0$. Right column: Signal-to-noise ratio $B_{\kappa}(\ell_1, \ell_2, \ell_3)/\sqrt{\text{cov}(\ell_1, \ell_2, \ell_3)}$ at $z=0$.}
\label{fig:FDMWeakLensingLoopBispectrum}
\end{figure}

Nevertheless, we compared against the signal-to-noise ratios of the loop-level lensing bispectra computed with CDM PT for FDM IC. Yet, these results are also also subject to substantial numerical uncertainty since the Monte Carlo-integration routine fails to provide error estimates.
Fig. \ref{fig:FDMWeakLensingBispectrum} visualises the angular dependence of the lensing bispectrum at tree-level for $m = 10^{-23}$ eV. The bottom plots reflect that the small-angular scales where the CDM and FDM models actually differ only have a comparatively small signal-to-noise ratio in a weak lensing survey.  In contrast, Fig. \ref{fig:FDMWeakLensingLoopBispectrum} shows the corresponding loop-level results approximated by CDM with FDM IC. We observe that the loop-level corrections significantly enhance the signal-to-noise ratio for multipole orders where CDM and FDM at $m=10^{-23}$ eV can be distinguished.

% --- subsection: differentiability --- %
\subsection{differentiability between fuzzy dark matter and standard cold dark matter}
We can also give an estimate of whether we can distinguish CDM and FDM experimentally via a weak lensing survey. We assume that the true spectra are given by the CDM spectra and compute the $\chi^2$-functionals for measuring the noise-weighted mismatch between the true CDM and the wrongly assumed FDM spectra:
\begin{alignat}{3}
    \chi^2_C(\ell) &=\quad &&\smashoperator[l]{\sum_{\ell_1=\ell_{min}}^{\ell}}&&\frac{(C_{\kappa}^{CDM} - C_{\kappa}^{FDM})^2(\ell_1)}{\mathrm{cov}(C_{\kappa}^{CDM})(\ell_1, \ell_1)},
\label{eq:SpectrumChiSquared}\\
    \chi^2_B(\ell) &=\quad &&\smashoperator[l]{\sum_{\ell_1, \ell_2, \ell_3=\ell_{min}}^{\ell}}&&\frac{(B_{\kappa}^{CDM} - B_{\kappa}^{FDM})^2(\ell_1, \ell_2, \ell_3)}{\mathrm{cov}(B_{\kappa}^{CDM})(\ell_1, \ell_2, \ell_3)}, \label{eq:BispectrumChiSquared}\\
    \chi^2_T(\ell) &=\quad &&\smashoperator[l]{\sum_{\ell_1, \ell_2, \ell_3, \ell_4 = \ell_{min}}^{\ell}}&&\frac{(T_{\kappa}^{CDM} - T_{\kappa}^{FDM})^2(\ell_1, \ell_2, \ell_3, \ell_4)}{\mathrm{cov}(T_{\kappa}^{CDM})(\ell_1, \ell_2, \ell_3, \ell_4)}\label{eq:TrispectrumChiSquared}.
\end{alignat}

\begin{figure}
\centering
\includegraphics[width=.47\textwidth]{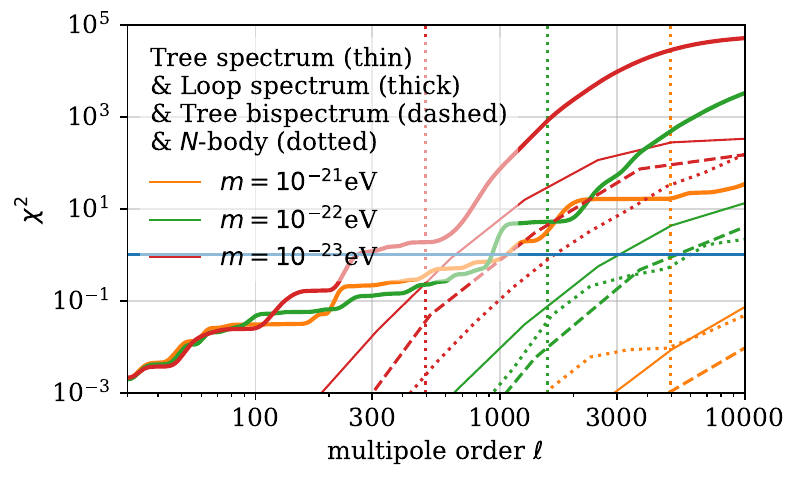}
\caption{Differentiability between cold dark matter and fuzzy dark matter in terms of the $\chi^2$-functional for weak lensing spectra and bispectra as a function of the maximum multipole order $\ell$ according to Eqs. \eqref{eq:SpectrumChiSquared} and \eqref{eq:BispectrumChiSquared}. The vertical, dotted lines correspond to $0.1\times\ell_J$, where the quantum Jeans multipole order $\ell_J$ is defined in Eq. \eqref{eq:JeansMultipoleOrder}. The horizontal blue line corresponds to $\chi^2 = 1$.}
\label{fig:FDMCDMChiSquared}
\end{figure}

Fig. \ref{fig:FDMCDMChiSquared} shows the $\chi^2$-functionals for distinguishing CDM and FDM computed according to Eqs. \eqref{eq:SpectrumChiSquared}, \eqref{eq:BispectrumChiSquared} where all sums are again expressed as integrals and calculated using the \textsc{CUBA}-library \citep{Hahn2004}. 

\begin{figure}
\centering
\includegraphics[width=.47\textwidth]{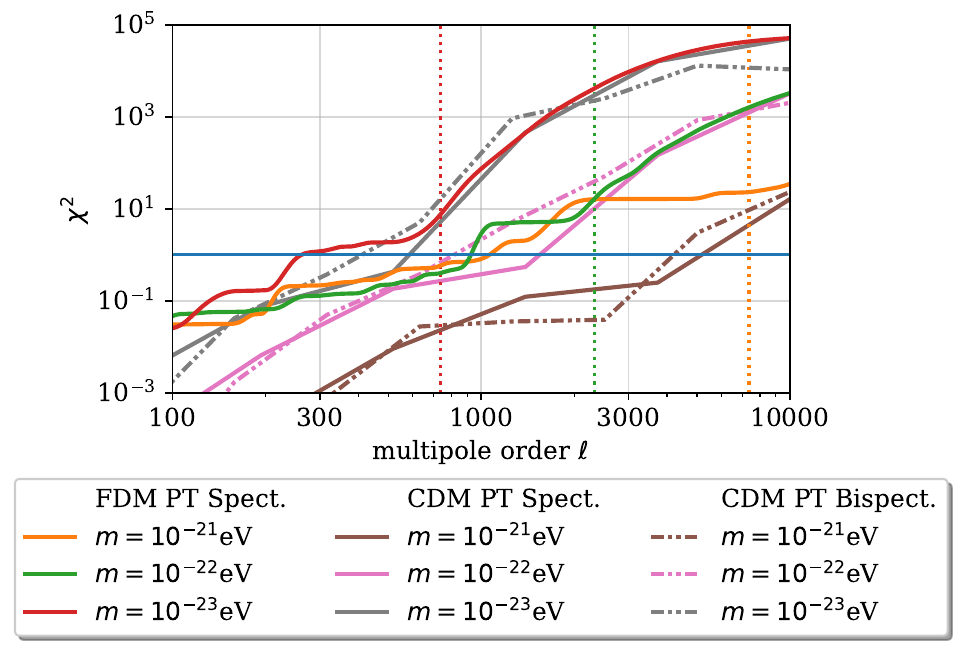}
\caption{Differentiability between cold dark matter and fuzzy dark matter in terms of the $\chi^2$-functional for weak lensing spectra and bispectra as a function of the maximum multipole order $\ell$ according to Eqs. \eqref{eq:SpectrumChiSquared}, \eqref{eq:BispectrumChiSquared} and \eqref{eq:TrispectrumChiSquared}. Both FDM PT and FDM dynamics approximated by CDM PT with FDM IC are shown. The vertical, dotted lines correspond to $0.1\times\ell_J$, where the quantum Jeans multipole order $\ell_J$ is defined in Eq. \eqref{eq:JeansMultipoleOrder}.}
\label{fig:FDMCDMChiSquaredFDMIC2}
\end{figure}

\begin{figure}
\centering
\includegraphics[width=.47\textwidth]{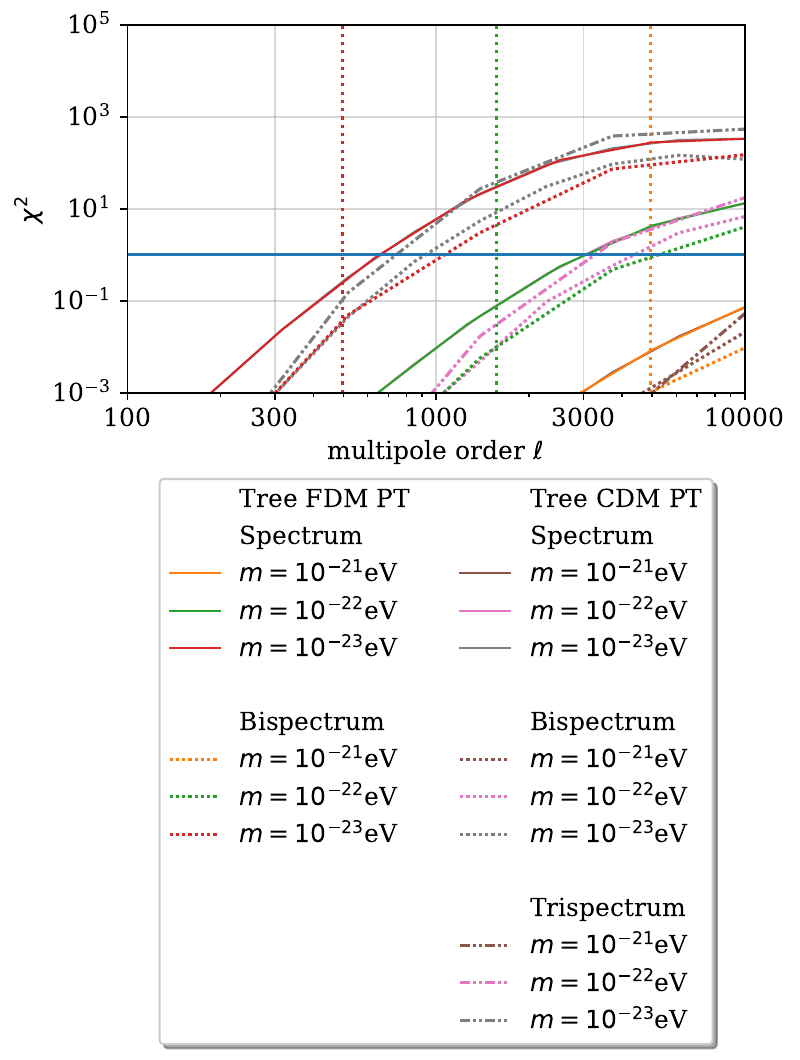}
\caption{Differentiability between cold dark matter and fuzzy dark matter in terms of the $\chi^2$-functional for weak lensing spectra, bispectra and trispectra as a function of the maximum multipole order $\ell$ according to Eqs. \eqref{eq:SpectrumChiSquared}, \eqref{eq:BispectrumChiSquared} and \eqref{eq:TrispectrumChiSquared}. Both FDM PT and FDM dynamics approximated by CDM PT with FDM IC are shown. The vertical, dotted lines correspond to $0.1\times\ell_J$, where the quantum Jeans multipole order $\ell_J$ is defined in Eq. \eqref{eq:JeansMultipoleOrder}.}
\label{fig:FDMCDMChiSquaredFDMIC}
\end{figure}

\highlighttext{We observe that the $\chi^2$-values obtained from the $N$-body simulation in the $L=30$ Mpc$/h$ box from $k$-modes between $6$ $h$/Mpc and $30$ $h$/Mpc are significantly lower than the PT predictions at loop- and even at tree-level. While we expected linear PT to provide a conservative estimate of the distinguishing power of the model lensing survey, it seems that the restoration of small scale power through late-time nonlinear CDM dynamics might cause the real signal to be smaller predicted by PT. 
Further contrasting FDM PT and CDM PT in Fig.  \ref{fig:FDMCDMChiSquaredFDMIC2}, we conclude that the difference in late-time dynamics between FDM and CDM is negligible for the $\chi^2$-values computed via PT. The only obvious difference at high $\ell$ arises for the $m = 10^{-21}$ eV loop-level PT prediction. Since the other spectra agree well and the FDM PT loop lensing prediction is computed using splines, we conclude that the respective FDM PT $\chi^2$ functional is dominated by noise up to high $\ell$. Fig. \ref{fig:FDMCDMChiSquaredFDMIC} displaying the tree-level $\chi^2$-functionals for FDM PT and CDM PT with FDM IC underscores this conclusion.}

% --- section: summary --- %
\section{summary and discussion}\label{sect_summary}
In this work, we studied structure formation in the cold dark matter and fuzzy dark matter models and their possible distinction through a Euclid-like weak lensing survey. We extended Eulerian perturbation theory to account for genuine quantum mechanical effects on the de Broglie scale of the dark matter particle. For sufficiently light elementary particles, such as axions and axion-like particles, this scale can be set to be relevant for cosmological structures on the scale of galaxies and below, typically for masses in the range of $10^{-23\ldots -21}$~eV. As a consequence of the Madelung transform of the Sch{\"o}dinger equation, the fluid mechanical equations acquire a quantum-pressure term, counteracting structure formation on scales smaller than the de Broglie scale.
We draw the following conclusions:
\begin{enumerate}
\item{Evolving the density and velocity fields using Eulerian perturbation theory including a quantum pressure term with suitable Gaussian initial conditions leads us to perturbative expressions for the bi- and trispectra, as well as nonlinear corrections to the spectra themselves. We can consistently compute corrections to the spectra and the bispectra at loop-level using FDM perturbation theory and CDM perturbation theory with FDM IC for comparison. Limitations in the numerical evaluation of the resulting integrals via adaptive integration schemes restrict us to tree-level evaluation of the trispectra. \highlighttext{In all computations, nonlinear structure formation as predicted by PT does not make up for the lack of initial power on small scales whereas the $N$-body prediction restores small-scale power on mildly nonlinear scales almost completely. In both CDM and FDM, non-linear effects leads to increased spectral amplitudes on scales larger than the de Broglie scale.}}

\item{Limber projection in the flat-sky approximation with lensing efficiency functions incorporating a Euclid-like source redshift distribution yields lensing spectra, bispectra and trispectra. Again, the lensing quantities are evaluated using adaptive integration. We choose lensing as an observational channel in order to be independent of any biasing assumption typical for galaxy surveys. The respective shear spectra can be found through a correlation analysis on galaxy shapes. As expected, we find a loss in power on the de Broglie scale in the FDM models compared to the CDM model. \highlighttext{However, the loss of power predicted by the $N$-body simulations is significantly smaller than the loss of power predicted by FDM PT.} For reference, we convert the de Broglie scale into an angular scale with a typical comoving distance corresponding to the median redshift. Measurements of bi- and even trispectra are clearly within reach of Euclid, with a significance of a few hundred $\sigma$: In these estimates we use a full configuration and scale integration with shape noise and a nonlinear covariance.}

\item{With a similar numerical computation, we can estimate whether the spectra, bispectra and trispectra for the CDM and FDM cases are distinguishable: For this purpose, we compute the $\chi^2$-functional for a given Gaussian-approximated nonlinear CDM covariance as a function of the FDM particle mass. As in the case of the signal to noise-computations, the $\chi^2$-functionals follow from a complete configuration space integration up to a limiting multipole, and we consider $\Delta\chi^2 = 1$ as a rough criterion of distinguishability: For FDM at tree-level with $m = 10^{-23}$~eV this limit is reached at $\ell \sim 700$. \highlighttext{The $N$-body lensing prediction indicates that this limit is reached at $\ell \sim 1000 - 2000$. Yet, for $m = 10^{-23}$~eV the late-time impact of the quantum pressure on the scales considered is non-negligible which is why this result suffers from a substantial uncertainty.} For masses $m = 10^{-22}$~eV and $m = 10^{-21}$~eV, the respective signals in our weak lensing survey are too weak to distinguish CDM and FDM below $\ell = 3\times 10^3$; this is the maximum multipole order accessible in a Euclid-like survey. The lensing bispectrum gives a lower $\chi^2$-functional than the lensing spectrum at tree-level, but might still serve as an important cross-validation.}

\item{\highlighttext{CDM loop-level perturbative results are found to agree reasonably well with the CDM $N$-body prediction while FDM loop-level PT significantly underestimates the power on scales $k \sim 1 - 100$ $h$/Mpc when compared to the the $N$-body prediction. Unless, the additional suppression of small-scale power through late-time FDM dynamics significantly enhances the $\chi^2$-functionals, the values $\chi^2 = 1$ at $\ell \sim 600$ and $\ell \sim 1500$ for $m = 10^{-23}$~eV and $m = 10^{-22}$~eV by the loop-level lensing spectra are found to be unreliable. In any case, the signal for $m = 10^{-21}$~eV is still too weak to be measurable in a realistic weak lensing survey. Considering PT at loop-level, the lensing bispectrum gives slightly higher $\chi^2$-values than the lensing spectrum. Again, these values are likely unreliable and overestimate the potential of a Euclid-like weak lensing survey.}}

\item{The main uncertainty in our perturbative approach stems from the modelling of nonlinear structure formation via perturbation theory. \highlighttext{In general, the extent to which the FDM and CDM models can be compared depends on the times and scales considered. As the analysis of the impact of a high-$k$ cutoff on the loop-level matter power spectra demonstrates, the bulk of the lensing signal comes from modes with $k > 3$ $h$/Mpc for all masses considered. At these scales, loop-level PT is known to underestimate the true power spectra at late times. However, as the comparison with the $N$-body simulations highlights, full nonlinear CDM dynamics make up for the initial lack of small-scale power while loop-level PT does not. As a consequence, PT seems to significantly overestimate the attainable $\chi^2$-functionals and is therefore not the right tool to distinguish the considered FDM models from the CDM model in our weak lensing survey. In contrast, PT might be better suited for other observational channels, such as neutral hydrogen surveys for probing the cosmic dawn. They are confined to probing the cosmic state before reionisation at $z \sim 6$ where PT still provides a better estimate for the nonlinearity at smaller scales.} Technical and numerical details of the FDM perturbation theory are described in detail in appendices~\ref{appendix:NLEPT} to~\ref{appendix:InfraredSafeIntegrands}, along with the representation of the coupling kernels in terms of Feynman diagrams.}
\item{\highlighttext{The estimation of the power spectra from $N$-body simulations is subject to a number of uncertainties: We only run one simulation per box size and initial condition. Ideally, one averages over a statistically representative sample of simulations with different initial conditions and average the resulting spectra to obtain error estimates. The large-scale power suffers estimated from runs in small simulation volumes is underestimated and suffers from high variance because of a lack of samples. Most importantly, a consistent FDM simulation in a sufficiently large simulation box would be required to reliably estimate the late-time power spectrum on small-scales and as a also consequence the lensing spectra for the FDM masses $m = 10^{-22}$ eV and $m = 10^{-23}$ eV considered here. }}
\end{enumerate}

% --- section: acknowledgements --- %
\section*{acknowledgements}
We would like to thank Guan-Ming Su for providing power spectra from numerical FDM simulations to crosscheck the perturbation theoretical results.  This work made significant use of many open source software packages, including \textsc{Python}, \textsc{IPython}, \textsc{NumPy}, \textsc{SciPy}, \textsc{Matplotlib}, \textsc{YT}, \textsc{Cuba} and \textsc{GSL}. These are products of collaborative effort by many independent developers from numerous institutions around the world. Their commitment to open science has helped to make this work possible.

% --- section: data availability --- %
\section*{data availability statement}
The code that computes the matter spectra, bispectra and trispectra in CDM and FDM is available under 
\href{https://github.com/KunkelAlexander/fdm-eulerpt}{https://github.com/KunkelAlexander/fdm-eulerpt}.

% --- bibliography --- %
\bibliographystyle{mnras}
\bibliography{references.bib, references3.bib}

\begin{thebibliography}{}
\makeatletter
\relax
\def\mn@urlcharsother{\let\do\@makeother \do\$\do\&\do\#\do\^\do\_\do\%\do\~}
\def\mn@doi{\begingroup\mn@urlcharsother \@ifnextchar [ {\mn@doi@}
  {\mn@doi@[]}}
\def\mn@doi@[#1]#2{\def\@tempa{#1}\ifx\@tempa\@empty \href
  {http://dx.doi.org/#2} {doi:#2}\else \href {http://dx.doi.org/#2} {#1}\fi
  \endgroup}
\def\mn@eprint#1#2{\mn@eprint@#1:#2::\@nil}
\def\mn@eprint@arXiv#1{\href {http://arxiv.org/abs/#1} {{\tt arXiv:#1}}}
\def\mn@eprint@dblp#1{\href {http://dblp.uni-trier.de/rec/bibtex/#1.xml}
  {dblp:#1}}
\def\mn@eprint@#1:#2:#3:#4\@nil{\def\@tempa {#1}\def\@tempb {#2}\def\@tempc
  {#3}\ifx \@tempc \@empty \let \@tempc \@tempb \let \@tempb \@tempa \fi \ifx
  \@tempb \@empty \def\@tempb {arXiv}\fi \@ifundefined
  {mn@eprint@\@tempb}{\@tempb:\@tempc}{\expandafter \expandafter \csname
  mn@eprint@\@tempb\endcsname \expandafter{\@tempc}}}

\bibitem[\protect\citeauthoryear{Amorisco \& Loeb}{Amorisco \&
  Loeb}{2018}]{Amorisco2018}
Amorisco N.~C.,  Loeb A.,  2018, First constraints on Fuzzy Dark Matter from
  the dynamics of stellar streams in the Milky Way,
  \mn@doi{10.48550/ARXIV.1808.00464}

\bibitem[\protect\citeauthoryear{Armengaud, Palanque-Delabrouille, Y{\`{e}}che,
  Marsh  \& Baur}{Armengaud et~al.}{2017}]{Armengaud2017}
Armengaud E.,  Palanque-Delabrouille N.,  Y{\`{e}}che C.,  Marsh D. J.~E.,
  Baur J.,  2017, \mn@doi [Monthly Notices of the Royal Astronomical Society]
  {10.1093/mnras/stx1870}, 471, 4606

\bibitem[\protect\citeauthoryear{Baldauf, Mercolli, Mirbabayi  \&
  Pajer}{Baldauf et~al.}{2014}]{Baldauf2014}
Baldauf T.,  Mercolli L.,  Mirbabayi M.,   Pajer E.,  2014, \mn@doi [Journal of
  Cosmology and Astroparticle Physics] {10.1088/1475-7516/2015/05/007}, 2015,
  007

\bibitem[\protect\citeauthoryear{Bar, Blas, Blum  \& Sibiryakov}{Bar
  et~al.}{2018}]{Bar2018}
Bar N.,  Blas D.,  Blum K.,   Sibiryakov S.,  2018, \mn@doi [Physical Review D]
  {10.1103/PhysRevD.98.083027}, 98

\bibitem[\protect\citeauthoryear{Bartelmann, Fabis, Berg, Kozlikin, Lilow  \&
  Viermann}{Bartelmann et~al.}{2016}]{Bartelmann2016b}
Bartelmann M.,  Fabis F.,  Berg D.,  Kozlikin E.,  Lilow R.,   Viermann C.,
  2016, \mn@doi [New Journal of Physics] {10.1088/1367-2630/18/4/043020}, 18,
  043020

\bibitem[\protect\citeauthoryear{{Bird}}{{Bird}}{2017}]{GenPK}
{Bird} S.,  2017, {GenPK: Power spectrum generator}, Astrophysics Source Code
  Library, record ascl:1706.006 (\mn@eprint {ascl} {1706.006})

\bibitem[\protect\citeauthoryear{Calabrese \& Spergel}{Calabrese \&
  Spergel}{2016}]{Calabrese2016}
Calabrese E.,  Spergel D.~N.,  2016, \mn@doi [Monthly Notices of the Royal
  Astronomical Society] {10.1093/mnras/stw1256}, 460, 4397

\bibitem[\protect\citeauthoryear{Carrasco, Foreman, Green  \&
  Senatore}{Carrasco et~al.}{2013}]{Carrasco2013}
Carrasco J. J.~M.,  Foreman S.,  Green D.,   Senatore L.,  2013, \mn@doi
  [Journal of Cosmology and Astroparticle Physics]
  {10.1088/1475-7516/2014/07/056}, 2014, 056

\bibitem[\protect\citeauthoryear{Carrasco, Foreman, Green  \&
  Senatore}{Carrasco et~al.}{2014}]{Carrasco2014}
Carrasco J. J.~M.,  Foreman S.,  Green D.,   Senatore L.,  2014, \mn@doi
  [Journal of Cosmology and Astroparticle Physics]
  {10.1088/1475-7516/2014/07/057}, 2014, 057

\bibitem[\protect\citeauthoryear{Chavanis}{Chavanis}{2012}]{Chavanis2011b}
Chavanis P.-H.,  2012, \mn@doi [Astronomy {\&} Astrophysics]
  {10.1051/0004-6361/201116905}, 537, A127

\bibitem[\protect\citeauthoryear{Chen, Schive  \& Chiueh}{Chen
  et~al.}{2017}]{Chen2016}
Chen S.-R.,  Schive H.-Y.,   Chiueh T.,  2017, \mn@doi [Monthly Notices of the
  Royal Astronomical Society] {10.1093/mnras/stx449}, 468, 1338

\bibitem[\protect\citeauthoryear{Church, Mocz  \& Ostriker}{Church
  et~al.}{2019}]{Church2019}
Church B.~V.,  Mocz P.,   Ostriker J.~P.,  2019, \mn@doi [Monthly Notices of
  the Royal Astronomical Society] {10.1093/mnras/stz534}, 485, 2861

\bibitem[\protect\citeauthoryear{Dalal \& Kravtsov}{Dalal \&
  Kravtsov}{2022}]{Dalal2022}
Dalal N.,  Kravtsov A.,  2022, \mn@doi [Physical Review D]
  {10.1103/physrevd.106.063517}, 106

\bibitem[\protect\citeauthoryear{Davies \& Mocz}{Davies \&
  Mocz}{2020}]{Davies2020}
Davies E.~Y.,  Mocz P.,  2020, \mn@doi [Monthly Notices of the Royal
  Astronomical Society] {10.1093/mnras/staa202}, 492, 5721

\bibitem[\protect\citeauthoryear{Dentler, Marsh, Hlo{\v{z}}ek, Laguë, Rogers
  \& Grin}{Dentler et~al.}{2022}]{Dentler2021}
Dentler M.,  Marsh D. J.~E.,  Hlo{\v{z}}ek R.,  Laguë A.,  Rogers K.~K.,
  Grin D.,  2022, \mn@doi [Monthly Notices of the Royal Astronomical Society]
  {10.1093/mnras/stac1946}, 515, 5646

\bibitem[\protect\citeauthoryear{Dome, Fialkov, Mocz, Schäfer, Boylan-Kolchin
  \& Vogelsberger}{Dome et~al.}{2022}]{Dome2022}
Dome T.,  Fialkov A.,  Mocz P.,  Schäfer B.~M.,  Boylan-Kolchin M.,
  Vogelsberger M.,  2022, \mn@doi [Monthly Notices of the Royal Astronomical
  Society] {10.1093/mnras/stac3766}, 519, 4183

\bibitem[\protect\citeauthoryear{Feix, Frank, Pargner, Reischke, Sch\"afer  \&
  Schwetz}{Feix et~al.}{2019}]{Feix:2019lpo}
Feix M.,  Frank J.,  Pargner A.,  Reischke R.,  Sch\"afer B.~M.,   Schwetz T.,
  2019, \mn@doi [JCAP] {10.1088/1475-7516/2019/05/021}, 05, 021

\bibitem[\protect\citeauthoryear{Feix, Hagstotz, Pargner, Reischke, Sch\"afer
  \& Schwetz}{Feix et~al.}{2020}]{Feix:2020txt}
Feix M.,  Hagstotz S.,  Pargner A.,  Reischke R.,  Sch\"afer B.~M.,   Schwetz
  T.,  2020, \mn@doi [JCAP] {10.1088/1475-7516/2020/11/046}, 11, 046

\bibitem[\protect\citeauthoryear{Fry}{Fry}{1984}]{Fry1984}
Fry J.~N.,  1984, \mn@doi [The Astrophysical Journal] {10.1086/161913}, 279,
  499

\bibitem[\protect\citeauthoryear{Gonz{\'{a}}lez-Morales, Marsh,
  Pe{\~{n}}arrubia  \& Ure{\~{n}}a-L{\'{o}}pez}{Gonz{\'{a}}lez-Morales
  et~al.}{2017}]{GonzalezMorales2017}
Gonz{\'{a}}lez-Morales A.~X.,  Marsh D. J.~E.,  Pe{\~{n}}arrubia J.,
  Ure{\~{n}}a-L{\'{o}}pez L.~A.,  2017, \mn@doi [Monthly Notices of the Royal
  Astronomical Society] {10.1093/mnras/stx1941}, 472, 1346

\bibitem[\protect\citeauthoryear{Goroff, Grinstein, Rey  \& Wise}{Goroff
  et~al.}{1986}]{Goroff1986}
Goroff M.~H.,  Grinstein B.,  Rey S.-J.,   Wise M.~B.,  1986, \mn@doi [The
  Astrophysical Journal] {10.1086/164749}, 311, 6

\bibitem[\protect\citeauthoryear{Hahn}{Hahn}{2005}]{Hahn2004}
Hahn T.,  2005, \mn@doi [Computer Physics Communications]
  {10.1016/j.cpc.2005.01.010}, 168, 78

\bibitem[\protect\citeauthoryear{Hahn \& Abel}{Hahn \& Abel}{2011}]{Hahn2011}
Hahn O.,  Abel T.,  2011, \mn@doi [Monthly Notices of the Royal Astronomical
  Society] {10.1111/j.1365-2966.2011.18820.x}, 415, 2101

\bibitem[\protect\citeauthoryear{Hlozek, Grin, Marsh  \& Ferreira}{Hlozek
  et~al.}{2014}]{axionCAMB}
Hlozek R.,  Grin D.,  Marsh D. J.~E.,   Ferreira P.~G.,  2014, \mn@doi
  [Physical Review D] {10.1103/PhysRevD.91.103512}, 91

\bibitem[\protect\citeauthoryear{Hlo{\v{z}}ek, Marsh  \& Grin}{Hlo{\v{z}}ek
  et~al.}{2018}]{Hlozek2018}
Hlo{\v{z}}ek R.,  Marsh D. J.~E.,   Grin D.,  2018, \mn@doi [Monthly Notices of
  the Royal Astronomical Society] {10.1093/mnras/sty271}, 476, 3063

\bibitem[\protect\citeauthoryear{Hotinli, Marsh  \& Kamionkowski}{Hotinli
  et~al.}{2022}]{Hotinli2022}
Hotinli S.~C.,  Marsh D.~J.,   Kamionkowski M.,  2022, \mn@doi [Physical Review
  D] {10.1103/physrevd.106.043529}, 106

\bibitem[\protect\citeauthoryear{Hu, Barkana  \& Gruzinov}{Hu
  et~al.}{2000}]{Hu2000}
Hu W.,  Barkana R.,   Gruzinov A.,  2000, \mn@doi [Physical Review Letters]
  {10.1103/PhysRevLett.85.1158}, 85, 1158

\bibitem[\protect\citeauthoryear{Hui, Ostriker, Tremaine  \& Witten}{Hui
  et~al.}{2017}]{Hui2017}
Hui L.,  Ostriker J.~P.,  Tremaine S.,   Witten E.,  2017, \mn@doi [Physical
  Review D] {10.1103/physrevd.95.043541}, 95

\bibitem[\protect\citeauthoryear{Iršič, Viel, Haehnelt, Bolton  \&
  Becker}{Iršič et~al.}{2017}]{Irsic2017}
Iršič V.,  Viel M.,  Haehnelt M.~G.,  Bolton J.~S.,   Becker G.~D.,  2017,
  \mn@doi [Physical Review Letters] {10.1103/PhysRevLett.119.031302}, 119,
  031302

\bibitem[\protect\citeauthoryear{Jain \& Bertschinger}{Jain \&
  Bertschinger}{1994}]{Jain1993}
Jain B.,  Bertschinger E.,  1994, \mn@doi [The Astrophysical Journal]
  {10.1086/174502}, 431, 495

\bibitem[\protect\citeauthoryear{Kaiser}{Kaiser}{1998}]{Kaiser1996}
Kaiser N.,  1998, \mn@doi [The Astrophysical Journal] {10.1086/305515}, 498, 26

\bibitem[\protect\citeauthoryear{Khlopov, Malomed  \& Zeldovich}{Khlopov
  et~al.}{1985}]{10.1093/mnras/215.4.575}
Khlopov M.~Y.,  Malomed B.~A.,   Zeldovich Y.~B.,  1985, \mn@doi [Monthly
  Notices of the Royal Astronomical Society] {10.1093/mnras/215.4.575}, 215,
  575

\bibitem[\protect\citeauthoryear{Kobayashi, Murgia, Simone, Iršič  \&
  Viel}{Kobayashi et~al.}{2017}]{Kobayashi2017}
Kobayashi T.,  Murgia R.,  Simone A.~D.,  Iršič V.,   Viel M.,  2017, \mn@doi
  [Physical Review D] {10.1103/PhysRevD.96.123514}, 96

\bibitem[\protect\citeauthoryear{Kozlikin, Lilow, Fabis  \&
  Bartelmann}{Kozlikin et~al.}{2021}]{Kozlikin:2020exj}
Kozlikin E.,  Lilow R.,  Fabis F.,   Bartelmann M.,  2021, \mn@doi [JCAP]
  {10.1088/1475-7516/2021/06/035}, 06, 035

\bibitem[\protect\citeauthoryear{Laguë, Bond, Hložek, Marsh  \&
  Söding}{Laguë et~al.}{2020}]{Lague2020}
Laguë A.,  Bond J.~R.,  Hložek R.,  Marsh D. J.~E.,   Söding L.,  2020,
  \mn@doi [Monthly Notices of the Royal Astronomical Society]
  {10.1093/mnras/stab601}, 504, 2391

\bibitem[\protect\citeauthoryear{Laureijs et~al.,}{Laureijs
  et~al.}{2011}]{Laureijs2011}
Laureijs R.,  et~al., 2011, arXiv preprint arXiv:1110.3193

\bibitem[\protect\citeauthoryear{Lewis, Challinor  \& Lasenby}{Lewis
  et~al.}{2000}]{CAMB}
Lewis A.,  Challinor A.,   Lasenby A.,  2000, \mn@doi [The Astrophysical
  Journal] {10.1086/309179}, 538, 473

\bibitem[\protect\citeauthoryear{Li, Hui  \& Bryan}{Li et~al.}{2019}]{Li2018}
Li X.,  Hui L.,   Bryan G.~L.,  2019, \mn@doi [Physical Review D]
  {10.1103/PhysRevD.99.063509}, 99

\bibitem[\protect\citeauthoryear{Lidz \& Hui}{Lidz \& Hui}{2018}]{Lidz2018}
Lidz A.,  Hui L.,  2018, \mn@doi [Physical Review D]
  {10.1103/physrevd.98.023011}, 98

\bibitem[\protect\citeauthoryear{Limber}{Limber}{1953}]{Limber1953}
Limber D.~N.,  1953, \mn@doi [The Astrophysical Journal] {10.1086/145672}, 117,
  134

\bibitem[\protect\citeauthoryear{Littek}{Littek}{2018}]{Littek:2018ljy}
Littek C.,  2018, PhD thesis, U. Heidelberg (main),
  \mn@doi{10.11588/heidok.00025108}

\bibitem[\protect\citeauthoryear{Madelung}{Madelung}{1927}]{Madelung1927}
Madelung E.,  1927, \mn@doi [Zeitschrift für Physik] {10.1007/bf01400372}, 40,
  322

\bibitem[\protect\citeauthoryear{Makino, Sasaki  \& Suto}{Makino
  et~al.}{1992}]{Makino1992}
Makino N.,  Sasaki M.,   Suto Y.,  1992, \mn@doi [Physical Review D]
  {10.1103/physrevd.46.585}, 46, 585

\bibitem[\protect\citeauthoryear{Maleki, Baghram  \& Rahvar}{Maleki
  et~al.}{2020}]{Maleki2019}
Maleki A.,  Baghram S.,   Rahvar S.,  2020, \mn@doi [Physical Review D]
  {10.1103/PhysRevD.101.023508}, 101

\bibitem[\protect\citeauthoryear{Marsh}{Marsh}{2016}]{Marsh2016}
Marsh D.~J.,  2016, \mn@doi [Physics Reports] {10.1016/j.physrep.2016.06.005},
  643, 1

\bibitem[\protect\citeauthoryear{Marsh \& Niemeyer}{Marsh \&
  Niemeyer}{2019}]{Marsh2018}
Marsh D. J.~E.,  Niemeyer J.~C.,  2019, \mn@doi [Physical Review Letters]
  {10.1103/PhysRevLett.123.051103}, 123

\bibitem[\protect\citeauthoryear{Marsh, Macaulay, Trebitsch  \& Ferreira}{Marsh
  et~al.}{2012}]{Marsh2011}
Marsh D. J.~E.,  Macaulay E.,  Trebitsch M.,   Ferreira P.~G.,  2012, \mn@doi
  [Physical Review D] {10.1103/PhysRevD.85.103514}, 85

\bibitem[\protect\citeauthoryear{May \& Springel}{May \&
  Springel}{2021}]{May2021}
May S.,  Springel V.,  2021, \mn@doi [Monthly Notices of the Royal Astronomical
  Society] {10.1093/mnras/stab1764}, 506, 2603

\bibitem[\protect\citeauthoryear{May \& Springel}{May \&
  Springel}{2022}]{May2022}
May S.,  Springel V.,  2022, \mn@doi [Monthly Notices of the Royal Astronomical
  Society] {10.1093/mnras/stad2031}, 524, 4256

\bibitem[\protect\citeauthoryear{Nadler et~al.,}{Nadler
  et~al.}{2021}]{Nadler2021}
Nadler E.,  et~al., 2021, \mn@doi [Physical Review Letters]
  {10.1103/physrevlett.126.091101}, 126

\bibitem[\protect\citeauthoryear{Nori \& Baldi}{Nori \& Baldi}{2018}]{Nori2018}
Nori M.,  Baldi M.,  2018, \mn@doi [Monthly Notices of the Royal Astronomical
  Society] {10.1093/mnras/sty1224}, 478, 3935

\bibitem[\protect\citeauthoryear{Nori, Murgia, Iršič, Baldi  \& Viel}{Nori
  et~al.}{2018}]{Nori2018a}
Nori M.,  Murgia R.,  Iršič V.,  Baldi M.,   Viel M.,  2018, \mn@doi [Monthly
  Notices of the Royal Astronomical Society] {10.1093/mnras/sty2888}, 482, 3227

\bibitem[\protect\citeauthoryear{Pietroni}{Pietroni}{2008}]{Pietroni:2008jx}
Pietroni M.,  2008, \mn@doi [JCAP] {10.1088/1475-7516/2008/10/036}, 10, 036

\bibitem[\protect\citeauthoryear{Powell, Vegetti, McKean, White, Ferreira, May
  \& Spingola}{Powell et~al.}{2023}]{Powell2023}
Powell D.~M.,  Vegetti S.,  McKean J.~P.,  White S. D.~M.,  Ferreira E. G.~M.,
  May S.,   Spingola C.,  2023, \mn@doi [Monthly Notices of the Royal
  Astronomical Society: Letters] {10.1093/mnrasl/slad074}, 524, L84

\bibitem[\protect\citeauthoryear{Rogers \& Peiris}{Rogers \&
  Peiris}{2021}]{Rogers2021}
Rogers K.~K.,  Peiris H.~V.,  2021, \mn@doi [Physical Review Letters]
  {10.1103/physrevlett.126.071302}, 126, 071302

\bibitem[\protect\citeauthoryear{Rogers, Hlo{\v{z}}ek, Laguë, Ivanov, Philcox,
  Cabass, Akitsu  \& Marsh}{Rogers et~al.}{2023}]{Rogers2023}
Rogers K.~K.,  Hlo{\v{z}}ek R.,  Laguë A.,  Ivanov M.~M.,  Philcox O.~H.,
  Cabass G.,  Akitsu K.,   Marsh D.~J.,  2023, \mn@doi [Journal of Cosmology
  and Astroparticle Physics] {10.1088/1475-7516/2023/06/023}, 2023, 023

\bibitem[\protect\citeauthoryear{Safarzadeh \& Spergel}{Safarzadeh \&
  Spergel}{2020}]{Safarzadeh_2020}
Safarzadeh M.,  Spergel D.~N.,  2020, \mn@doi [The Astrophysical Journal]
  {10.3847/1538-4357/ab7db2}, 893, 21

\bibitem[\protect\citeauthoryear{Sarkar, Pandey  \& Sethi}{Sarkar
  et~al.}{2021}]{Sarkar2021}
Sarkar A.~K.,  Pandey K.~L.,   Sethi S.~K.,  2021, \mn@doi [Journal of
  Cosmology and Astroparticle Physics] {10.1088/1475-7516/2021/10/077}, 2021,
  077

\bibitem[\protect\citeauthoryear{Schive, Chiueh  \& Broadhurst}{Schive
  et~al.}{2014}]{Schive2014}
Schive H.-Y.,  Chiueh T.,   Broadhurst T.,  2014, \mn@doi [Nature Physics]
  {10.1038/nphys2996}, 10, 496

\bibitem[\protect\citeauthoryear{Schive, Chiueh, Broadhurst  \& Huang}{Schive
  et~al.}{2016}]{Schive2015}
Schive H.-Y.,  Chiueh T.,  Broadhurst T.,   Huang K.-W.,  2016, \mn@doi [The
  Astrophysical Journal] {10.3847/0004-637X/818/1/89}, 818, 89

\bibitem[\protect\citeauthoryear{Schive, Chiueh  \& Broadhurst}{Schive
  et~al.}{2020}]{Schive2020}
Schive H.-Y.,  Chiueh T.,   Broadhurst T.,  2020, \mn@doi [Physical Review
  Letters] {10.1103/physrevlett.124.201301}, 124

\bibitem[\protect\citeauthoryear{Scoccimarro}{Scoccimarro}{1998}]{Scoccimarro1998}
Scoccimarro R.,  1998, \mn@doi [Monthly Notices of the Royal Astronomical
  Society] {10.1046/j.1365-8711.1998.01845.x}, 299, 1097

\bibitem[\protect\citeauthoryear{Scoccimarro}{Scoccimarro}{2006}]{Scoccimarro2006}
Scoccimarro R.,  2006, \mn@doi [Annals of the New York Academy of Sciences]
  {10.1111/j.1749-6632.2001.tb05618.x}, 927, 13

\bibitem[\protect\citeauthoryear{Scoccimarro, Zaldarriaga  \& Hui}{Scoccimarro
  et~al.}{1999}]{Scoccimarro1999}
Scoccimarro R.,  Zaldarriaga M.,   Hui L.,  1999, \mn@doi [The Astrophysical
  Journal] {10.1086/308059}, 527, 1

\bibitem[\protect\citeauthoryear{Springel}{Springel}{2005}]{Springel2005}
Springel V.,  2005, \mn@doi [Monthly Notices of the Royal Astronomical Society]
  {10.1111/j.1365-2966.2005.09655.x}, 364, 1105

\bibitem[\protect\citeauthoryear{Su{\'{a}}rez \& Chavanis}{Su{\'{a}}rez \&
  Chavanis}{2015}]{Chavanis2015}
Su{\'{a}}rez A.,  Chavanis P.-H.,  2015, \mn@doi [Physical Review D]
  {10.1103/PhysRevD.92.023510}, 92, 023510

\bibitem[\protect\citeauthoryear{Takada \& Jain}{Takada \&
  Jain}{2004}]{Takada2003}
Takada M.,  Jain B.,  2004, \mn@doi [Monthly Notices of the Royal Astronomical
  Society] {10.1111/j.1365-2966.2004.07410.x}, 348, 897

\bibitem[\protect\citeauthoryear{Wolfram~Research}{Wolfram~Research}{2021}]{Mathematica}
Wolfram~Research I.,  2021, Mathematica, {V}ersion 12.3.1, \url
  {https://www.wolfram.com/mathematica}

\bibitem[\protect\citeauthoryear{Woo \& Chiueh}{Woo \& Chiueh}{2009}]{Woo2009}
Woo T.-P.,  Chiueh T.,  2009, \mn@doi [The Astrophysical Journal]
  {10.1088/0004-637X/697/1/850}, 697, 850

\bibitem[\protect\citeauthoryear{Zhang, Liu  \& Chu}{Zhang
  et~al.}{2019}]{Zhang2019}
Zhang J.,  Liu H.,   Chu M.-C.,  2019, \mn@doi [Frontiers in Astronomy and
  Space Sciences] {10.3389/fspas.2018.00048}, 5

\bibitem[\protect\citeauthoryear{Ünal, Pacucci  \& Loeb}{Ünal
  et~al.}{2021}]{Uenal2021}
Ünal C.,  Pacucci F.,   Loeb A.,  2021, \mn@doi [Journal of Cosmology and
  Astroparticle Physics] {10.1088/1475-7516/2021/05/007}, 2021, 007

\makeatother
\end{thebibliography}

% --- appendix --- %
\appendix

% --- section: time-dependent perturbation theory --- %
\section{time-dependent perturbation theory}
\label{appendix:NLEPT}
In this section, we develop a general framework for time-dependent Eulerian PT that we then apply to FDM. The CDM case follows by setting $\hbar = 0$, i.e. the formal transition from quantum to classical mechanics. We start with a system of two coupled equations of the form
\begin{align}
    \partial_{\tau} \delta  &= - \nabla \cdot \left((1+\delta)\bm{\upsilon}\right), \label{eq:GeneralContinuity}\\
    \partial_{\tau} \theta &= f(\theta, \delta) \label{eq:GeneralEuler}
\end{align}
where $\nabla \cdot \bm{\upsilon} = \theta$, $f$ is a well-behaved functions that is allowed to depend on $\theta$ and $\delta$ and their spatial derivatives and we work in the conformal time $\tau$. Hence, Eq. \eqref{eq:GeneralEuler} includes both the ideal fluid equations with pressure if we neglect the vorticity degrees of freedom as well as the Madelung equations. In the latter case, $f(\theta, \delta)$ includes the full nonlinear quantum pressure term. By later expanding this expression in terms of the density contrast $\delta$ and truncating the resulting series at a given order in $\delta$, we perturbatively include the effects of the nonlinearities of the quantum pressure.
Next, we introduce the two-component vector
\begin{equation}
\label{eq:VectorPsi}
    \Psi = \left[\delta \quad \theta\right]^\top
\end{equation}
such that the density and velocity fields can be treated on equal footing. We now expand Eq. \eqref{eq:GeneralEuler} in terms of powers of $\delta$ and $\theta$ and then Fourier transform it:
\begin{equation}
\label{eq:CompactFluidEquations}
\begin{split}
    &\partial_\tau \Psi_a(\bm{k}) + \Omega_{ab}(\bm{k}, \tau) \Psi_b(\bm{k}) =\\
    &\sum_{n=2}^\infty \delta_D(\bm{k} - \bm{k}_{1\ldots n}) \Gamma^n_{a,i_1\ldots i_n}(\bm{k}, \bm{k}_1, \ldots, \bm{k}_n, \tau)\\
    &\times \Psi_{i_1}(\bm{k}_1, \tau) \times \ldots \times \Psi_{i_n}(\bm{k}_{n}, \tau)
\end{split}
\end{equation}
where we introduced the time- and scale-dependent mode-coupling matrices  $\Omega_{ab}$ and $\Gamma_{a, i_1,\ldots,i_n}$ for the indices $a,b,i_1,\ldots,i_n \in \{1, 2\}$ and we employ the Einstein sum convention. Moreover, we employ a convention where integration over momenta $\bm{k}_i, \bm{k}_j$ with equal indices $i, j$ is understood, e.g.:
\begin{equation}
\begin{split}
\delta_D&(\bm{k} - \bm{k}_{12}) \Gamma(\bm{k}_1, \bm{k}_2) \psi(\bm{k}_1) \psi(\bm{k}_2) \equiv\\
&\int \frac{\mathrm{d}^3\bm{k}_1\mathrm{d}^3\bm{k}_2}{(2\pi)^6}  \delta_D(\bm{k} - \bm{k}_{12}) \Gamma(\bm{k}_1, \bm{k}_2) \psi(\bm{k}_1) \psi(\bm{k}_2).
\end{split}
\end{equation}
The matrix $\Omega_{ab}$ encodes the linearised fluctuations, whereas the matrices $\Gamma_{a, i_1\ldots i_n}$ encode all nonlinearities. The meaning of the indices can be easily understood: The index $a$ tells us whether we are looking at a contribution to the density contrast $\delta$ or the velocity divergence $\theta$. The indices $i$ tells us which fields couple to one another. The continuity equation Eq. \eqref{eq:GeneralContinuity} gives $\Omega_{11} = 0$ and $\Omega_{12} = 1$. We can therefore substitute
\begin{equation}
\begin{split}
    &\Psi_2(\bm{k}, \tau) = -\partial_{\tau} \Psi_1(\bm{k}, \tau)\\
    &+\sum_{n=2}^\infty \delta_D(\bm{k} - \bm{k}_{1\ldots n}) \Gamma^n_{1,i_1\ldots i_n}(\bm{k}, \bm{k}_1, \ldots, \bm{k}_n, \tau)\\
    &\times \Psi_{i_1}(\bm{k}_1, \tau) \times \ldots \times \Psi_{i_n}(\bm{k}_{n}, \tau)
\end{split}
\end{equation}
into Eq. \eqref{eq:GeneralEuler} for $a = 2$ to obtain
\begin{equation}
\label{eq:InhomogeneousGrowthEquation}
\partial^2_\tau \Psi_1 +\Omega_{22} \partial_{\tau} \Psi_1 - \Omega_{21}\Psi_1 = g(\bm{k}, \tau)
\end{equation}
where we omitted time and momentum variables and the inhomogeneity $g(\bm{k}, \tau)$ is defined as 
\begin{equation}
\label{eq:Inhomogeneity}
\begin{split}
    g(\bm{k}, \tau) \equiv &\sum_{n=2}^\infty \delta_D(\bm{k} - \bm{k}_{1\ldots n})\\
&\Bigl[\partial_{\tau} \bigl[\Gamma^n_{1,i_1\ldots i_n} \Psi_{i_1}(\bm{k}_1, \tau) \times \ldots \times \Psi_{i_n}(\bm{k}_n, \tau)\bigl]\\
&+ (\Omega_{22} - 1) \Gamma^n_{2,i_1\ldots i_n}  \Psi_{i_1}(\bm{k}_1, \tau) \times \ldots \times \Psi_{i_n}(\bm{k}_n, \tau)\Bigl].
\end{split}
\end{equation}
Linearising Eq. \eqref{eq:InhomogeneousGrowthEquation} gives the homogeneous, linear second-order ODE
\begin{equation}
    \partial^2_\tau \Psi_1^{(1)} +\Omega_{22} \partial_{\tau} \Psi_1^{(1)} - \Omega_{21}\Psi_1^{(1)}  = 0
\end{equation}
whose solutions are the linear growth factors $D_+$ and $D_-$ if we make the ansatz $\Psi_1^{(1)}(\bm{k}, \tau) = \delta(\bm{k}, \tau_0) D(\tau, \tau_0)$. 
A particular solution of an inhomogeneous second-order ODE can be found by the convolution of the inhomogeneity $g(k, a)$ with the Green's function $G_k(s, a)$:
\begin{equation}
\label{eq:Convolution}
    u(\bm{k}, \tau) = \int_{\tau_0}^\tau \mathrm{d}s\:g(\bm{k}, s) G_k(s, \tau)
\end{equation}
where the Green's function $G_k(s, \tau)$ of the ODE is given by a combination of two linearly independent solutions, i.e. $D_+$ and $D_-$, of the homogeneous equation
\begin{equation}
\label{eq:GreensFunction}
G(s, \tau) = \frac{D_-(s) D_+(\tau) - D_-(\tau) D_+(s)}{D_-(s) \partial_s D_+(s) - \partial_s D_-(s) D_+(s)}.
\end{equation}
The Green's function has the following properties:
\begin{equation*}
    G_k(\tau, \tau) = 0, \qquad
    \partial_{\tau} G_k(s, \tau)|_{s = \tau} = 1, \qquad
    \partial_{s} G_k(s, \tau)|_{s = \tau} = -1.
\end{equation*}

The crucial idea of the derivation of the coupling kernels is as follows: Writing down the convolution in Eq. \eqref{eq:Convolution} for the inhomogeneity defined in Eq. \eqref{eq:Inhomogeneity}, one obtains an integral equation for $\Psi_a$. This integral equation can be brought into the form of Eqs. \eqref{eq:DefinitionFn} and \eqref{eq:DefinitionGn}. Finding the coupling kernels $F_i$ and $G_i$ then boils down to comparing integrands. 
Pursuing this idea, we express the solution $\Psi_1$ of the full theory as sum of the homogeneous solution $\Psi^{(1)}_1$ and a particular solution
\begin{equation}
    \Psi_1(\bm{k}, \tau) = \Psi_1^{(1)}(\bm{k}, \tau) + \int_{\tau_0}^\tau \mathrm{d}s\,g(\bm{k}, s) G_k(s, \tau).
\end{equation}
We now parameterise the inhomogeneous solution in terms of the vertex couplings $C^{(n)}_{a, i_1\ldots i_n}(\bm{k}, \bm{k}_1, \ldots, \bm{k}_n, s, \tau)$ and write
\begin{equation}
\label{eq:InhomogeneityModeCoupling}
\begin{split}
    \Psi_a(\bm{k}, \tau) &= \Psi_a^{(1)}(\bm{k}, \tau) + \sum_{n=2}^\infty \delta_D(\bm{k} - \bm{k}_{1\ldots n})\\
    &\times \int_{\tau_0}^\tau \mathrm{d}s\, C^{(n)}_{a, i_1\ldots i_n}(\bm{k}, \bm{k}_1, \ldots, \bm{k}_n, s, \tau)\\
    &\times \Psi_{i_1}(\bm{k}_1, s) \times \ldots \times \Psi_{i_n}(\bm{k}_n, s).
\end{split}
\end{equation}
The vertex couplings for $\Psi_1$ can be immediately derived from Eq. \eqref{eq:Inhomogeneity} by partial integration of the derivative term $\partial_{s} \bigl[\Gamma_{1,i_1\ldots i_n} \Psi_{i_1}(\bm{k}_1, s) \times \ldots \times \Psi_{i_n}(\bm{k}_n, s)\bigl]$ using $G_k(\tau, \tau) = 0$:
\begin{equation}
\label{eq:C1CouplingKernels}
\begin{split}
    C^{(n)}_{1, i_1, \ldots, i_n}&(\bm{k}, \bm{k}_1, \ldots, \bm{k}_n, s, \tau) =\\ 
    &- \Gamma^n_{1,i_1\ldots i_n}(\bm{k}, \bm{k}_1, \ldots, \bm{k}_n, s) \partial_s G_k(s, \tau)\\
    &+ \left[\Omega_{22}(\bm{k}, \bm{k}_1, \ldots, \bm{k}_n, s) - 1\right]\\
    &\times \Gamma^n_{2,i_1\ldots i_n}(\bm{k}, \bm{k}_1, \ldots, \bm{k}_n, s) G_k(s, \tau).
\end{split}
\end{equation}

\begin{figure}
\centering
\includegraphics[page=2,width =.47\textwidth]{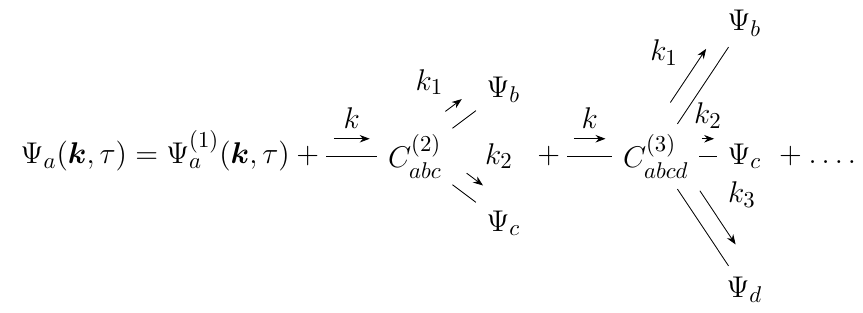}
%\begin{equation*}
%\Psi_a(\bm{k}, \tau)
%=
%\Psi_a^{(1)}(\bm{k}, \tau)
%+
%\feynmandiagram [small, baseline=\plusheight, %horizontal=a to b, tree layout] {
%  a [] -- [momentum = \(k\)] b %[particle=\(C^{(2)}_{abc}\)] -- [momentum = %\(k_1\)] c [particle = \(\Psi_b\)],
%  b -- [momentum = \(k_2\)] d [particle = %\(\Psi_c\)]
%};
%+
%\feynmandiagram [small, baseline=\plusheight, %horizontal=a to b, tree layout] {
%  a [] -- [momentum = \(k\)] b %[particle=\(C^{(3)}_{abcd}\)],
%  b -- [momentum = \(k_1\)] c [particle = %\(\Psi_b\)],
%  b -- [momentum = \(k_2\)] d [particle = %\(\Psi_c\)],
%  b -- [momentum = \(k_3\)] e [particle = %\(\Psi_d\)]
%};
%+
%\ldots.
%\end{equation*}
\caption{Diagrammatic representation of Eq. \eqref{eq:InhomogeneityModeCoupling} in terms of trees. Outgoing momentum arrows express momentum integrations. The Dirac-Delta functions enforce momentum conservation at the vertices and we employ the Einstein sum convention for repeated indices. The depth of the tree, i.e. $1$, denotes the number of time integrations.}
\label{fig:DiagrammaticalRepresentation}
\end{figure}

\begin{figure} 
%\begin{equation*}
%\feynmandiagram [small, baseline = %(a.base), horizontal=a to b, tree %layout] {
%  a [] -- [scalar, momentum=\(k\)] b %[particle = \(C^{(2)}_{abc}\)],
%  b -- [scalar, momentum=\(k_1\)] c %[particle = \(C^{(2)}_{bde}\)],
%  c -- [scalar, momentum=\(k_2\)] fd %[particle=\(f_d\)],
%  c-- [scalar, momentum=\(k_3\)] fe %[particle=\(f_e\)],
%  b -- [scalar] fc %[particle=\(f_c\)]
%};, \qquad
%%\feynmandiagram [small, baseline = %(a.base), horizontal=a to b, tree %%layout] {
%%  a [] -- [scalar, momentum=\(k\)] %b [particle = \(C^{(2)}_{abc}\)],
%%  b -- [scalar, momentum=\(k_1\)] %fb [particle=\(f_b\)], 
%%  b -- [scalar] c [particle = %\(C^{(2)}_{cde}\)],
%%  c -- [scalar, momentum=\(k_2\)] %fd [particle=\(f_d\)],
%%  c -- [scalar, momentum=\(k_3\)] %fe [particle=\(f_e\)]
%%};
%%\end{equation*}
%%\begin{equation*}
%\feynmandiagram [small, baseline = %(a.base), horizontal=a to b, tree %layout] {
%  a [] -- [scalar, momentum=\(k\)] b %[particle = \(C^{(3)}_{abcd}\)],
%  b -- [scalar, momentum=\(k_1\)] fa %[particle=\(f_b\)], 
%  b -- [scalar, momentum=\(k_2\)] fb %[particle=\(f_c\)],
%  b -- [scalar, momentum=\(k_3\)] fc %[particle=\(f_d\)]
%};
%\end{equation*}
\centering
\includegraphics[page=4, height=4cm]{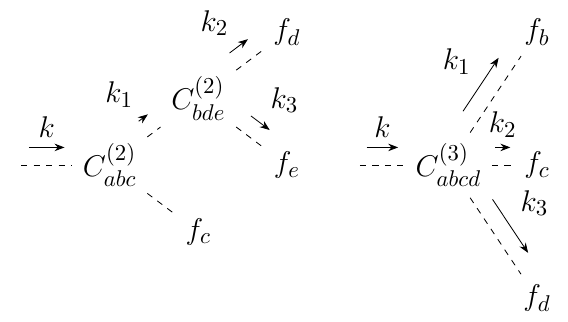}
\caption{Diagrammatic representation of contributions to PT kernels $F_3$ and $G_3$. The left diagram describes two contributions with the permutation $b \leftrightarrow c$.
%and corresponds to the Eq. \eqref{eq:I3}. 
The depth of the tree is $2$. Accordingly, it requires two time integrations. %The right diagrams corresponds to the expression Eq. \eqref{eq:J3}.
}
\label{fig:I3Diagram}
\end{figure}

\begin{figure}
\centering
\includegraphics[page=5, height=3cm]{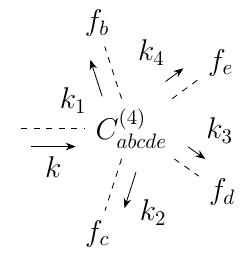}
%\begin{equation*}
%\feynmandiagram [small, horizontal=a %to b] {
%  a -- [scalar, momentum'=\(k\)] b %[particle = \(C^{(4)}_{abcde}\)],
%  b -- [scalar, momentum=\(k_1\)] fb %[particle=\(f_b\)], 
%  b -- [scalar, momentum=\(k_2\)] fc %[particle=\(f_c\)],
%  b -- [scalar, momentum=\(k_3\)] fd %[particle=\(f_d\)],
%  b -- [scalar, momentum=\(k_4\)] fe %[particle=\(f_e\)]
%};
%\end{equation*}
\caption{Diagrammatic representation of contributions to PT kernels $F_4$ and $G_4$ involving only the fourth-order vertex coupling $C^{(4)}$.}
\label{fig:W4Diagram}
\end{figure}

\begin{figure}
\centering
\includegraphics[page=6, height=4.5cm]{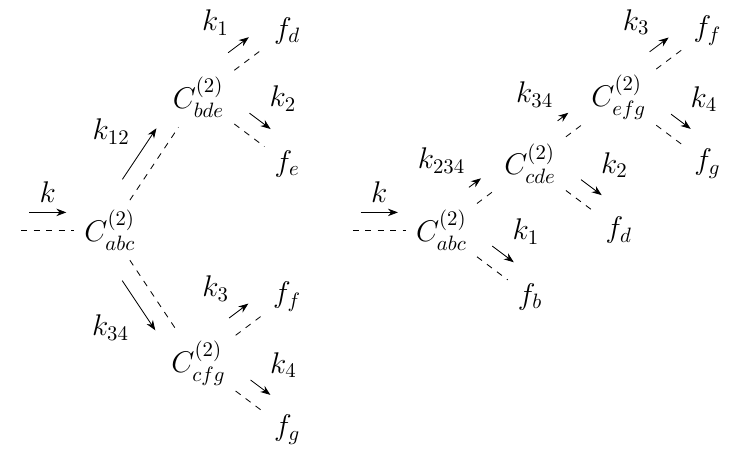}
%\begin{equation*}
%\feynmandiagram [small, baseline = %(a.base), horizontal=a to b, tree %layout] {
%  a [] -- [scalar, momentum=\(k\)] b %[particle = \(C^{(2)}_{abc}\)],
%  b -- [scalar, momentum=\(k_{12}\)] %c1 [particle=\(C^{(2)}_{bde}\)], 
%  b -- [scalar, momentum=\(k_{34}\)] %c2 [particle=\(C^{(2)}_{cfg}\)],
%  c1 -- [scalar, momentum=\(k_1\)] %fd [particle=\(f_d\)],
%  c1 -- [scalar, momentum=\(k_2\)] %fe [particle=\(f_e\)],
%  c2 -- [scalar, momentum=\(k_3\)] %ff [particle=\(f_f\)],
%  c2 -- [scalar, momentum=\(k_4\)] %fg [particle=\(f_g\)],
%};\qquad 
%\feynmandiagram [small, baseline = %(a.base), horizontal=a to b, tree %layout] {
%  a [] -- [scalar, momentum=\(k\)] b %[particle = \(C^{(2)}_{abc}\)],
%  b -- [scalar, %momentum=\(k_{234}\)] c1 %[particle=\(C^{(2)}_{cde}\)], 
%  c1 -- [scalar, %momentum=\(k_{34}\)] c2 %[particle=\(C^{(2)}_{efg}\)],
%  b  -- [scalar, momentum=\(k_1\)] %fb [particle=\(f_b\)],
%  c1 -- [scalar, momentum=\(k_2\)] %fe [particle=\(f_d\)],
%  c2 -- [scalar, momentum=\(k_3\)] %ff [particle=\(f_f\)],
%  c2 -- [scalar, momentum=\(k_4\)] %fg [particle=\(f_g\)]
%};
%\end{equation*}
\caption{Diagrammatic representation of contributions to PT kernels $F_4$ and $G_4$ involving only the second-order vertex coupling $C^{(2)}$. There are one additional permutation $b \leftrightarrow c$ for the left diagram as well as three additional permutations $b \leftrightarrow c$ and $d \leftrightarrow e$ for the diagram on the right.}
\label{fig:IJ4Diagram}
\end{figure}

\begin{figure}
%\begin{equation*}
%\feynmandiagram [small, baseline = %(a.base), horizontal=a to b, tree %layout] {
%  a [] -- [scalar, momentum=\(k\)] b %[particle = \(C^{(2)}_{abc}\)],
%  b -- [scalar, %momentum=\(k_{234}\)] c1 %[particle=\(C^{(3)}_{cdef}\)], 
%  b -- [scalar, momentum=\(k_1\)] fb %[particle=\(f_b\)],
%  c1 -- [scalar, momentum=\(k_2\)] %fd [particle=\(f_d\)],
%  c1 -- [scalar, momentum=\(k_3\)] %fe [particle=\(f_e\)],
%  c1 -- [scalar, momentum=\(k_4\)] %ff [particle=\(f_f\)],
%};\qquad 
%\feynmandiagram [small, baseline = %(a.base), horizontal=a to b, tree %layout] {
%  a [] -- [scalar, momentum=\(k\)] b %[particle = \(C^{(3)}_{abcd}\)],
%  b -- [scalar, momentum=\(k_{34}\)] %c1 [particle=\(C^{(2)}_{bef}\)], 
%  b -- [scalar, momentum=\(k_1\)] fc %[particle=\(f_c\)],
%  b -- [scalar, momentum=\(k_2\)] fd %[particle=\(f_d\)],
%  c1 -- [scalar, momentum=\(k_3\)] %fe [particle=\(f_e\)],
%  c1 -- [scalar, momentum=\(k_4\)] %ff [particle=\(f_f\)],
%};
%\end{equation*}
\centering
\includegraphics[page=7, height=4.5cm]{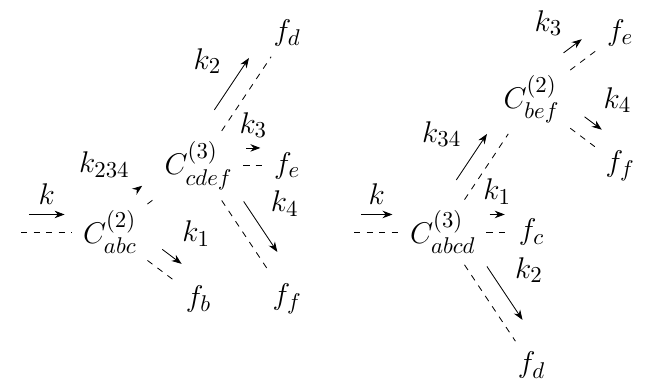}
\caption{Diagrammatic representation of contributions to PT kernels $F_4$ and $G_4$ involving both the second- and third-order vertex couplings $C^{(2)}$ and $C^{(3)}$. There are one additional permutation $b \leftrightarrow c$ for the diagram on the left and two additional permutations $b \leftrightarrow c \leftrightarrow d$ for the diagram on the right. }
\label{fig:HK4Diagram}
\end{figure}

The vertex couplings $C^{(n)}$ owe their name to the diagrammatic representation of Eq. \eqref{eq:InhomogeneityModeCoupling} shown in Fig. \ref{fig:DiagrammaticalRepresentation}.
In the following, we make use of this diagrammatic language to compute the higher-order coupling kernels. Eq. \eqref{eq:InhomogeneityModeCoupling} can be used to obtain the perturbative solution iteratively: We substitute $\Psi = \Psi^{(1)} + \mathcal{O}(\delta^2, \upsilon^2, \delta \upsilon)$ into the RHS of Eq. \eqref{eq:InhomogeneityModeCoupling} which gives an equation whose solution is $\Psi = \Psi^{(1)} + \Psi^{(2)} + \text{third-order terms}$. Substituting this expression back into the RHS of Eq. \eqref{eq:InhomogeneityModeCoupling} gives an equation whose solution is $\Psi = \Psi^{(1)} + \Psi^{(2)} + \Psi^{(3)} + \text{fourth-order terms}$ and so on. 
The following section gives expressions for $\Psi_a(\bm{k}, \tau)$ up to quartic order in order to compute the kernels $F_2$, $F_3$ and $F_4$.
The linear velocity divergence fluctuations are related to the density fluctuations via 
\begin{align}
    \Psi_1^{(1)}(\bm{k}, \tau) &= D_k(\tau)\Psi_1^{(1)}(\bm{k}, \tau_0),\\ \Psi_2^{(1)}(\bm{k}, \tau) &= -\partial_{\tau} D_k(\tau)\Psi_1^{(1)}(\bm{k}, \tau_0).
\end{align}
We therefore define the two-component vector $\bm{f}$ that relates the linear density and velocity divergence fluctuations at time $s$ with the linear density fluctuations at time $\tau$: 
\begin{align}
    \bm{f}(k, s, \tau) &= \left[\frac{D_k(s)}{D_k(\tau)}\quad -\frac{\partial_s D_k(s)}{D_k(\tau)}\right]^\top.
\end{align}
With this definition in place, $\Psi_a^{(2)}(\bm{k}, \tau)$ can be expressed as
\begin{equation}
\begin{split}
    \Psi_a^{(2)}(\bm{k}, \tau) &= \delta_D(\bm{k} - \bm{k}_{12}) \delta(\bm{k}_1, \tau) \delta(\bm{k}_2, \tau)\\
    &\times \int_{\tau_0}^\tau \mathrm{d}s\, C^{(2)}_{abc}(\bm{k}, \bm{k}_1, \bm{k}_2, s, \tau) f_b(k_1, s, \tau) f_c(k_2, s, \tau).
\end{split}
\end{equation}
\begin{figure} 
\centering
\includegraphics[page=3, height=3cm]{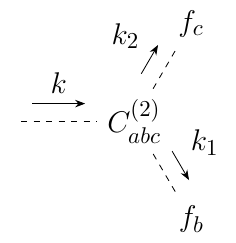}
%\begin{equation*}
%\feynmandiagram [small, horizontal=a %to b] {
%  a [] -- [scalar, momentum=\(k\)] b %[particle=\(C^{(2)}_{abc}\)] -- %[scalar, momentum=\(k_1\)] fb %[particle=\(f_b\)],
%  b -- [scalar, momentum=\(k_2\)] fc %[particle=\(f_c\)]
%};
%\end{equation*}
\caption{Diagrammatic representation of the coupling kernels $F_2$ and $G_2$. There is one free index $a$ that describes the density coupling kernel $F_2(\bm{k}, \bm{k}_1, \bm{k}_2)$ for $a=1$ and the velocity coupling kernel $G_2(\bm{k}, \bm{k}_1, \bm{k}_2)$ for $a=2$. The tree has depth $1$ and therefore requires a single time integration.}
\label{fig:DiagrammaticalF2}
\end{figure}
Comparing this expression to the definition of $F_n$ in Eq. \eqref{eq:DefinitionFn}, we can finally read off the coupling kernels. The coupling kernel $F_2$ can be expressed diagrammatically as shown in Fig. \ref{fig:DiagrammaticalF2} or explicitly written as
\begin{align}
    F_2(\bm{k}_1, \bm{k}_2, \tau) &= \int_{\tau_0}^\tau \mathrm{d}s\, C^{(2)}_{1bc}(\bm{k}_{12}, \bm{k}_1, \bm{k}_2, s, \tau)\nonumber \\
    &\times f_b(k_1, s, \tau) f_c(k_2, s, \tau). \label{eq:GeneralF2}
\end{align}
Note that all the kernels derived in this section have yet to be symmetrised w.r.t. exchange of momenta. We now proceed by deriving the kernels at third and fourth order in diagrammatic representation. 
At third order, the diagrammatic representation of the coupling kernels $F_3$ and $G_3$ is given by the two types of diagrams shown in Fig. \ref{fig:I3Diagram}. Finally, at fourth order, we find the diagrams shown in Figs. \ref{fig:W4Diagram}, \ref{fig:IJ4Diagram} and \ref{fig:HK4Diagram}.
%Explicitly, we find that the time-dependent coupling kernels %$F_3$ and $G_3$  are given by $F_3 = I^{(3)}_1 + J^{(3)}_1$ %and $G_3 = I^{(3)}_2 + J^{(3)}_2$ with
%\begin{alignat}{3}
%     I^{(3)}_a &= &\Bigl(&\int_{\tau_0}^\tau %\mathrm{d}s\,C^{(2)}_{abc}(\bm{k}, \bm{k}_1, %\bm{k}_{23}, s, \tau) \times f_b(\bm{k}_1, s, \tau)
%\label{eq:I3}\\
%     & &\times&\int_{\tau_0}^s \mathrm{d}s_1 %C^{(2)}_{cde}(\bm{k}_{23}, \bm{k}_2, \bm{k}_3, s_1, s) %f_d(\bm{k}_2, s_1, s) f_e(\bm{k}_3, s_1, %s)\Bigl)\nonumber\\
%     &+& &\Bigl(\int_{\tau_0}^\tau %\mathrm{d}s\,C^{(2)}_{abc}(\bm{k}, \bm{k}_1, %\bm{k}_{23}, s, \tau) \times f_c(\bm{k}_1, s, %\tau)\nonumber\\
%     & &\times&\int_{\tau_0}^s %\mathrm{d}s_1\,C^{(2)}_{bde}(\bm{k}_{23}, \bm{k}_2, %\bm{k}_3, s_1, s) f_d(\bm{k}_2, s_1, s) f_e(\bm{k}_3, %s_1, s)\Bigl),\nonumber\\ 
%     J^{(3)}_a &= & &\int_{\tau_0}^\tau %\mathrm{d}s\,C^{(3)}_{abcd}(\bm{k}, \bm{k}_1, \bm{k}_2, %\bm{k}_3, s, \tau)
%\label{eq:J3}\\
%     & &\times&f_b(\bm{k}_1, s, \tau)f_c(\bm{k}_2, s, %\tau)f_d(\bm{k}_3, s, \tau)\nonumber.
%\end{alignat}

% --- section: FDM for EdS --- %
\subsection{FDM for EdS cosmology}
We now apply the time-dependent PT framework to the FDM case.  First, we review the Madelung equations \eqref{eq:MadelungEquations} in terms of the conformal time $\tau$
\begin{align}
    \partial_{\tau} \delta + \nabla \cdot \left((1+\delta)\bm{\upsilon}\right) &= 0,\\
    \partial_{\tau} \bm{\upsilon} + \mathcal{H} \bm{\upsilon} + (\bm{\upsilon} \cdot \nabla)\bm{\upsilon} + \nabla \Phi &= -\frac{\hbar^2}{2m^2a^2} \nabla \left(\frac{\left(\Delta \sqrt{1 + \delta}\right)}{\sqrt{1 + \delta}}\right),\\
    \Delta \Phi - \frac{3}{2} \Omega_{m, 0} \mathcal{H}^2 \delta &= 0.
\end{align}
Next, we Fourier transform the Madelung equations. The Fourier transform of the quantum pressure term is computed by the Taylor expansion up to eighth order using \textsc{Mathematica} \citep{Mathematica}. We obtain the same continuity equation as in the CDM case
\begin{equation}
\label{eq:FourierFDMContinuity}
    \partial_{\tau}\delta(\bm{k}) + \theta(\bm{k}) = -  \delta_D(\bm{k} - \bm{k}_{12}) \frac{\bm{k}\cdot\bm{k}_2}{k_2^2} \delta(\bm{k}_1) \theta(\bm{k}_2)
\end{equation}
as well as the Euler equation with quantum pressure corrections
\begin{equation}
\label{eq:FourierFDMEuler}
\begin{split}
    \partial_{\tau}\theta(\bm{k}) &+ \mathcal{H} \theta(\bm{k}) + \frac{3}{2} \Omega_{m, 0} \mathcal{H}^2 \delta(\bm{k}) - \frac{k^4}{4 a^2 m^2} \delta(\bm{k}) =  \\
    &- \delta_D(\bm{k} - \bm{k}_{12}) \left(\frac{1}{2} k^2 \frac{\bm{k}_1\cdot\bm{k}_2}{k_1^2 k_2^2}\theta(\bm{k}_1) \theta(\bm{k}_2)\right)\\ 
    &- \delta_D(\bm{k} - \bm{k}_{12}) \delta(\bm{k}_1) \delta(\bm{k}_2) \frac{k^4}{16a^2m^2}\\
    &\times \left(1 + \frac{\sum_{i}^2 k_i^2}{k^2}\right)\\ 
    &+ \delta_D(\bm{k} - \bm{k}_{123}) \delta(\bm{k}_1)\delta(\bm{k}_2)\delta(\bm{k}_3) \frac{k^4}{32a^2m^2}\\
    &\times \left(1 + \frac{\sum_{i}^3 k_i^2}{k^2} + \frac{\sum_{i, j, i < j}^3 \bm{k}_{ij}^2}{3 k^2}\right)\\
    &- \delta_D(\bm{k} - \bm{k}_{1234}) \delta(\bm{k}_1)\delta(\bm{k}_2)\delta(\bm{k}_3)\delta(\bm{k}_4)\frac{3 k^4}{128a^2m^2}\\
    &\times \left(1 + \frac{2}{3}\frac{\sum_{i}^4 k_i^2}{k^2} + \frac{1}{3}\frac{\sum_{i, j, i < j}^4\bm{k}_{ij}^2}{k^2}\right)\\
    &+ \mathcal{O}(\delta^5)
\end{split}
\end{equation}
where we omitted the time dependence of all quantities and integrations over momenta with repeated indices are understood.
Comparing Eqs. \eqref{eq:FourierFDMContinuity} and \eqref{eq:FourierFDMEuler} to Eq. \eqref{eq:CompactFluidEquations}, we can now determine the mode coupling matrices. Following the convention in \citep{Li2018} we provide them in terms of the time $\eta \equiv 2 \sqrt{a} = H_0 \tau$. The linear mode coupling matrix $\Omega_{ab}$ reads
\begin{align}
\Omega(\bm{k}) &=
\begin{pmatrix}
0 & 1\\
\frac{6}{\eta^2} - \frac{b(k)^2}{\eta^4} & \frac{2}{\eta}
\end{pmatrix}
\end{align}
where the characteristic FDM scale $b(k)$ is defined as
\begin{equation}
    b(k) = \frac{2 k^2 \hbar}{m H_0 \Omega_m^{\frac{1}{2}}}.
\end{equation}
$\Omega_{21}$ vanishes when $k$ equals the comoving quantum Jeans scale $k_J$.
We compute the nonlinear mode coupling matrices $\Gamma_{a, i_1, \ldots, i_n}$ up to $n = 4$:
At second order, the matrices $\Gamma_{a,i_1, i_2}$ include the nonlinear contribution from the convection term as well as the second-order mode coupling from the quantum pressure term:
\begin{equation}
\label{eq:FDMVertexCoupling2}
\begin{split}
\Gamma_{112}(\bm{k}, \bm{k}_1, \bm{k}_2) &= - \alpha(\bm{k}, \bm{k}_2),\\
\Gamma_{121}(\bm{k}, \bm{k}_1, \bm{k}_2) &= - \alpha(\bm{k}, \bm{k}_1),\\
\Gamma_{211}(\bm{k}, \bm{k}_1, \bm{k}_2) &= - \frac{b(k)^2}{\eta^4}\frac{1}{4}\left(1 + \frac{\sum_{i}^2 k_i^2}{k^2}\right),\\
\Gamma_{222}(\bm{k}, \bm{k}_1, \bm{k}_2) &= - \beta(\bm{k}, \bm{k}_1, \bm{k}_2)\\
\end{split}
\end{equation}
where $\alpha(\bm{k}, \bm{k}_1)$ and $\beta(\bm{k}, \bm{k}_1, \bm{k}_2)$ are defined as
\begin{align}
     \alpha(\bm{k}_1, \bm{k}_2) &\equiv     \begin{cases}
       \frac{\bm{k}_{12}\cdot\bm{k}_1}{k_1^2}, &\quad \text{if } k_1 \neq 0,\\
       0 ,&\quad \text{otherwise},\\ 
     \end{cases} \label{eq:AlphaModeCoupling}\\
    \beta(\bm{k}_1, \bm{k}_2) &\equiv     \begin{cases}
        \frac{k_{12}^2(\bm{k}_1\cdot\bm{k}_2)}{2 k_1^2 k_2^2}, &\quad \text{if } k_1 \neq 0 \text{ and } k_2 \neq 0,\\
       0, &\quad \text{otherwise},\\ 
     \end{cases} \label{eq:BetaModeCoupling}
\end{align}
and all other components vanish. All higher-order contributions to $\Gamma_{a, i_1\ldots i_n}$ stem solely from the quantum pressure term and therefore represent self-interactions of the density field:
\begin{align}
    \Gamma_{2111} &(\bm{k}, \bm{k}_1, \bm{k}_2, \bm{k}_3) =\nonumber\\ 
    &+\frac{b(k)^2}{\eta^4}\frac{1}{8}\left(1 + \frac{\sum_{i}^3 k_i^2}{k^2} + \frac{\sum_{i, j, i < j}^3 \bm{k}_{ij}^2}{3 k^2}\right), \label{eq:FDMVertexCoupling3}\\
    \Gamma_{21111} &(\bm{k}, \bm{k}_1, \bm{k}_2, \bm{k}_3, \bm{k}_4)= \nonumber \\
    &- \frac{b(k)^2}{\eta^4}\frac{3}{32}\left(1 + \frac{2}{3}\frac{\sum_{i}^4 k_i^2}{k^2} + \frac{1}{3}\frac{\sum_{i, j, i < j}^4\bm{k}_{ij}^2}{k^2}\right)\label{eq:FDMVertexCoupling4},
\end{align}
and all other contributions at third and fourth order vanish. Next is the Green's function for the linear growth equation \eqref{eq:FDMLinearGrowthScaleFactor}. 
Using the analytical growth factors give in Eq. \eqref{eq:FDMLinearGrowthASolution}, we find the Green's function
\begin{equation}
    \label{eq:FDMGreensFunction}
    \begin{split}
    &G_k(s, \eta) =\\
    &\frac{\pi s^{3/2}}{2\eta^{1/2}} \left[J_{5/2}\left(\frac{b(k)}{s}\right) J_{-5/2}\left(\frac{b(k)}{\eta}\right) - J_{5/2}\left(\frac{b(k)}{\eta}\right) J_{-5/2}\left(\frac{b(k)}{s}\right) \right].
    \end{split}
\end{equation}
Note that the Green's function is free of divergences because the denominators of the growth functions at the initial time cancel. Using equation \eqref{eq:C1CouplingKernels}, the vertex couplings are given by
\begin{equation}
\begin{split}
C^{(2)}_{1bc}&(\bm{k}, \bm{k}_1, \bm{k}_2, s, \eta) =\\
&-\Gamma_{1bc}(\bm{k}, \bm{k}_1, \bm{k}_2, s)\partial_s G_k(s, \eta)\\
&- \Bigl(\Gamma_{2bc}(\bm{k}, \bm{k}_1, \bm{k}_2, s) - \frac{2}{s}\Gamma_{1bc}(\bm{k}, \bm{k}_1, \bm{k}_2, s)\Bigl) G_k(s, \eta)
\end{split}
\end{equation}
at second order as well as 
\begin{equation}
\begin{split}
C^{(3)}_{1111}&(\bm{k}, \bm{k}_1, \bm{k}_2, \bm{k}_3, s, \eta) =\\
&-\Gamma_{2111}(\bm{k}, \bm{k}_1, \bm{k}_2, \bm{k}_3) G_k(s, \eta),
\end{split}
\end{equation}
\begin{equation}
\begin{split}
C^{(4)}_{11111}&(\bm{k}, \bm{k}_1, \bm{k}_2, \bm{k}_3, \bm{k}_4, s, \eta) =\\
&-\Gamma_{21111}(\bm{k}, \bm{k}_1, \bm{k}_2, \bm{k}_3, \bm{k}_4) G_k(s, \eta)
\end{split}
\end{equation}
at third and fourth order where all other vertex couplings vanish. This is because above order two there are only self-interactions of the density field stemming from the Taylor expansion of the quantum pressure term. This simplifies the computation and the symmetrisation of the kernels $F_3^{FDM}$ and $F_4^{FDM}$ significantly. A more detailed discussion can be found in appendix \ref{appendix:KernelSymmetrisation}.

Since $\gamma_{abc}(\bm{k}, \bm{k}_1, \bm{k}_2, s, \eta)$ is symmetric w.r.t. exchange of $\bm{k}_1$ and $\bm{k}_2$, $C^{(2)}_{1bc}(\bm{k}, \bm{k}_1, \bm{k}_2, s, \eta)$ inherits this property. As a consequence $F_2^{FDM}$ as given by Eq. \eqref{eq:GeneralF2}  is already symmetric under exchange of $\bm{k}_1$ and $\bm{k}_2$.

% --- section: explicit symmetrisation --- %
\section{explicit symmetrisation of FDM PT kernels}
\label{appendix:KernelSymmetrisation}
In this section, we explicitly derive the symmetrised FDM PT kernels $F_3^{(s)}$ and $F_4^{(s)}$ by making use of the symmetries of the vertex couplings in FDM. We start with the third-order kernel $F_3$ depicted in Fig. \ref{fig:I3Diagram}. It can be explicitly expressed as $F_3 = I^{(3)}_1 + J^{(3)}_1$ with
\begin{alignat}{3}
     I^{(3)}_a &= &\Bigl(&\int_{\eta_0}^\eta \mathrm{d}s\,C^{(2)}_{abc}(\bm{k}, \bm{k}_1, \bm{k}_{23}, s, \eta) \times f_b(\bm{k}_1, s, \eta)
\label{eq:I3}\\
     & &\times&\int_{\eta_0}^s \mathrm{d}s_1 C^{(2)}_{cde}(\bm{k}_{23}, \bm{k}_2, \bm{k}_3, s_1, s)\nonumber\\
     & & &\times f_d(\bm{k}_2, s_1, s) f_e(\bm{k}_3, s_1, s)\Bigl)\nonumber\\
     &+ & &\Bigl(\int_{\eta_0}^\eta \mathrm{d}s\,C^{(2)}_{abc}(\bm{k}, \bm{k}_1, \bm{k}_{23}, s, \eta) \times f_c(\bm{k}_1, s, \eta)\nonumber\\
     & &\times&\int_{\eta_0}^s \mathrm{d}s_1\,C^{(2)}_{bde}(\bm{k}_{23}, \bm{k}_2, \bm{k}_3, s_1, s) f_d(\bm{k}_2, s_1, s) f_e(\bm{k}_3, s_1, s)\Bigl),\nonumber\\ 
     J^{(3)}_a &= & &\int_{\eta_0}^\eta \mathrm{d}s\,C^{(3)}_{abcd}(\bm{k}, \bm{k}_1, \bm{k}_2, \bm{k}_3, s, \eta)
\label{eq:J3}\\
     & &\times& f_b(\bm{k}_1, s, \eta)f_c(\bm{k}_2, s, \eta)f_d(\bm{k}_3, s, \eta)\nonumber.
\end{alignat}
Since $\Gamma_{1bc}$ = $\Gamma_{1cb}$, as follows from Eq. \eqref{eq:FDMVertexCoupling2}, Eq. \eqref{eq:I3} reduces to 
\begin{equation}
\begin{split}
     I^{(3)}_1&(\bm{k}_1, \bm{k}_2, \bm{k}_3, \eta) =\\
     &2 \int_{\eta_0}^\eta \mathrm{d}s\,\Gamma^{(2)}_{1bc}(\bm{k}, \bm{k}_1, \bm{k}_{23}, s, \eta) \int_{\eta_0}^s \mathrm{d}s_1 \Gamma^{(2)}_{cde}(\bm{k}_{23}, \bm{k}_2, \bm{k}_3, s_1, s)\\
     &\times f_b(\bm{k}_1, s, \eta) f_d(\bm{k}_2, s_1, s) f_e(\bm{k}_3, s_1, s).
\end{split}
\end{equation}
Since $I^{(3)}_1$ already enjoys symmetry under $\bm{k}_2 \leftrightarrow \bm{k}_3$ and $\bm{k}_{23} \leftrightarrow \bm{k}_1$, it can be symmetrised as
\begin{equation}
\begin{split}
    I^{(3,s)}_1\bigl(\bm{k}_1, \bm{k}_2, \bm{k}_3, \eta) =  &\frac{1}{3}(I_1^{(3)}(\bm{k}_1, \bm{k}_2, \bm{k}_3, \eta)\\
    &+ I_1^{(3)}(\bm{k}_2, \bm{k}_1, \bm{k}_3, \eta)\\
    &+I_1^{(3)}(\bm{k}_3, \bm{k}_2, \bm{k}_1, \eta)\bigl).
\end{split}
\end{equation}
Further, $J^{(3)}_1$ is already symmetric under exchange of momenta because the only nonvanishing element of the vertex coupling $\Gamma$ at fourth order is the $(1111)$-component given in Eq. \eqref{eq:FDMVertexCoupling3} that stems from the quantum pressure term and is already symmetric under exchange of momenta $J^{(3, s)}_1(\bm{k}_1, \bm{k}_2, \bm{k}_3, \eta) = J^{(3)}_1(\bm{k}_1, \bm{k}_2, \bm{k}_3, \eta)$.
We conclude $F_3^{(s)} = J^{(3, s)}_1(\bm{k}_1, \bm{k}_2, \bm{k}_3, \eta) + I^{(3,s)}_1(\bm{k}_1, \bm{k}_2, \bm{k}_3, \eta)$.
At fourth order, we distinguish five contributions to the coupling kernel:
\begin{equation}
\begin{split}
    \Psi_2^{(4)}(\bm{k}, \eta) = &\delta(\bm{k}_1, \eta) \delta(\bm{k}_2, \eta) \delta(\bm{k}_3, \eta) \delta(\bm{k}_4, \eta) \delta_D(\bm{k} - \bm{k}_{1234})\\
    &\bigl(J_4(\bm{k}_1, \bm{k}_2, \bm{k}_3, \bm{k}_4, \eta) \\
    &+ K_4(\bm{k}_1, \bm{k}_2, \bm{k}_3, \bm{k}_4, \eta)\\
    &+ H_4(\bm{k}_1, \bm{k}_2, \bm{k}_3, \bm{k}_4, \eta) \\
    &+ W_4(\bm{k}_1, \bm{k}_2, \bm{k}_3, \bm{k}_4, \eta) \\
    &+ I_4(\bm{k}_1, \bm{k}_2, \bm{k}_3, \bm{k}_4, \eta)\bigl). \\
\end{split}
\end{equation}
The contribution $W_4$ corresponds to the diagram in Fig. \ref{fig:W4Diagram} and takes the simple form
\begin{equation}
\begin{split}
    W_{4}&(\bm{k}_1, \bm{k}_2, \bm{k}_3, \bm{k}_4, \eta) =\\ &\int_{\eta_0}^\eta \Gamma^{(4)}_{11111}(\bm{k}_{1234}, \bm{k}_1, \bm{k}_2, \bm{k}_3, \bm{k}_4, s, \eta)\\
    &\times f_1(\bm{k}_1, s, \eta)f_1(\bm{k}_2, s, \eta)f_1(\bm{k}_3, s, \eta)f_1(\bm{k}_4, s, \eta).
\end{split}
\end{equation}
Like in the third-order case, $\Gamma^{(4)}_{11111}$ stems from the quantum pressure term and is therefore symmetric under exchange momenta. As a consequence $W_{4}$ is symmetric under exchange of momenta. 
The contribution $I_4$ corresponds to the left diagram in Fig. \ref{fig:IJ4Diagram}:
\begin{equation}
\begin{split}
&I_{4}(\bm{k}_1, \bm{k}_2, \bm{k}_3, \bm{k}_4, \eta) =\\ &\int_{\eta_0}^\eta \mathrm{d}s \, \Gamma^{(2)}_{1bc}(\bm{k}_{1234}, \bm{k}_{12}, \bm{k}_{34}, s, \eta)\\
&\times \int_{\eta_0}^s \mathrm{d}s_1  \int_{\eta_0}^s \mathrm{d}s_2 \Gamma^{(2)}_{bde}(\bm{k}_{12}, \bm{k_{1}}, \bm{k_{2}}, s_1, s) \Gamma^{(2)}_{cfg}(\bm{k}_{34}, \bm{k_{3}}, \bm{k_{4}}, s_2, s)\\ 
&\times f_d(\bm{k}_1, s_1, s) f(\bm{k}_2, s_1, s)   f_f(\bm{k}_3, s_2, s) f_g(\bm{k}_4, s_2, s)
\end{split}
\end{equation}
which can be symmetrised by using the fact that it is already invariant under $\bm{k}_1 \leftrightarrow \bm{k}_2$; $\bm{k}_3 \leftrightarrow \bm{k}_4$; $\bm{k}_1 , \bm{k}_2 \leftrightarrow \bm{k}_3, \bm{k}_4$:
\begin{equation}
    \begin{split}
        I^s_4(\bm{k}_1, \bm{k}_2, \bm{k}_3, \bm{k}_4, \eta) = &\frac{1}{3} \bigl(I_4(\bm{k}_1, \bm{k}_2, \bm{k}_3, \bm{k}_4, \eta) \\
        &+I_4(\bm{k}_1, \bm{k}_3, \bm{k}_2, \bm{k}_4, \eta)\\ &+I_4(\bm{k}_1, \bm{k}_4, \bm{k}_3, \bm{k}_2, \eta)\bigl).
    \end{split}
\end{equation}
The contribution $J_4$ corresponds to the right diagram in Fig. \ref{fig:IJ4Diagram}:
\begin{equation}
\begin{split}
J_4&(\bm{k}_1, \bm{k}_2, \bm{k}_3, \bm{k}_4, \eta) =\\
&4 \int_{\eta_0}^\eta \mathrm{d}s W_{2bc}(\bm{k}_{1234}, \bm{k_{1}}, \bm{k}_{234}, s, \eta) f_b(\bm{k}_1, s) \\
&\times \int_{\eta_0}^s \mathrm{d}s_1 W_{cde}(\bm{k}_{234}, \bm{k}_2, \bm{k}_{34}, s_1, s)\\
&\times \int_{\eta_0}^{s_1} \mathrm{d}s_2 W_{efg}(\bm{k}_{34}, \bm{k}_3, \bm{k}_4, s_2, s_1)\\
&\times f_d(\bm{k}_2, s_1, s) f_f(\bm{k}_3, s_2, s_1) f_g(\bm{k}_4, s_2, s_1)
\end{split}
\end{equation}
and enjoys symmetry under $\bm{k}_1 \leftrightarrow \bm{k}_{234}$; $\bm{k}_2 \leftrightarrow \bm{k}_{34}$; $\bm{k}_{3} \leftrightarrow \bm{k}_4$. The factor $4$ follows from the three additional permutations of the diagram. Its fully symmetric form is given by
\begin{equation}
\begin{split}
J_4^s&(\bm{k}_1, \bm{k}_2, \bm{k}_3, \bm{k}_4, \eta) =\\
&\frac{1}{12}
 (J_4(\bm{k}_1, \bm{k}_2, \bm{k}_3, \bm{k}_4, \eta) 
+ J_4(\bm{k}_1, \bm{k}_3, \bm{k}_2, \bm{k}_4, \eta)\\
&+ J_4(\bm{k}_1, \bm{k}_4, \bm{k}_3, \bm{k}_2, \eta) 
+ J_4(\bm{k}_2, \bm{k}_3, \bm{k}_1, \bm{k}_4, \eta)\\
&+ J_4(\bm{k}_2, \bm{k}_4, \bm{k}_1, \bm{k}_3, \eta) 
+ J_4(\bm{k}_3, \bm{k}_4, \bm{k}_1, \bm{k}_2, \eta)\\
&+ J_4(\bm{k}_2, \bm{k}_1, \bm{k}_3, \bm{k}_4, \eta) 
+ J_4(\bm{k}_3, \bm{k}_1, \bm{k}_2, \bm{k}_4, \eta)\\
&+ J_4(\bm{k}_4, \bm{k}_1, \bm{k}_3, \bm{k}_2, \eta) 
+ J_4(\bm{k}_3, \bm{k}_2, \bm{k}_1, \bm{k}_4, \eta)\\
&+ J_4(\bm{k}_4, \bm{k}_2, \bm{k}_1, \bm{k}_3, \eta) 
+ J_4(\bm{k}_4, \bm{k}_3, \bm{k}_1, \bm{k}_2, \eta)).\\
\end{split}
\end{equation}
The contribution $K_4$ corresponds to the left diagram in Fig. \ref{fig:HK4Diagram}:
\begin{equation}
\begin{split}
 K_4&(\bm{k}_1, \bm{k}_2, \bm{k}_3, \bm{k}_4, \eta) = \\
 &2 \int_{\eta_0}^\eta \mathrm{d}s\, \Gamma^{(2)}_{1b2}(\bm{k}_{1234}, \bm{k_{1}}, \bm{k}_{234}, s, \eta) f_b(\bm{k}_1, s, \eta)\\
 &\times\int_{\eta_0}^{s} \mathrm{d}s_1 (\Gamma^{(3)}_{1111}(\bm{k}_{234}, \bm{k}_2, \bm{k}_3, \bm{k}_4, s_1, s)\\
 &\times f_1(\bm{k}_2, s_1, \eta)f_1(\bm{k}_3, s_1, s)f_1(\bm{k}_4, s_1, s)).
 \end{split}
\end{equation}
where the additional permutation together with the symmetry properties of $\Gamma^{(2)}_{abc}$ give the factor of $2$. Again, we recognise that the expression is invariant under permutations of $\bm{k}_2, \bm{k}_3$ and $\bm{k}_4$. Its symmetric form is given by
\begin{equation}
\begin{split}
 K_4^{(s)}&(\bm{k}_1, \bm{k}_2, \bm{k}_3, \bm{k}_4, \eta) =\\
 &\frac{1}{4} \bigl[K_4(\bm{k}_1, \bm{k}_2, \bm{k}_3, \bm{k}_4, \eta) +  K_4(\bm{k}_2, \bm{k}_1, \bm{k}_3, \bm{k}_4, \eta)\\
 &+  K_4(\bm{k}_3, \bm{k}_2, \bm{k}_1, \bm{k}_4, \eta) + K_4(\bm{k}_4, \bm{k}_2, \bm{k}_3, \bm{k}_1. \eta)\bigl]
 \end{split}
\end{equation}
The last remaining contribution $H_4$ corresponds to the right diagram in Fig. \ref{fig:HK4Diagram}:
\begin{equation}
\begin{split}
 H_4&(\bm{k}_1, \bm{k}_2, \bm{k}_3, \bm{k}_4, \eta) =\\ &\int_{\eta_0}^\eta \Gamma^{(3)}_{1111}(\bm{k}_{1234}, \bm{k}_{14}, \bm{k}_2, \bm{k}_3, s, \eta) f_1(\bm{k}_2, s, \eta)f_1(\bm{k}_3, s, \eta)\\
&\times \int_{\eta_0}^s \mathrm{d}s_1 \Gamma^{(2)}_{1ef}(\bm{k}_{14}, \bm{k}_1, \bm{k}_4, s_1, \eta)f_e(\bm{k}_1, s_1, s)f_f(\bm{k}_4, s_1, s)\\
&+ \text{permutations with } \bm{k}_1 \leftrightarrow \bm{k}_2, \bm{k}_1 \leftrightarrow \bm{k}_3.
\end{split}
\end{equation}
A symmetrised form of $H_4$ is given by  
\begin{equation}
\begin{split}
    H_4^{(s)}&(\bm{k}_1, \bm{k}_2, \bm{k}_3, \bm{k}_4, \eta) = \\
    &\frac{1}{6} \bigl[(H_4(\bm{k}_1, \bm{k}_2, \bm{k}_3, \bm{k}_4, \eta) + H_4(\bm{k}_1, \bm{k}_3, \bm{k}_2, \bm{k}_4, \eta)\\
    &+H_4(\bm{k}_1, \bm{k}_4, \bm{k}_3, \bm{k}_2, \eta) +H_4(\bm{k}_2, \bm{k}_3, \bm{k}_1, \bm{k}_4, \eta)\\
    &+H_4(\bm{k}_2, \bm{k}_4, \bm{k}_1, \bm{k}_3, \eta) +H_4(\bm{k}_3, \bm{k}_4, \bm{k}_1, \bm{k}_2, \eta)\bigl].
\end{split}
\end{equation}
It follows that 
\begin{equation}
\begin{split}
    F_4^{(s)}&(\bm{k}_1, \bm{k}_2, \bm{k}_3, \bm{k}_4, \eta) =\\ &\bigl[J_4^{(s)}(\bm{k}_1, \bm{k}_2, \bm{k}_3, \bm{k}_4, \eta)
    + H_4^{(s)}(\bm{k}_1, \bm{k}_2, \bm{k}_3, \bm{k}_4, \eta)\\
    &+ W_4^{(s)}(\bm{k}_1, \bm{k}_2, \bm{k}_3, \bm{k}_4, \eta)
    + I_4^{(s)}(\bm{k}_1, \bm{k}_2, \bm{k}_3, \bm{k}_4, \eta)\\
    &+ K_4^{(s)}(\bm{k}_1, \bm{k}_2, \bm{k}_3, \bm{k}_4, \eta)\bigl].
\end{split}
\end{equation}

% --- section: computation of lensing integrals --- %
\section{computation of lensing integrals}
\label{appendix:InfraredSafeIntegrands}
In the following, we describe how to perform the loop-level, line-of-sight, signal-to-noise and $\chi^2$ integrations.  In the beginning, we consider the numerical loop-level integrations to the matter spectra. The relevant integrals are given by Eqs.~\eqref{eq:PowerSpectrumCorrections} and~\eqref{eq:BispectrumCorrections}. Both integrands exhibit IR divergences. \cite{Carrasco2013} argue that if succeeds in rewriting the integrands such that the leading IR divergences cancel, all sub-leading IR divergences are guaranteed to cancel as well. They provide such an expression for the power spectrum:
\begin{equation}
\label{eq:IRSafePowerSpectrum}
\begin{split}
    &P^{(1)}_\mathrm{IR-safe} = \int \frac{\mathrm{d}^3q}{(2\pi)^3} \Bigl[6 F_3^{(s)}(\bm{k}, \bm{q}, -\bm{q}) P^{(0)}(k) P^{(0)}(q)\\
    &+ 2 [F_2^{(s)}(\bm{k} - \bm{q}, \bm{q})]^2 P^{(0)}(|\bm{k} - \bm{q}|) P^{(0)}(q) \theta(|\bm{k} - \bm{q}| - q)\\
    &+ 2 [F_2^{(s)}(\bm{k} + \bm{q}, -\bm{q})]^2 P^{(0)}(|\bm{k} + \bm{q}|) P^{(0)}(q) \theta(|\bm{k} + \bm{q}| - q)\Bigl].
\end{split}
\end{equation}
Eq.~\eqref{eq:IRSafePowerSpectrum} can be applied to the FDM case since the momentum dependence of the FDM mode coupling functions does not introduce any additional divergences compared to the CDM case. The same holds true for the IR safe version of the bispectrum corrections that are divergent just as in the case of the one-loop power spectrum.

\cite{Baldauf2014} provide an IR-safe expression for the bispectrum contributions: The integrands of the contributions $B_{321}^{II}$ and $B_{411}$ as defined in Eq. \eqref{eq:BispectrumCorrections} only exhibit divergences at $q = 0$. The integrand of $B_{321}^{II}$ exhibits a divergence at $\bm{q} = \bm{k}_2$ which can be mapped to a divergence at $q = 0$ by writing
\begin{equation}
    \begin{split}
        \int_q \tilde{b}_{321}^I =& \int_{q < |\bm{k_2 - q}|} \mathrm{d}^3 q\: b_{321}^{I}(\bm{q}, \bm{k}_2, \bm{k}_3)\\
        &+ \int_{q \geq |\bm{k_2 - q}|} \mathrm{d}^3\bm{q}\; b_{321}^{I}(\bm{q}, \bm{k}_2, \bm{k}_3)\\
        &+ 5 \text{ permutations}\\
        =&~6 \int \mathrm{d}^3 q\: b_{321}^{I}(\bm{q}, \bm{k}_2, \bm{k}_3) \theta(|\bm{k}_2 - \bm{q}| - q)\\
        &+ b_{321}^{I}(-\bm{q}, \bm{k}_2, \bm{k}_3) \theta(|\bm{k}_2 + \bm{q}| - q)\\
        &+ 5 \text{ permutations}.
    \end{split}
\end{equation}
Similarly, one finds the following expression for the integrand of $B_{222}$:
\begin{equation}
    \begin{split}
        \int_q \tilde{b}_{222} =& \frac{1}{2} \int \mathrm{d}^3 q\: \Bigl([ b_{222}(\bm{q}, \bm{k}_1, \bm{k}_2)\\
        &\times \theta(|\bm{k_1} + \bm{q}| - q) \theta(|\bm{k}_2 - \bm{q}| - q) \\
        &+ b_{222}(-\bm{q}, \bm{k}_1, \bm{k}_2)\\
        &\times \theta(|\bm{k}_1 - \bm{q}| - q) \theta(|\bm{k}_2 + \bm{q}| - q) ]\\
        & +  [\bm{k}_1 \leftrightarrow \bm{k}_3] + [\bm{k}_2 \leftrightarrow \bm{k}_3]\Bigl).
    \end{split}
\end{equation}
The full one-loop bispectrum in a form where all IR-divergences cancel is then given by
\begin{equation}
\begin{split}
    B_{SPT} &= B_{112} + \int_q \tilde{b}_{222} + \tilde{b}_{321}^I + [b_{321}^{II} + 5 \text{ perm.}]\\
    &+ [b_{411} + 2 \text{ cyclic perm.}].
\end{split}
\end{equation}

We integrate the above expressions for the IR-safe loop-level corrections to the matter power spectrum and bispectrum using the \textsc{CUBA}-library \citep{Hahn2004}. For the loop-corrections to the CDM power spectrum, the four integration algorithms Vegas, Cuhre, Divonne and Suave all provide consistently good results. The FDM integrations are much more problematic because they involve time integrations for the PT kernels in addition to the the momentum integrations. In addition, the PT kernels and the growth factor both exhibit strong oscillations for large $k$. As a consequence, integrations take as long as $5$ CPU hours for the Monte Carlo integration of the one-loop contribution to the FDM bispectrum for a single triangle configuration with a relative error of $10$ \%. For computing the power spectrum loop corrections, we use the Vegas algorithm that stores the integrand structure between different integration runs and thus accelerates integration for different $k$ values. For the bispectrum loop corrections, we use the Divonne algorithm. We tested our integration routines by comparison against analytical solutions in the CDM case derived by \cite{Makino1992}. Further, we numerically verified that the FDM PT kernels reduced to the correct analytical CDM expressions for small $k$. Moreover, we implemented the PT code independently in Python and C++ and crosschecked results.

For the numerical integration of the line-of-sight integrals, we combine the PT kernel integrations, the loop integrations and the line-of-sight integrations into higher-dimensional integrals that we integrate using the CUBA-library. This proves advantageous since the line-of-sight integrations smooth oscillations and make the loop integrations more numerically tractable. The signal-to-noise sums and $\chi^2$-functionals are also computed as integrals using the \textsc{CUBA}-library. This leads to difficulties during the evaluation of the FDM loop-level bispectrum signal-to-noise sums and $\chi^2$-functionals. The corresponding PT integrals are six-dimensional and the line-of-sight integral to compute the lensing bispectra makes them seven-dimensional. The computation of a loop-level signal-to-noise ratio therefore requires a three-dimensional integral of a seven-dimensional integral which does not compute in our tests. As an alternative, we combine the signal-to-noise integrals and the perturbative integrals to obtain a seventeen-dimensional integral which again proves computationally intractable. In the CDM case, this approach yields an eleven-dimensional integral that can be evaluated numerically. We use this approach to estimate the loop-level bispectrum signal-to-noise ratios and $\chi^2$-functionals.

% --- section: N-body --- %
\section{N-body simulations}
\label{appendix:NBody}
\begin{figure}
\centering
\includegraphics[width=0.47\textwidth]{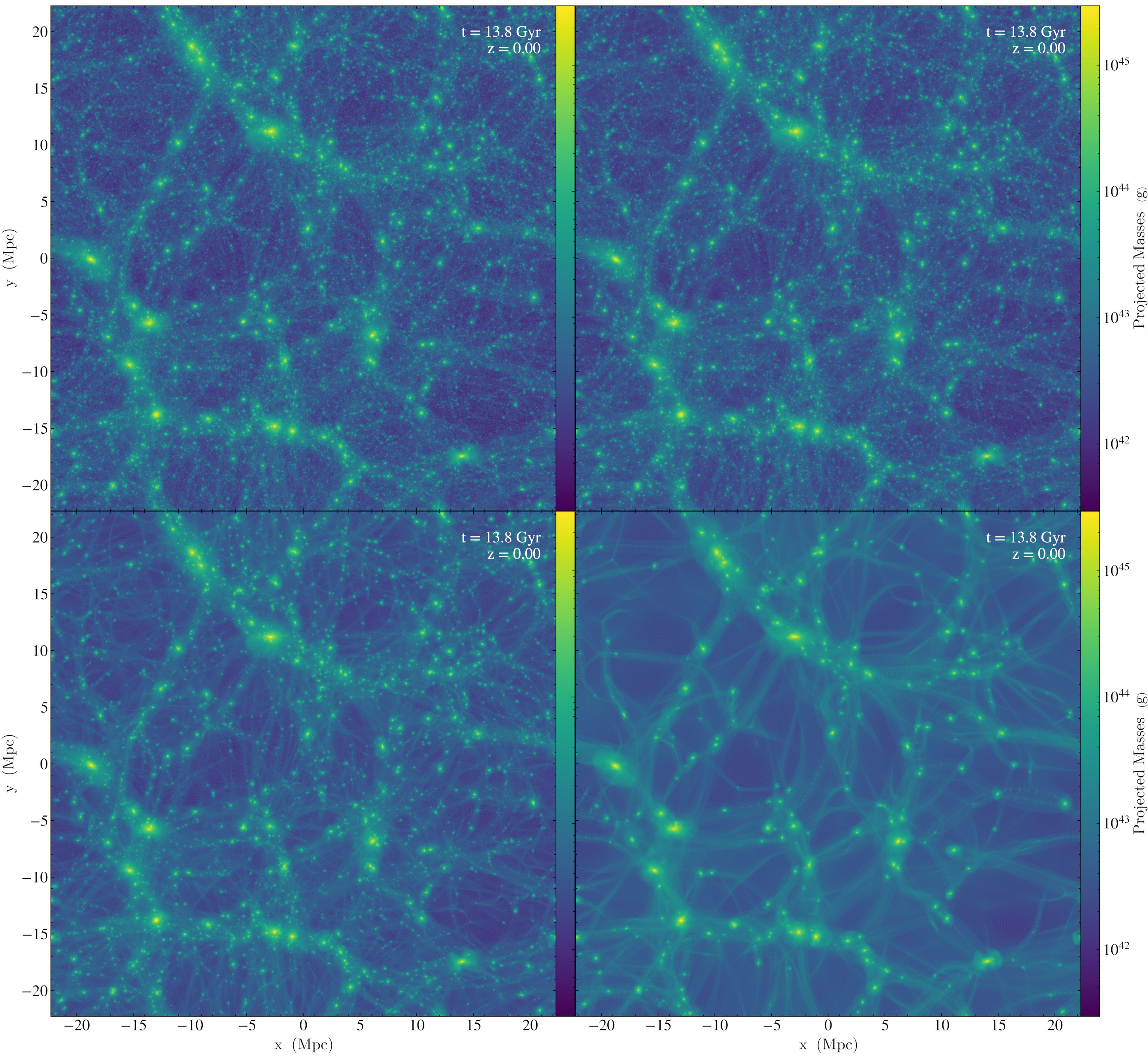}
\caption{\highlighttext{Mass projections of $N$-body simulations for CDM (upper left) and FDM IC with $m=10^{-21}$ eV (upper right), $m=10^{-22}$ eV (lower left) and $m=10^{-23}$ eV (lower right) in $L = 30$ Mpc/$h$ box with $512^3$ particles at $z=0$.}}
\label{fig:NBodyProj}
\end{figure}

\highlighttext{
We run a total of $8$ simulations with $N = 512^3$ particles with the box sizes $L = 30$ Mpc/$h$ and $L = 256$ Mpc/$h$. They serve as an estimate of the full nonlinear evolution of the CDM and FDM models. The initial conditions are created from the respective CAMB and axionCAMB spectra using the initial condition generator \textsc{MUSIC} \citep{Hahn2011}. The simulations themselves are run from $z=99$ to $z=0$ with a total of $99$ snapshots and a gravitational softening length of a comoving $14$ kpc/$h$ and $1.7$ kpc/$h$ respectively. We use the tool GenPK to compute the matter power spectra from the snapshots \citep{GenPK}. The lensing spectra are integrated using a 2D spline fitted to the matter spectra in time and k-space. 
Fig. \ref{fig:NBodyProj} shows the mass projections of the simulations used for the lensing predictions. The $m=10^{-23}$ eV plot in the lower right corner highlights another problem that $N$-body simulations with a suppression of small-scale initial power suffer from: the formation of spurious halos arranged like beads on a string along the filaments, for instance \citep{Schive2015}. These spurious collapsed objects likely also lead to an overestimation of small-scale power at late-times in the $N$-body runs. 

}

\bsp
\label{lastpage}
\end{document}